\documentclass[12pt]{article}

\usepackage[british]{babel}
\usepackage[utf8]{inputenc}
\usepackage[T1]{fontenc}
\usepackage{graphicx}
\usepackage{amsmath}
\usepackage{amsfonts}
\usepackage{amsbsy}
\usepackage{bm}
\usepackage{longtable}

\usepackage{amsthm}

\theoremstyle{plain}
\newtheorem{theorem}{Theorem}

\theoremstyle{remark}
\newtheorem{remark}{Remark}

\DeclareMathOperator{\tr}{tr}

\begin{document}

\title{A random field formulation of Hooke's law in all elasticity classes\thanks{This material is based upon the research partially supported by the NSF under grants CMMI-1462749 and IP-1362146 (I/UCRC on Novel High Voltage/Temperature Materials and Structures.}}

\author{Anatoliy Malyarenko\thanks{M\"{a}lardalen University, Sweden} \and Martin Ostoja-Starzewski\thanks{University of Illinois at Urbana-Champaign, USA}}

\date{\today}

\maketitle

\begin{abstract}
For each of the $8$ symmetry classes of elastic materials, we consider a
homogeneous random field taking values in the fixed point set $\mathsf{V}$ of the corresponding class, that is isotropic with respect to the natural orthogonal representation of a group lying between the isotropy group of the class and its normaliser. We find the general form of the correlation tensors of orders $1$ and $2$ of such a field, and the field's spectral expansion.
\end{abstract}

\tableofcontents
\listoffigures
\listoftables

\section{Introduction}

Microstructural randomness is present in just about all solid materials.
When dominant (macroscopic) length scales are large relative to microscales,
one can safely work with deterministic homogeneous continuum models.
However, when the separation of scales does not hold and spatial randomness
needs to be accounted for, various concepts of continuum mechanics need to
be re-examined and new methods developed. This involves: (1) being able to
theoretically model and simulate any such randomness, and (2) using such
results as input into stochastic field equations. In this paper, we work in
the setting of linear elastic random media that are statistically wide-sense
homogeneous and isotropic.

\begin{figure}[htbp]
  \centering
  \includegraphics[width=\columnwidth]{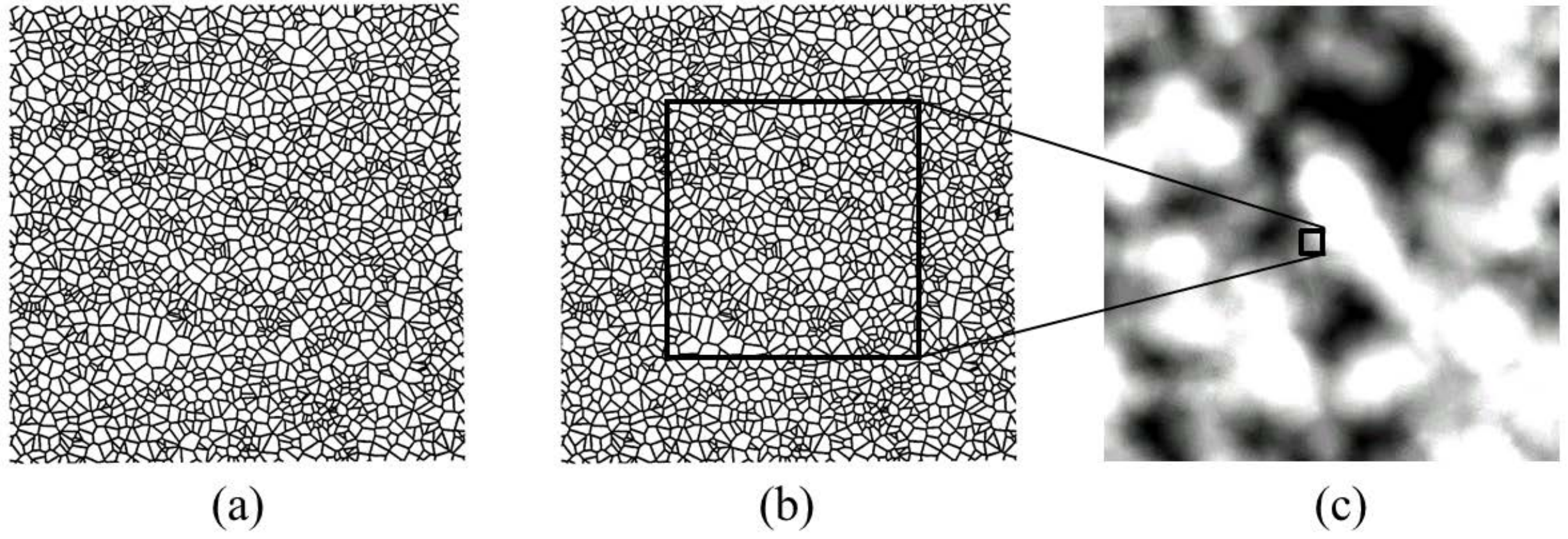}
  \caption{(a) A realisation of a Voronoi tesselation (or mosaic); (b) placing a mesoscale window leads, via upscaling, to a mesoscale
  random continuum approximation in (c).}\label{fig:1}
\end{figure}

Regarding the modelling motivation (1), two basic issues are considered in
this study: (i) type of anisotropy, and (ii) type of correlation structure.
Now, with reference to Fig.~\ref{fig:1} showing a planar Voronoi tessellation of $E^2$ which serves as a planar geometric model of a polycrystal
(although the same arguments apply in $E^3$), each cell may be
occupied by a differently oriented crystal, with all the crystals belonging
to any specific crystal class. The latter include:

\begin{itemize}
\item transverse isotropy modelling, say, sedimentary rocks at long
wavelengths;

\item tetragonal modelling, say, wulfenite (PbMoO$_{\text{4}}$);

\item trigonal modelling, say, dolomite (CaMg(CO$_{\text{3}}$)$_{\text{2}}$);

\item orthotropic modelling, say wood or orthoclase feldspar;

\item triclinic modelling, say, microcline feldspar.
\end{itemize}

Thus, we need to be able to model rank~$4$ tensor random fields,
point-wise taking values in any crystal class. While the crystal orientations
from grain to grain are random, in general they are not spatially independent of each other~--- the assignment of crystal properties over the tessellation is not white noise. This is precisely where the two-point characterisation of the random field of elasticity tensor is needed, so as to account for any mathematically admissible correlation structures as dictated by the statistically wide-sense homogeneous and isotropic assumption. A specific correlation can then be fitted to physical measurements.

Regarding the modelling motivation (1), it may also be of interest to work
with a mesoscale random continuum approximation defined by placing a
mesoscale window at any spatial position as shown in Fig.~\ref{fig:1}(b). Clearly, the larger is the mesoscale window, the weaker are the random fluctuations in the mesoscale elasticity tensor: this is the trend to homogenise the material when upscaling from a statistical volume element (SVE) to a representative volume element (RVE), e.g. \cite{MR2341287,MR2590496}. A simple paradigm of this upscaling, albeit only in terms of a scalar random field, is the opacity of a sheet of paper held against light: the further away is the sheet from our eyes, the more homogeneous it appears. Similarly, in the case of upscaling of elastic properties, on any finite scale there is (almost surely) an anisotropy and this anisotropy, with mesoscale increasing, tends to zero hand-in-hand with the fluctuations and it is in the infinite mesoscale limit (i.e. RVE) that material isotropy is obtained as a consequence of the statistical isotropy.

Regarding the motivation (2) of this study, i.e. input of elasticity random
fields into stochastic field equations, there are two principal routes:
stochastic partial differential equations (SPDE) and stochastic finite
elements (SFE). The classical paradigm of SPDE \cite{MR3308418} can be written in terms of the anti-plane elastostatics (with $u\equiv u_{3}$):
\begin{equation}\label{eq:15new}
\bm{\nabla}\cdot\left(C\left(\mathbf{x},\omega\right)\bm{\nabla}u\right) =0,\qquad\mathbf{x}\in E^2,\quad \omega\in\Omega
\end{equation}
with $C\left(\mathbf{\cdot},\omega \right)$ being spatial realisations of
a scalar RF. In view of the foregoing discussion, \eqref{eq:15new} is well justified for a piecewise-constant description of realisations of a random medium such as a multiphase composite made of locally isotropic grains. However, in the case of a boundary value problem set up on coarser (i.e. mesoscales) scales, a rank~$2$ tensor random field (TRF) of material properties would be much more appropriate, Fig.~\ref{fig:1}(b). The field equation should then read
\begin{equation}\label{eq:16new}
\bm{\nabla}\cdot\left(\mathsf{C}\left(\mathbf{x},\omega\right)
\cdot\bm{\nabla}u\right)=0,\qquad\mathbf{x}\in E^2,\quad \omega\in\Omega,
\end{equation}
where $\mathsf{C}$\ is the rank~$2$ tensor random field. Indeed, this type of upscaling is sorely needed in the stochastic finite element (SFE) method, where, instead of assuming the local isotropy of the elasticity tensor for each and every material volume (and, hence, finite element), full triclinic-type anisotropy is needed \cite{MR2870261}.

Moving to the in-plane or 3d elasticity, the starting point is the Navier
equation of motion (written in symbolic and tensor notations)
\begin{equation}\label{Navier-equation}
\mu \nabla ^{2}\mathbf{u}+\left( \lambda +\mu \right) \bm{\nabla }\left(
\bm{\nabla }\cdot \mathbf{u}\right) =\rho \mathbf{\ddot{u}}\quad \text{or}\quad \mu u_{i},_{jj}+\left( \lambda +\mu \right) u_{j},_{ji}=\rho \ddot{u}_{i}.
\end{equation}
Here $\mathbf{u}$\ is the displacement field, $\lambda$ and $\mu$ are
two Lam\'{e} constants, and $\rho$ is the mass density. This equation is
often (e.g. in stochastic wave propagation) used as an Ansatz, typically
with the pair $\left(\lambda,\mu\right) $ taken \emph{ad hoc} as a
``vector'' random field with some simple correlation structure for both
components. However, in order to properly introduce the smooth randomness in
$\lambda$ and $\mu$, one has to go one step back in derivation of \eqref{Navier-equation} and write
\begin{equation}\label{pre-Navier-equation}
\mu \nabla ^{2}\mathbf{u}+\left( \lambda +\mu \right) \bm{\nabla }\left(
\bm{\nabla }\cdot \mathbf{u}\right) +\bm{\nabla }\mu \left( \bm{\nabla u}+\left( \bm{\nabla}\mathbf{u}\right) ^{\top}\right) +\bm{\nabla }\lambda \bm{\nabla }\cdot \mathbf{u}=\rho \ddot{\mathbf{u}}
\end{equation}
or
\[
\mu u_{i},_{jj}+\left( \lambda +\mu \right) u_{j},_{ji}+\mu ,_{j}\left(
u_{j},_{i}+u_{i},_{j}\right) +\lambda ,_{i}u_{j},_{j}=\rho \ddot{u}_{i}.
\]
While two extra terms are now correctly present on the left-hand side, this
equation still suffers from the drawback (just as \eqref{eq:15new}) of local isotropy so that, again by micromechanics upscaling arguments, it should be replaced by
\begin{equation}\label{3d-displacement-equation}
\bm{\nabla \cdot }\left( \mathsf{C}\bm{\nabla}\cdot \mathbf{u}\right) ^{\top }=\rho \mathbf{\ddot{u}}\quad \text{or}\quad \left(
C_{ijkl}u_{(k},_{l)}\right) ,_{j}=\rho \ddot{u}_{i}.
\end{equation}
Here $\mathsf{C}$\ ($=C_{ijkl}\mathbf{e}_{i}\otimes \mathbf{e}_{j}\otimes \mathbf{e}%
_{k}\otimes \mathbf{e}_{l}$), which, at any scale finitely larger than the
microstructural scale, is almost surely (a.s.) anisotropic. Clearly, instead
of \eqref{pre-Navier-equation} one should work with the SPDE \eqref{3d-displacement-equation} for $\mathbf{u}$.

While the mathematical theory of SPDEs with anisotropic realisations is not
developed, one powerful way to numerically solve such equations is through
stochastic finite elements (SFE). However, the SFE, just like the SPDE,
require a general representation of the random field $\mathsf{C}$ \cite{MR2870261}, so it can be fitted to micromechanics upscaling studies, as well as its spectral expansion. Observe that each and every material volume (and, hence, the finite element) is an SVE of Fig.~\ref{fig:1}(b), so that a full triclinic-type anisotropy is needed: all the entries of the rank~$4$ stiffness tensor $\mathsf{C}$ are non-zero with probability one. While a micromechanically consistent procedure for upscaling has been discussed in \cite{Sena2013131} and references cited there, general forms of the correlation tensors are sorely needed.

In this paper we develop second-order TRF models of linear hyperelastic
media in each of the eight elasticity classes. That is, for each class, the
fourth-rank elasticity tensor is taken as an isotropic and homogeneous
random field in a three-dimensional Euclidean space, for which the one-point
(mean) and two-point correlation functions need to be explicitly specified.
The simplest case is that of an isotropic class, which implies that two Lam\'{e} constants are random fields. Next, we develop representations of seven higher crystal classes: cubic, transversely isotropic, trigonal, tetragonal, orthotropic, monoclinic, and triclinic. We also find the general form of field's spectral expansion for each of the eight isotropy classes.

\section{The formulation of the problem}\label{sec:problem}

Let $E=E^3$ be a three-dimensional Euclidean point space, and let $V$ be the translation space of $E$ with an inner product $(\boldsymbol{\cdot},\boldsymbol{\cdot})$. Following \cite{MR1162744}, the elements $A$ of $E$ are called the \emph{places} in $E$. The symbol $B-A$ is the vector in $V$ that translates $A$ into $B$.

Let $\mathcal{B}\subset E$ be a deformable body. The \emph{strain tensor} $\varepsilon(A)$, $A\in\mathcal{B}$, is a \emph{configuration variable} taking values in the symmetric tensor square $\mathsf{S}^2(V)$ of dimension~$6$. Following \cite{Olive2013}, we call this space a \emph{state tensor space}.

The \emph{stress tensor} $\sigma(A)$ also takes values in $\mathsf{S}^2(V)$. This is a \emph{source variable}, it describes the source of a field \cite{MR3113882}.

We work with materials obeying \emph{Hooke's law} linking the configuration
variable $\varepsilon(A)$ with the source variable $\sigma(A)$ by
\[
\sigma(A)=\mathsf{C}(A)\varepsilon(A),\qquad A\in\mathcal{B}.
\]
Here the \emph{elastic modulus} $\mathsf{C}$ is a linear map $\mathsf{C}(A)\colon\mathsf{S}^{2}(V)\to\mathsf{S}^{2}(V)$. In linearised hyperelasticity, the map $\mathsf{C}(A)$ is symmetric, i.e., an element of a \emph{constitutive tensor space} $\mathsf{V}=\mathsf{S}^{2}(\mathsf{S}^{2}(V))$ of dimension~$21$.

We assume that $\mathsf{C}(A)$ is a single realisation of a \emph{random field}. In other words, denote by $\mathfrak{B}(\mathsf{V})$ the $\sigma$-field of Borel subsets of $\mathsf{V}$. There is a probability space $(\Omega,\mathfrak{F},\mathsf{P})$ and a mapping $\mathsf{C}\colon\mathcal{B}\times\Omega\to\mathsf{V}$ such that for any $A_0\in\mathcal{B}$ the mapping $\mathsf{C}(A_0,\omega)\colon\Omega\to\mathsf{V}$ is $(\mathfrak{F},\mathfrak{B}(\mathsf{V}))$-measurable.

Translate the whole body $\mathcal{B}$ by a vector $\mathbf{x}\in V$. The random fields $\mathsf{C}(A+\mathbf{x})$ and $\mathsf{C}(A)$ have the same finite-dimensional distributions. It is therefore convenient to assume that there is a random field defined \emph{on all of} $E$ such that its restriction to $\mathcal{B}$ is equal to $\mathsf{C}(A)$. For brevity, denote the new field by the same symbol $\mathsf{C}(A)$ (but this time $A\in E$). The random field $\mathsf{C}(A)$ is \emph{strictly homogeneous}, that is, the random fields $\mathsf{C}(A+\mathbf{x})$ and $\mathsf{C}(A)$ have the same finite-dimensional distributions. In other words, for each positive integer $n$, for each $\mathbf{x}\in V$, and for all distinct places $A_1$, \dots, $A_n\in E$ the random elements $\mathsf{C}(A_1)\oplus\cdots\oplus\mathsf{C}(A_n)$ and $\mathsf{C}(A_1+\mathbf{x})\oplus\cdots\oplus\mathsf{C}(A_n+\mathbf{x})$ of the direct sum on $n$ copies of the space $\mathsf{V}$ have the same probability distribution.

Let $K$ be the material symmetry group of the body $\mathcal{B}$ acting in $V$. The group $K$ is a subgroup of the orthogonal group $\mathrm{O}(V)$. Fix a place $O\in\mathcal{B}$ and identify $E$ with $V$ by the map $f$ that maps $A\in E$ to $A-O\in V$. Then $K$ acts in $E$ and rotates the body $\mathcal{B}$ by
\[
g\cdot A=f^{-1}gfA,\qquad g\in K,\quad A\in\mathcal{B}.
\]
Let  $A_0\in\mathcal{B}$. Under the above action of $K$ the point $A_0$ becomes $g\cdot A_0$. The random tensor $\mathsf{C}(A_0)$ becomes $\mathsf{S}^{2}(\mathsf{S}^{2}(g))\mathsf{C}(A_0)$. The random fields $\mathsf{C}(g\cdot A)$ and $\mathsf{S}^{2}(\mathsf{S}^{2}(g))\mathsf{C}(A)$ must have the same finite-dimensional distributions, because $g\cdot A_0$ is the same material point in a different place. Note that this property does not depend on a particular choice of the place $O$, because the field is strictly homogeneous.

To formalise the non-formal considerations of the above paragraph, note that the map $g\mapsto\mathsf{S}^{2}(\mathsf{S}^{2}(g))$ is an \emph{orthogonal representation} of the group $K$, that is, a continuous map from $K$ to the orthogonal group $\mathrm{O}(\mathsf{V})$ that respects the group operations:
\[
\mathsf{S}^{2}(\mathsf{S}^{2}(g_1g_2))=\mathsf{S}^{2}(\mathsf{S}^{2}(g_1))
\mathsf{S}^{2}(\mathsf{S}^{2}(g_2)),\qquad g_1,g_2\in K.
\]
Let $U$ be an arbitrary orthogonal representation of the group $K$ in a real finite-dimensional linear space $\mathsf{V}$ with an inner product $(\boldsymbol{\cdot},\boldsymbol{\cdot})$, and let $O$ be a place in $E$. A $\mathsf{V}$-valued field $\mathsf{C}(A)$ is called \emph{strictly isotropic} with respect to $O$ if for any $g\in K$ the random fields $\mathsf{C}(g\cdot A)$ and $U(g)\mathsf{C}(A)$ have the same finite-dimensional distributions. If in addition the random field  $\mathsf{C}(A)$ is strictly homogeneous, then it is strictly isotropic with respect to any place.

Assume that the random field $\mathsf{C}(A)$ is \emph{second-order}, that is
\[
\mathsf{E}[\|\mathsf{C}(A)\|^2]<\infty,\qquad A\in E.
\]
Define the \emph{one-point correlation tensor} of the field $\mathsf{C}(\mathbf{x})$ by
\[
\langle\mathsf{C}(A)\rangle=\mathsf{E}[\mathsf{C}(A)]
\]
and its \emph{two-point correlation tensor} by
\[
\langle\mathsf{C}(A),\mathsf{C}(B)\rangle=\mathsf{E}[(\mathsf{C}(A)
-\langle\mathsf{C}(A)\rangle)\otimes(\mathsf{C}(B)
-\langle\mathsf{C}(B)\rangle)].
\]
Assume that the field $\mathsf{C}(\mathbf{x})$ is \emph{mean-square continuous}, that is, its two-point correlation tensor $\langle\mathsf{C}(A),\mathsf{C}(B)\rangle\colon E\times E\to\mathsf{V}\otimes\mathsf{V}$ is a continuous function. If the field $\mathsf{C}(A)$ is strictly homogeneous, then its one-point correlation tensor is a constant tensor in $\mathsf{V}$, while its two-point correlation tensor is a function of the vector $B-A$, i.e., a function on $V$. Call such a field \emph{wide-sense homogeneous}.

Similarly, if the field $\mathsf{C}(A)$ is strictly isotropic, then we have
\[
\begin{aligned}
\langle\mathsf{C}(g\cdot A)\rangle&=U(g)\langle\mathsf{C}(A)\rangle,\\
\langle\mathsf{C}(g\cdot A),\mathsf{C}(g\cdot B)\rangle
&=(U\otimes U)(g)\langle\mathsf{C}(A),\mathsf{C}(B)\rangle.
\end{aligned}
\]
Call such a field \emph{wide-sense isotropic}. In what follows, we consider only wide-sense homogeneous and isotropic random fields and omit the words ``wide-sense''.

For simplicity, identify the field $\{\,\mathsf{C}(A)\colon A\in E\,\}$ defined on $E$ with the field $\{\,\mathsf{C}'(\mathbf{x})\colon\mathbf{x}\in V\,\}$ defined by $\mathsf{C}'(\mathbf{x})=\mathsf{C}(O+\mathbf{x})$. Introduce the Cartesian coordinate system $(x,y,z)$ in $V$. Use the introduced system to identify $V$ with the coordinate space $\mathbb{R}^3$ and $\mathrm{O}(V)$ with $\mathrm{O}(3)$. The action of $\mathrm{O}(3)$ on $\mathbb{R}^3$ is the matrix-vector multiplication.

Forte and Vianello \cite{MR1405284} proved the existence of $8$ symmetry classes of elasticity tensors, or \emph{elasticity classes}. In other words, consider the action
\[
g\cdot\mathsf{C}=\mathsf{S}^{2}(\mathsf{S}^{2}(g))\mathsf{C}
\]
of the group $K=\mathrm{O}(3)$ in the space $\mathsf{V}=\mathsf{S}^{2}(\mathsf{S}^{2}(\mathbb{R}^3))$. The symmetry group of an elasticity tensor $\mathsf{C}\in\mathsf{V}$ is
\[
K(\mathsf{C})=\{\,g\in\mathrm{O}(V)\colon g\cdot\mathsf{C}=\mathsf{C}\,\}.
\]
Note that the symmetry group $K(g\cdot\mathsf{C})$ is conjugate through $g$ to $K(\mathsf{C})$:
\begin{equation}\label{eq:1new}
K(g\cdot\mathsf{C})=\{\,ghg^{-1}\colon h\in K(\mathsf{C})\,\}.
\end{equation}
Whenever two bodies can be rotated so that their symmetry groups coincide, they share the same symmetry class. Mathematically, two elasticity tensors $\mathsf{C}_1$ and $\mathsf{C}_2$ are equivalent if and only if there is $g\in\mathrm{O}(3)$ such that $K(\mathsf{C}_1)=K(g\cdot\mathsf{C}_2)$. In view of \eqref{eq:1new}, $\mathsf{C}_1$ and $\mathsf{C}_2$ are equivalent if and only if their symmetry groups are conjugate. The equivalence classes of the above relation are called the \emph{elasticity classes}.

The first column of Table~\ref{tab:1} adapted from \cite{MR3165064}, contains the name of an elasticity class. The second column represents a collection of subgroups $H$ of $\mathrm{O}(3)$ such that $H$ is conjugate to a symmetry group of any elasticity tensor of the given class. In other words, the above symmetry group lies in the conjugacy class $[H]$ of the group $H$. The third column contains the notation for the normaliser $N(H)$:
\[
N(H)=\{\,g\in\mathrm{O}(3)\colon gHg^{-1}=H\,\}.
\]

\begin{table}[htbp]
\caption{Elasticity classes}
\label{tab:1}
\begin{tabular}{lll}
\hline\noalign{\smallskip}
Elasticity class & $H$ & $N(H)$  \\
\noalign{\smallskip}\hline\noalign{\smallskip}
Triclinic & $Z^c_2$ & $\mathrm{O}(3)$ \\
Monoclinic & $Z_2\times Z^c_2$ & $\mathrm{O}(2)\times Z^c_2$ \\
Orthotropic & $D_2\times Z^c_2$ & $\mathcal{O}\times Z^c_2$ \\
Trigonal & $D_3\times Z^c_2$ & $D_6\times Z^c_2$ \\
Tetragonal & $D_4\times Z^c_2$ & $D_8\times Z^c_2$ \\
Transverse isotropic & $\mathrm{O}(2)\times Z^c_2$ & $\mathrm{O}(2)\times Z^c_2$ \\
Cubic & $\mathcal{O}\times Z^c_2$ & $\mathcal{O}\times Z^c_2$ \\
Isotropic & $\mathrm{O}(3)$ & $\mathrm{O}(3)$ \\
\noalign{\smallskip}\hline
\end{tabular}
\end{table}

Here $Z^c_2=\{I,-I\}$, where $I$ is the $3\times 3$ identity matrix, $Z_n$ is generated by the rotation about the $z$-axis with angle $2\pi/n$, $\mathrm{O}(2)$ is generated by rotations about the $z$-axis with angle $\theta$, $0\leq\theta<2\pi$ and the rotation about the $x$-axis with angle $\pi$, $D_n$ is the \emph{dihedral group} generated by $Z_n$ and the rotation about the $x$-axis with angle $\pi$, and $\mathcal{O}$ is the \emph{octahedral group} which fixes an octahedron. See also \cite[Appendix~B]{MR3208052} for the correspondence between the above notation and notation of Hermann--Mauguin \cite{Hermann,Mauguin} and Sch\"{o}nflie{\ss} \cite{Schoenflies}.

The importance of the group $N(H)$ can be clarified as follows. Consider the \emph{fixed point set} of $H$:
\[
\mathsf{V}^H=\{\,\mathsf{C}\in\mathsf{V}\colon g\cdot\mathsf{C}=\mathsf{C}\quad\text{for all}\quad g\in H\,\}.
\]
By \cite[Lemma~3.1]{MR3165064}, if $H$ is the symmetry group of some tensor $\mathsf{C}\in\mathsf{V}$, then $N(H)$ is the maximal subgroup of $\mathrm{O}(3)$ which leaves $\mathsf{V}^H$ invariant. In the language of the representation theory, $\mathsf{V}^H$ is an \emph{invariant subspace} of the representation $g\mapsto\mathsf{S}^{2}(\mathsf{S}^{2}(g))$ of any group $K$ that lies between $H$ and $N(H)$, that is, $\mathsf{S}^{2}(\mathsf{S}^{2}(g))\mathsf{C}\in\mathsf{V}^H$ for all $g\in K$ and for all $\mathsf{C}\in\mathsf{V}^H$. Denote by $U(g)$ the restriction of the above representation to $\mathsf{V}^H$.

The problem is formulated as follows. For each elasticity class $[H]$ and for each group $K$ that lies between $H$ and $N(H)$, consider an $\mathsf{V}^H$-valued homogeneous random field $\mathsf{C}(\mathbf{x})$ on $\mathbb{R}^3$. Assume that $\mathsf{C}(\mathbf{x})$ is isotropic with respect to $U$:
\begin{equation}\label{eq:3new}
\begin{aligned}
\langle\mathsf{C}(g\mathbf{x})\rangle&=U(g)\langle\mathsf{C}(\mathbf{x})\rangle,\\
\langle\mathsf{C}(g\mathbf{x}),\mathsf{C}(g\mathbf{y})\rangle
&=(U\otimes U)(g)\langle\mathsf{C}(\mathbf{x}),\mathsf{C}(\mathbf{y})\rangle.
\end{aligned}
\end{equation}
We would like to \textbf{find the general form of the one- and two-point correlation tensors of such a field and the spectral expansion of the field itself in terms of stochastic integrals}.

To explain what we mean consider the simplest case when the answer is known. Put $K=H=\mathrm{O}(3)$, $\mathsf{V}^H=\mathbb{R}^1$, and $U(g)=1$, the \emph{trivial representation} of $K$. Recall that a measure $\Phi$ on the $\sigma$-field of Borel sets of a Hausdorff topological space $X$ is called \emph{tight} if for any Borel set $B$, $\Phi(B)$ is the supremum of $\Phi(K)$ over all compact subsets $K$ of $B$. A measure $\Phi$ is called \emph{locally finite} if every point of $X$ has a neighbourhood $U$ for which $\Phi(U)$ is finite. A measure $\Phi$ is called a \emph{Radon measure} if it is tight and locally finite. In what follows we consider only Radon measures and call them just measures.

Schoenberg \cite{MR1503439} proved that the equation
\[
\langle\tau(\mathbf{x}),\tau(\mathbf{y})\rangle=\int^{\infty}_0
\frac{\sin(\lambda\|\mathbf{y}-\mathbf{x}\|)}
{\lambda\|\mathbf{y}-\mathbf{x}\|}\,\mathrm{d}\Phi(\lambda)
\]
establishes a one-to-one correspondence between the class of two-point correlation tensors of homogeneous and isotropic random fields $\tau(\mathbf{x})$ and the class of finite measures on $[0,\infty)$.

Let $L^2_0(\Omega)$ be the Hilbert space of centred complex-valued random variables with finite variance. Let $Z$ be a $L^2_0(\Omega)$-valued measure on the $\sigma$-field of Borel sets of a Hausdorff topological space $X$. A measure $\Phi$ is called the \emph{control measure} for $Z$, if for any Borel sets $B_1$ and $B_2$ we have
\[
\mathsf{E}[\overline{Z(B_1)}Z(B_2)]=\Phi(B_1\cap B_2).
\]
Yaglom \cite{MR0146880} and independently M.\u{I}.~Yadrenko in his unpublished PhD thesis proved that the field $\tau(\mathbf{x})$ has the form
\[
\tau(\rho,\theta,\varphi)=C+\pi\sqrt{2}\sum^{\infty}_{\ell=0}
\sum^{\ell}_{m=-\ell}S^m_{\ell}(\theta,\varphi)
\int^{\infty}_0\frac{J_{\ell+1/2}(\lambda\rho)}{\sqrt{\lambda\rho}}
\,\mathrm{d}Z^m_{\ell}(\lambda),
\]
where $C=\langle\tau(\mathbf{x})\rangle\in\mathbb{R}^1$, $(\rho,\theta,\varphi)$ are spherical coordinates in $\mathbb{R}^3$, $S^m_{\ell}(\theta,\varphi)$ are real-valued spherical harmonics, $J_{\ell+1/2}(\lambda\rho)$ are the Bessel functions of the first kind of order $\ell+1/2$, and $Z^m_{\ell}$ is a sequence of centred uncorrelated real-valued orthogonal random measures on $[0,\infty)$ with the measure $\Phi$ as their common control measure.

Other known results include the case of $\mathsf{V}^H=\mathbb{R}^3$, and $U(g)=g$. Yaglom \cite{MR0033702} found the general form of the two-point correlation tensor. Malyarenko and Ostoja-Starzewski \cite{Malyarenko2016} found the spectral expansion of the field. In the same paper, they found both the general form of the two-point correlation tensor and the spectral expansion of the field for the case of $\mathsf{V}^H=\mathsf{S}^2(\mathbb{R}^3)$, and $U(g)=\mathsf{S}^2(g)$. In \cite{Malyarenko2014a} they solved one of the cases for two-dimensional elasticity, when $V=\mathbb{R}^2$, $K=\mathrm{O}(2)$, $\mathsf{V}^H=\mathsf{S}^2(\mathsf{S}^2(\mathbb{R}^2))$, and $U(g)=\mathsf{S}^2(\mathsf{S}^2(g))$.

\begin{remark}

The set of possible values of elasticity tensors is a proper subset of $\mathsf{V}^H$, namely, the intersection of $\mathsf{V}^H$ with the cone $K$ of symmetric nonnegative operators in $\mathsf{S}^2(V)$. The complete description of homogeneous and isotropic random fields taking values in $\mathsf{V}^H\cap K$ is not known even in the simplest case, when $\mathsf{V}^H=\mathbb{R}^1$ and $K=[0,\infty)$. It is possible to construct various particular classes of such random fields using the ideas of Guilleminot and Soize \cite{MR2859613,MR2913312,MR3036127,MR3091362,MR3284577}. The advantage of their approach is that the random field depends on a few real parameters and may be easily simulated and calibrated. Our approach is based on general spectral expansions, whereby the above questions become more complicated and will be considered in forthcoming publications.

\end{remark}

\section{A general result}

The idea of this Section is as follows. Let $\mathsf{V}$ be a finite-dimensional real linear space, let $K$ be a closed subgroup of the group $\mathrm{O}(3)$, and let $U$ be an orthogonal representation of the group $K$ in the space $\mathsf{V}$. Consider a homogeneous and isotropic random field $\mathsf{C}(\mathbf{x})$, $\mathbf{x}\in\mathbb{R}^3$, and solve the problem formulated in Section~\ref{sec:problem}. In Section~\ref{sec:results}, apply general formulae to our cases. The resulting Theorems~\ref{th:1}--\ref{th:16} are particular cases of general Theorem~\ref{th:0}.

To obtain general formulae, we describe all homogeneous random fields taking values in $\mathsf{V}$ and throw away non-isotropic ones. The first obstacle here is as follows. The complete description of such fields is unknown. We use the following result instead.

Let $\mathsf{V}^{\mathbb{C}}$ be a \emph{complex} finite-dimensional linear space with an inner product $(\boldsymbol{\cdot},\boldsymbol{\cdot})$ that is linear in the \emph{second} argument, as is usual in physics. Let $J$ be a \emph{real structure} on $\mathsf{V}^{\mathbb{C}}$, that is, a map $J\colon\mathsf{V}^{\mathbb{C}}\to\mathsf{V}^{\mathbb{C}}$ satisfying the following conditions:
\[
\begin{aligned}
J(\alpha\mathsf{C}_1+\beta\mathsf{C}_2)&=\overline{\alpha}J(\mathsf{C}_1)
+\overline{\beta}J(\mathsf{C}_2),\\
J(J(\mathsf{C}))&=\mathsf{C}
\end{aligned}
\]
for all $\alpha$, $\beta\in\mathbb{C}$ and for all $\mathsf{C}_1$, $\mathsf{C}_2\in\mathsf{V}^{\mathbb{C}}$. In other words, $J$ is a multidimensional and coordinate-free generalisation of complex conjugation. The set of all eigenvectors of $J$ that correspond to eigenvalue $1$, constitute a \emph{real} linear space, denote it by $\mathsf{V}$. Let $\mathsf{H}$ be the real linear space of Hermitian linear operators in $\mathsf{V}^{\mathbb{C}}$. The real structure $J$ induces a linear operator $\mathsf{J}$ in $\mathsf{H}$. For any $A\in\mathsf{H}$, the operator $\mathsf{J}A$ acts by
\[
(\mathsf{J}A)\mathsf{C}=J(A\mathsf{C}),\qquad\mathsf{C}\in\mathsf{V}^{\mathbb{C}}.
\]
In coordinates, the operator $\mathsf{J}$ is just the transposition of a matrix.

The result by Cram{\'e}r \cite{MR0000920} in coordinate-free form is formulated as follows. Equation
\begin{equation}\label{eq:2new}
\langle\mathsf{C}(\mathbf{x}),\mathsf{C}(\mathbf{y})\rangle=\int_{\hat{\mathbb{R}}^3}
\mathrm{e}^{\mathrm{i}(\mathbf{p},\mathbf{y}-\mathbf{x})}\,\mathrm{d}F(\mathbf{p})
\end{equation}
establishes a one-to-one correspondence between the class of two-point correlation tensors of homogeneous mean-square continuous $\mathsf{V}^{\mathbb{C}}$-valued random fields $\mathsf{C}(\mathbf{x})$ and the class of Radon measures on the $\sigma$-field of Borel sets of the \emph{wavenumber domain} $\hat{\mathbb{R}}^3$ tasking values in the set of nonnegative-definite Hermitian linear operators in $\mathsf{V}^{\mathbb{C}}$. For $\mathsf{V}$-valued random fields, there is only a \emph{necessary condition}: if $\mathsf{C}(\mathbf{x})$ is $\mathsf{V}$-valued, then the measure $F$ satisfies
\[
F(-B)=\mathsf{J}F(B),\qquad B\in\mathfrak{B}(\hat{\mathbb{R}}^3),
\]
where $-B=\{\,-\mathsf{C}\colon\mathsf{C}\in B\,\}$.

Introduce the \emph{trace measure} $\mu$ by $\mu(B)=\tr F(B)$, $B\in\mathfrak{B}(\hat{\mathbb{R}}^3)$ and note that $F$ is absolutely continuous with respect to $\mu$. This means that Equation~\eqref{eq:2new} may be written as
\[
\langle\mathsf{C}(\mathbf{x}),\mathsf{C}(\mathbf{y})\rangle=\int_{\hat{\mathbb{R}}^3}
\mathrm{e}^{\mathrm{i}(\mathbf{p},\mathbf{y}-\mathbf{x})}f(\mathbf{p})\,\mathrm{d}\mu(\mathbf{p}),
\]
where $f(\mathbf{p})$ is a measurable function on the wavenumber domain taking values in the set of all nonnegative-definite Hermitian linear operators in $\mathsf{V}^{\mathbb{C}}$ with unit trace, that satisfies the following condition
\begin{equation}\label{eq:4new}
f(-\mathbf{p})=Jf(\mathbf{p}).
\end{equation}

Using representation theory, it is possible to prove the following. Let $\mathsf{C}_1$, $\mathsf{C}_2\in\mathsf{V}$. Let $L(\mathsf{C}_1\otimes\mathsf{C}_2)$ be the operator in $\mathsf{H}$ acting on a tensor $\mathsf{C}\in\mathsf{V}^{\mathbb{C}}$ by
\[
L(\mathsf{C}_1\otimes\mathsf{C}_2)\mathsf{C}=(J\mathsf{C}_1,\mathsf{C})\mathsf{C}_2.
\]
By linearity, this action may be extended to an isomorphism $L$ between $\mathsf{V}\otimes\mathsf{V}$ and $\mathsf{H}$. The orthogonal operators $LU\otimes U(g)L^{-1}$, $g\in K$, constitute an orthogonal representation of the group~$K$ in the space $\mathsf{H}$, \emph{equivalent} to the tensor square $U\otimes U$ of the representation $U$. The operator $L$ is an \emph{intertwining operator} between the spaces $\mathsf{V}\otimes\mathsf{V}$ and $\mathsf{H}$ where equivalent representations $U\otimes U$ and $LU\otimes UL^{-1}$ act. In what follows, we are working only with the latter representation, for simplicity denote it again by $U\otimes U$ and note that it acts in the space $\mathsf{H}$ by
\[
(U\otimes U)(g)A=U(g)AU^{-1}(g),\qquad A\in\mathsf{H}.
\]
Denote $\mathsf{H}_+=L\mathsf{S}^2(\mathsf{V})$. In coordinates, it is the subspace of Hermitian matrices with real-valued matrix entries. If $-I\in K$, then the second equation in \eqref{eq:3new} and Equation~\eqref{eq:4new} together are equivalent to the following conditions:
\begin{equation}\label{eq:5new}
\mu(gB)=\mu(B),\qquad B\in\mathfrak{B}(\hat{\mathbb{R}}^3)
\end{equation}
and
\begin{equation}\label{eq:6new}
f(\mathbf{p})\in\mathsf{H}_+,\qquad f(g\mathbf{p})=\mathsf{S}^2(U(g))f(\mathbf{p}).
\end{equation}

The description of all measures $\mu$ satisfying Equation~\eqref{eq:5new} is well known, see \cite{MR2098271}. There are finitely many, say $M$, orbit types for the action of $K$ in $\hat{\mathbb{R}}^3$ by
\[
(g\mathbf{p},\mathbf{x})=(\mathbf{p},g^{-1}\mathbf{x}).
\]
Denote by $(\hat{\mathbb{R}}^3/ K)_m$, $0\leq m\leq M-1$ the set of all orbits of the $m$th type. It is known, see \cite{MR3165064}, that all the above sets are manifolds. Assume for simplicity of notation that there are charts $\bm{\lambda}_m$ such that the domain of $\bm{\lambda}_m$ is dense in $(\hat{\mathbb{R}}^3/ K)_m$. The orbit of the $m$th type is the manifold $K/H_m$, where $H_m$ is a stationary subgroup of a point on the orbit. Assume that the domain of a chart $\bm{\varphi}_m$ is a dense set in $K/H_m$, and let $\mathrm{d}\bm{\varphi}_m$ be the unique probabilistic $K$-invariant measure on the $\sigma$-field of Borel sets of $K/H_m$. There are the unique measures $\Phi_m$ on the $\sigma$-fields of Borel sets in $(\hat{\mathbb{R}}^3/ K)_m$ such that
\[
\int_{\hat{\mathbb{R}}^3}
\mathrm{e}^{\mathrm{i}(\mathbf{p},\mathbf{y}-\mathbf{x})}f(\mathbf{p})\,\mathrm{d}\mu(\mathbf{p})
=\sum_{m=0}^{M-1}\int_{(\hat{\mathbb{R}}^3/ K)_m}\int_{K/H_m}
\mathrm{e}^{\mathrm{i}((\bm{\lambda}_m,\bm{\varphi}_m),\mathbf{y}-\mathbf{x})}
f(\bm{\lambda}_m,\bm{\varphi}_m)\,\mathrm{d}\bm{\varphi}_m\,\mathrm{d}\Phi_m(\bm{\lambda}_m).
\]

To find all functions $f$ satisfying Equation~\eqref{eq:6new}, proceed as follows. Fix an orbit $\bm{\lambda}_m$ and denote by $\bm{\varphi}_m^0$ the coordinates of the intersection of the orbit $\bm{\lambda}_m$ with the set $(\hat{\mathbb{R}}^3/ K)_m$. Let $U^m$ be the restriction of the representation $\mathsf{S}^2(U)$ to the group $H_m$. We have $g(\bm{\lambda}_m,\bm{\varphi}_m^0)=(\bm{\lambda}_m,\bm{\varphi}_m^0)$ for all $g\in H_m$, because $H_m$ is the stationary subgroup of the point $(\bm{\lambda}_m,\bm{\varphi}_m^0)$. For $g\in H_m$, Equation~\eqref{eq:6new} becomes
\begin{equation}\label{eq:7new}
f(\bm{\lambda}_m,\bm{\varphi}_m^0)=U^m(g)f(\bm{\lambda}_m,\bm{\varphi}_m^0).
\end{equation}

Any orthogonal representation of a compact topological group in a space $\mathsf{H}$ has at least two invariant subspaces: $\{\mathsf{0}\}$ and $\mathsf{H}$. The representation is called \emph{irreducible} if no other invariant subspaces exist. The space of any finite-dimensional orthogonal representation of a compact topological group can be uniquely decomposed into a direct sum of \emph{isotypic subspaces}. Each isotypic subspace is the direct sum of finitely many subspaces where the copies of the same irreducible representation act. Equation~\eqref{eq:7new} means that the operator $f(\bm{\lambda}_m,\bm{\varphi}_m^0)$ lies in the isotypic subspace $\mathsf{H}_m$ which corresponds to the trivial representation of the group~$H_m$. The intersection of this subspace with the convex compact set of all nonnegative-definite operators in $\mathsf{H}_+$ with unit trace is again a convex compact set, call it $\mathcal{C}_m$. As $\bm{\lambda}_m$ runs over $(\hat{\mathbb{R}}^3/ K)_m$, $f(\bm{\lambda}_m,\bm{\varphi}_m^0)$ becomes an arbitrary measurable function taking values in $\mathcal{C}_m$.

An irreducible orthogonal representation of the group~$K$ is called a \emph{representation of class}~$1$ with respect to the group~$H_m$ if the restriction of this representation to $H_m$ contains at least one copy of the trivial representation of $H_m$. Let $\mathsf{S}^2(U)_m$ be the restriction of the representation~$\mathsf{S}^2(U)$ to the direct sum of the isotypic subspaces of the irreducible representation of class~$1$ with respect to $H_m$. Let $g_{\bm{\varphi}_m}$ be an arbitrary element of $K$ such that $g_{\bm{\varphi}_m}(\bm{\varphi}_m^0)=\bm{\varphi}_m$. Two such elements differ by an element of $H_m$, therefore the second equation in \eqref{eq:6new} becomes
\[
f(\bm{\lambda}_m,\bm{\varphi}_m)=\mathsf{S}^2(U(g_{\bm{\varphi}_m}))_m
f(\bm{\lambda}_m,\bm{\varphi}_m^0).
\]
The two-point correlation tensor of the field takes the form
\begin{equation}\label{eq:8new}
\begin{aligned}
\langle\mathsf{C}(\mathbf{x}),\mathsf{C}(\mathbf{y})\rangle&=\sum_{m=0}^{M-1}
\int_{(\hat{\mathbb{R}}^3/ K)_m}\int_{K/H_m}
\mathrm{e}^{\mathrm{i}(g_{\bm{\varphi}_m}(\bm{\lambda}_m,\bm{\varphi}_m^0),
\mathbf{y}-\mathbf{x})}\mathsf{S}^2(U(g_{\bm{\varphi}_m}))_m\\
&\quad\times f(\bm{\lambda}_m,\bm{\varphi}_m^0)\,\mathrm{d}\bm{\varphi}_m
\,\mathrm{d}\Phi_m(\bm{\lambda}_m).
\end{aligned}
\end{equation}

Choose an orthonormal basis $\mathsf{T}^1$, \dots, $\mathsf{T}^{\dim\mathsf{V}}$ in the space~$\mathsf{V}$. The tensor square $\mathsf{V}\otimes\mathsf{V}$ has several orthonormal bases. The \emph{coupled basis} consists of tensor products $\mathsf{T}^i\otimes\mathsf{T}^j$, $1\leq i$, $j\leq\dim\mathsf{V}$. The $m$th \emph{uncoupled basis} is build as follows. Let $U^{m,1}$, \dots, $U^{m,k_m}$ be all non-equivalent irreducible orthogonal representations of the group $K$ of class~$1$ with respect to $H_m$ such that the representation $\mathsf{S}^2(U)$ contains isotypic subspaces where $c_{mk}$ copies of the representation $U^{m,k}$ act, and let the restriction of the representation $U^{m,k}$ to $H_m$ contains $d_{mk}$ copies of the trivial representation of $H_m$. Let $\mathsf{T}^{mkln}$, $1\leq l\leq d_{mk}$, $1\leq n\leq c_{mk}$ be an orthonormal basis in the space where the $n$th copy act. Complete the above basis to the basis $\mathsf{T}^{mkln}$, $1\leq l\leq\dim U^{m,k}$ and call this basis the $m$th uncoupled basis. The vectors of the coupled basis are linear combinations of the vectors of the $m$th uncoupled basis:
\[
\mathsf{T}^i\otimes\mathsf{T}^j=\sum_{k=1}^{k_m}\sum_{l=1}^{\dim U^{m,k}}\sum_{n=1}^{c_{mk}}c^{mkln}_{ij}\mathsf{T}^{mkln}+\cdots,
\]
where dots denote the terms that include the tensors in the basis of the space $\mathsf{S}^2(\mathsf{V})\ominus\mathsf{S}^2(\mathsf{V})_m$. In the introduced coordinates, Equation~\eqref{eq:8new} takes the form
\begin{equation}\label{eq:9new}
\begin{aligned}
\langle\mathsf{C}(\mathbf{x}),\mathsf{C}(\mathbf{y})\rangle_{ij}&=\sum_{m=0}^{M-1}
\sum_{k=1}^{k_m}\sum_{l=1}^{\dim U^{m,k}}\sum_{l'=1}^{d_{mk}}\sum_{n=1}^{c_{mk}}c^{mkln}_{ij}
\int_{(\hat{\mathbb{R}}^3/ K)_m}\int_{K/H_m}
\mathrm{e}^{\mathrm{i}(g_{\bm{\varphi}_m}(\bm{\lambda}_m,\bm{\varphi}_m^0),
\mathbf{y}-\mathbf{x})}\\
&\quad\times U^{m,k}_{ll'}(\bm{\varphi}_m)f_{l'n}(\bm{\lambda}_m,\bm{\varphi}_m^0)
\,\mathrm{d}\bm{\varphi}_m\,\mathrm{d}\Phi_m(\bm{\lambda}_m).
\end{aligned}
\end{equation}
The choice of bases inside the isotypic subspaces is not unique. One has to choose them in such a way that calculation of the transition coefficients $c^{mkln}_{ij}$ is as easy as possible.

To calculate the inner integrals, we proceed as follows. Consider the action of $K$ on $\mathbb{R}^3$ by matrix-vector multiplication. Let $(\mathbb{R}^3/ K)_m$, $0\leq m\leq M-1$ be the set of all orbits of the $m$th type. Let $\bm{\rho}_m$ be such a chart that its domain is dense in $(\mathbb{R}^3/ K)_m$. Let $\bm{\psi}_m$ be a chart in $K/H_m$ with a dense domain, and let $\mathrm{d}\bm{\psi}_m$ be the unique probabilistic $K$-invariant measure on the $\sigma$-field of Borel sets of $K/H_m$. It is known that the sets of orbits of one of the types, say $(\hat{\mathbb{R}}^3/ K)_{M-1}$ (resp. $(\mathbb{R}^3/ K)_{M-1}$), are dense in $\hat{\mathbb{R}}^3$ (resp. $\mathbb{R}^3$). Write the plane wave $\mathrm{e}^{\mathrm{i}(g_{\bm{\varphi}_{M-1}}(\bm{\lambda}_{M-1},\bm{\varphi}_{M-1}^0),
\mathbf{y}-\mathbf{x})}$ as
\[
\mathrm{e}^{\mathrm{i}(g_{\bm{\varphi}_{M-1}}(\bm{\lambda}_{M-1},\bm{\varphi}_{M-1}^0),
\mathbf{y}-\mathbf{x})}=\mathrm{e}^{\mathrm{i}(g_{\bm{\varphi}_{M-1}}(\bm{\lambda}_{M-1},\bm{\varphi}_{M-1}^0),
g_{\bm{\psi}_{M-1}}(\bm{\rho}_{M-1},\bm{\psi}_{M-1}^0))},
\]
and consider the plane wave as a function of two variables $\bm{\varphi}_{M-1}$ and $\bm{\psi}_{M-1}$ with domain $(K/H_{M-1})^2$. This function is $K$-invariant:
\[
\mathrm{e}^{\mathrm{i}(gg_{\bm{\varphi}_{M-1}}(\bm{\lambda}_{M-1},\bm{\varphi}_{M-1}^0),
gg_{\bm{\psi}_{M-1}}(\bm{\rho}_{M-1},\bm{\psi}_{M-1}^0))}
=\mathrm{e}^{\mathrm{i}(g_{\bm{\varphi}_{M-1}}(\bm{\lambda}_{M-1},\bm{\varphi}_{M-1}^0),
g_{\bm{\psi}_{M-1}}(\bm{\rho}_{M-1},\bm{\psi}_{M-1}^0))},\quad g\in K.
\]
Denote by $\hat{K}_{H_{M-1}}$ the set of all equivalence classes of irreducible representations of $K$ of class~$1$ with respect to $H_{M-1}$, and let the restriction of the representation $U^q\in\hat{K}_{H_{M-1}}$ to $H_{M-1}$ contains $d_q$ copies of the trivial representation of $H_{M-1}$. By the Fine Structure Theorem \cite{MR3114697}, there are some numbers $d'_q\leq d_q$ such that the set
\[
\{\,\dim U^q\cdot U^q_{ll'}(\bm{\varphi}_{M-1})U^q_{ll'}(\bm{\psi}_{M-1})\colon U^q\in\hat{K}_{H_{M-1}},1\leq l\leq\dim U^q,1\leq l'\leq d'_q\,\}
\]
is the orthonormal basis in the Hilbert space $L^2((K/H_{M-1})^2,\mathrm{d}\bm{\varphi}_{M-1}\,\mathrm{d}\bm{\psi}_{M-1})$. Let
\begin{equation}\label{eq:17}
\begin{aligned}
j^q_{ll'}(\bm{\lambda}_{M-1},\bm{\rho}_{M-1})&=\dim U^q\int_{(K/H_{M-1})^2}
\mathrm{e}^{\mathrm{i}(g_{\bm{\varphi}_{M-1}}(\bm{\lambda}_{M-1},\bm{\varphi}_{M-1}^0),
g_{\bm{\psi}_{M-1}}(\bm{\rho}_{M-1},\bm{\psi}_{M-1}^0))}\\
&\quad\times U^q_{ll'}(\bm{\varphi}_{M-1})
U^q_{ll'}(\bm{\psi}_{M-1})\,\mathrm{d}\bm{\varphi}_{M-1}\,\mathrm{d}\bm{\psi}_{M-1}
\end{aligned}
\end{equation}
be the corresponding Fourier coefficients. The uniformly convergent Fourier expansion takes the form
\begin{equation}\label{eq:10new}
\begin{aligned}
\mathrm{e}^{\mathrm{i}(g_{\bm{\varphi}_{M-1}}(\bm{\lambda}_{M-1},\bm{\varphi}_{M-1}^0),
g_{\bm{\psi}_{M-1}}(\bm{\rho}_{M-1},\bm{\psi}_{M-1}^0))}&=\sum_{U^q\in\hat{K}_{H_{M-1}}}
\sum_{l=1}^{\dim U^q}\sum_{l'=1}^{d'_q}\dim U^q\\
&\quad\times j^q_{ll'}(\bm{\lambda}_{M-1},\bm{\rho}_{M-1})U^q_{ll'}(\bm{\varphi}_{M-1})U^q_{ll'}(\bm{\psi}_{M-1}).
\end{aligned}
\end{equation}
This expansion is defined on the dense set
\[
(\hat{\mathbb{R}}^3/K)_{M-1}\times(K/H_{M-1})\times(\mathbb{R}^3/K)_{M-1}\times(K/H_{M-1})
\]
and may be extended to all of $\hat{\mathbb{R}}^3\times\mathbb{R}^3$ by continuity. Substituting the extended expansion to Equation~\eqref{eq:9new}, we obtain the expansion
\begin{equation}\label{eq:12new}
\begin{aligned}
\langle\mathsf{C}(\mathbf{x}),\mathsf{C}(\mathbf{y})\rangle_{ij}&=\sum_{m=0}^{M-1}
\sum_{k=1}^{k_m}\sum_{l=1}^{\dim U^{m,k}}\sum_{l'=1}^{d'_{mk}}\sum_{n=1}^{c_{mk}}c^{mkln}_{ij}
\int_{(\hat{\mathbb{R}}^3/ K)_m}j^q_{ll'}(\bm{\lambda}_m,\bm{\rho}_0)\\
&\quad\times   U^{m,k}_{ll'}(\bm{\psi}_m)f_{l'n}(\bm{\lambda}_m,\bm{\varphi}_m^0)
\,\mathrm{d}\Phi_m(\bm{\lambda}_m).
\end{aligned}
\end{equation}

\setcounter{theorem}{-1}

\begin{theorem}\label{th:0}
Let $-I\in K$. The one-point correlation tensor of a homogeneous and $(K,U)$-isotropic random field lies in the space of the isotypic component of the representation $U$ that corresponds to the trivial representation of $K$ and is equal to $\mathsf{0}$ if no such isotypic component exists. Its two-point correlation tensor is given by Equation~\emph{\eqref{eq:12new}}.
\end{theorem}

\begin{remark}
The results by \cite{Malyarenko2014a,Malyarenko2016,MR1503439,MR0033702,MR0146880} as well as Theorems~\ref{th:1}--\ref{th:16} below are particular cases of Theorem~\ref{th:0}. The expansion~\eqref{eq:12new} is the first necessary step in studying random fields connected to Hooke's law.
\end{remark}

Later we will see that it is easy to write the spectral expansion of the field directly if the group $K$ is finite. Otherwise, we write the Fourier expansion \eqref{eq:10new} for plane waves $\mathrm{e}^{\mathrm{i}(\mathbf{p},\mathbf{y})}$ and $\mathrm{e}^{-\mathrm{i}(\mathbf{p},\mathbf{x})}$ separately and substitute both expansions to Equation~\eqref{eq:9new}. As a result, we obtain the expansion of the two-point correlation tensor of the field in the form
\[
\langle\mathsf{C}(\mathbf{x}),\mathsf{C}(\mathbf{y})\rangle_{ij}=\int_{\Lambda}
\overline{u(\mathbf{x},\lambda)}u(\mathbf{y},\lambda)\,\mathrm{d}\Phi_{ij}(\lambda),
\]
where $\Lambda$ is a set, and where $F$ is a measure on a $\sigma$-field $\mathfrak{L}$ of subsets of $\Lambda$ taking values in the set of Hermitian nonnegative-definite operators on $\mathsf{V}^{\mathbb{C}}$. Moreover, the set $\{\,u(\mathbf{x},\lambda)\colon\mathbf{x}\in\mathbb{R}^3\,\}$ is \emph{total} in the Hilbert space $L^2(\Lambda,\Phi)$ of the measurable complex-valued functions on $\Lambda$ that are square-integrable with respect to the measure $\Phi$, that is, the set of finite linear combinations $\sum c_nu(\mathbf{x}_n,\lambda)$ is dense in the above space. By Karhunen's theorem \cite{MR0023013}, the field $\mathsf{C}(\mathbf{x})$ has the following spectral expansion:
\begin{equation}\label{eq:11new}
\mathsf{C}(\mathbf{x})=\mathsf{E}[\mathsf{C}(\mathbf{0})]+\int_{\Lambda}u(\mathbf{x},\lambda)
\,\mathrm{d}\mathsf{Z}(\lambda),
\end{equation}
where $\mathsf{Z}$ is a measure on the measurable space $(\Lambda,\mathfrak{L})$ taking values in the Hilbert space of random tensors $\mathsf{Z}\colon\Omega\to\mathsf{V}^{\mathbb{C}}$ with $\mathsf{E}[\mathsf{Z}]=\mathsf{0}$ and $\mathsf{E}[\|\mathsf{Z}\|^2]<\infty$. The measure $F$ is the \emph{control measure} of the measure $\mathsf{Z}$, i.e.,
\[
\mathsf{E}[J\mathsf{Z}(A)\mathsf{Z}^{\top}(B)]=\Phi(A\cap B),\qquad
A,B\in\mathfrak{L}.
\]
The components of the random tensor $\mathsf{Z}(A)$ are correlated, which creates difficulties when one tries to use Equation~\eqref{eq:11new} for computer simulation. It is possible to use Cholesky decomposition and to write the expansion of the field using uncorrelated random measures, see details in \cite{Malyarenko2016}.

\section{Preliminary calculations}

The possibilities for the group $K$ are as follows. In the triclinic class, there exist infinitely many groups between $Z^c_2$ and $\mathrm{O}(3)$, we put $K_1=Z^c_2$ and $K_2=\mathrm{O}(3)$. Similarly, for the monoclinic class put $K_3=Z_2\times Z^c_2$ and $K_4=\mathrm{O}(2)\times Z^c_2$. The possibilities for the orthotropic class are $K_5=D_2\times Z^c_2$, $K_6=D_4\times Z^c_2$, $K_7=D_6\times Z^c_2$, $K_8=\mathcal{T}\times Z^c_2$, and $K_9=\mathcal{O}\times Z^c_2$. Here $\mathcal{T}$ is the \emph{tetrahedral group} which fixes a tetrahedron. In the trigonal class, we have $K_{10}=D_3\times Z^c_2$ and $K_{11}=D_6\times Z^c_2$. In the tetragonal class, the possibilities are $K_{12}=D_4\times Z^c_2$ and $K_{13}=D_8\times Z^c_2$. In the three remaining classes, the possibilities are $K_{14}=\mathrm{O}(2)\times Z^c_2$, $K_{15}=\mathcal{O}\times Z^c_2$, and $K_{16}=\mathrm{O}(3)$. The intermediate groups were determined using \cite[Vol.~1, Fig.~10.1.3.2]{Brock2014}. For each group $K_i$, $1\leq i\leq 16$, we formulate Theorem number $i$ below.

\subsection{The structure of the representation $U$}

The notation for irreducible orthogonal representation is as follows. If $K_i$ is a \emph{finite} group, we use the \emph{Mulliken notation} \cite{Mulliken}, see also \cite[Chapter~14]{altmann1994point} to denote the irreducible \emph{unitary} representation of $K_i$. For an irreducible orthogonal representation, consider its complexification. A standard result of representation theory, see, for example, \cite[Proposition~4.8.4]{MR1738431}, states that there are three possibilities:
\begin{itemize}
  \item The complexification is irreducible, say $U$. Then, it is a sum of two equivalent orthogonal representations, and we denote each of them by $U$.
  \item The complexification is a direct sum of two mutually conjugate representation $U_1$ and $U_2$, that is, $U_2(g)=\overline{U_1(g)}$. We denote the orthogonal representation by $U_1\oplus U_2$.
  \item The complexification is a direct sum of two copies of an irreducible representation $U$. We denote the orthogonal representation by $U\oplus U$.
\end{itemize}

For infinite groups, the notation is as follows. For $K_2=\mathrm{O}(3)$, we denote the representations by $U^{\ell g}$ (the tensor product of the representation $U^{\ell}$ of the group $\mathrm{SO}(3)$ and the trivial representation $A_g$ of $Z^c_2$) and $U^{\ell u}$ (that of $U^{\ell}$ and the nontrivial representation $A_u$ of $Z^c_2$). For $K_4=K_{14}=\mathrm{O}(2)\times Z^c_2$ the notation is $U^{0gg}=U^{0g}\otimes A_g$, $U^{0gu}=U^{0g}\otimes A_u$, $U^{0ug}=U^{0u}\otimes A_g$, $U^{0uu}=U^{0u}\otimes A_u$, $U^{\ell g}=U^{\ell}\otimes A_g$, and $U^{\ell u}=U^{\ell}\otimes A_u$, where $U^{0g}$ is the trivial representation of $\mathrm{O}(2)$, $U^{0u}(g)=\det g$, and
\[
\begin{aligned}
U^{\ell}\left(
\begin{pmatrix}
  \cos(\varphi) & \sin(\varphi) \\
  -\sin(\varphi) & \cos(\varphi)
\end{pmatrix}
\right)&=
\begin{pmatrix}
  \cos(\ell\varphi) & \sin(\ell\varphi) \\
  -\sin(\ell\varphi) & \cos(\ell\varphi)
\end{pmatrix},\\
U^{\ell}\left(
\begin{pmatrix}
  \cos(\varphi) & \sin(\varphi) \\
  \sin(\varphi) & -\cos(\varphi)
\end{pmatrix}
\right)&=
\begin{pmatrix}
  \cos(\ell\varphi) & \sin(\ell\varphi) \\
  \sin(\ell\varphi) & -\cos(\ell\varphi)
\end{pmatrix}
.
\end{aligned}
\]

Fist, we determine the structure of the representation $g\mapsto g$ of the group $K_i$. For finite groups, the above structure is given in Table~$n$.10 in \cite{altmann1994point}, where $n$ in the number given in the second column of Table~\ref{tab:2}. For $K_2$ and $K_{16}$, this representation is $U^{1u}$, for $K_4$ and $K_{14}$ it is $U^{1u}\oplus U^{0uu}$. Then we determine the structure of the representations $\mathsf{S}^2(g)$ and $\mathsf{S}^2(\mathsf{S}^2(g))$. For finite groups, we use Table~$n$.8. For infinite groups, we use the following multiplication rules. The product of two isomorphic irreducible representations of $Z^c_2$ is $A_g$, that of two different representations is $A_u$. For $\mathrm{SO}(3)$, we have
\[
U^{\ell_1}\otimes U^{\ell_2}=\sum_{\ell=|\ell_1-\ell_2|}^{\ell_1+\ell_2}\oplus U^{\ell}.
\]
For $\mathrm{O}(2)$, we have $U^{\ell}\otimes U^{\ell}=U^{2\ell}\oplus U^{0g}\oplus U^{0u}$ and $U^{\ell_1}\otimes U^{\ell_2}=U^{\ell_1+\ell_2}\oplus U^{|\ell_1-\ell_2|}$ for $\ell_2\neq\ell_1$.

\begin{table}[htbp]
\caption{The structure of the representation $U$}
\label{tab:2}
\begin{tabular}{lll}
\hline\noalign{\smallskip}
$K_i$ & Table number & The structure of $U$  \\
\noalign{\smallskip}\hline\noalign{\smallskip}
$K_1=Z^c_2$ & 11 & $21A_g$ \\
$K_2=\mathrm{O}(3)$ & $-$ & $2U^{0g}\oplus 2U^{2g}\oplus U^{4g}$ \\
$K_3=Z_2\times Z^c_2$ & 60 & $13A_g$ \\
$K_4=\mathrm{O}(2)\times Z^c_2$ & $-$ & $5U^{0gg}\oplus 3U^{2g}\oplus U^{4g}$ \\
$K_5=D_2\times Z^c_2$ & 31 & $9A_g$ \\
$K_6=D_4\times Z^c_2$ & 33 & $6A_{1g}\oplus 3B_{1g}$ \\
$K_7=D_6\times Z^c_2$ & 35 & $5A_{1g}\oplus 2E_{2g}$ \\
$K_8=\mathcal{T}\times Z^c_2$ & 72 & $3A_g\oplus 3({}^1E_g\oplus{}^2E_g)$ \\
$K_9=\mathcal{O}\times Z^c_2$ & 71 & $3A_{1g}\oplus 3E_g$ \\
$K_{10}=D_3\times Z^c_2$ & 42 & $6A_{1g}$ \\
$K_{11}=D_6\times Z^c_2$ & 35 & $5A_{1g}\oplus B_{1g}$ \\
$K_{12}=D_4\times Z^c_2$ & 33 & $6A_{1g}$ \\
$K_{13}=D_8\times Z^c_2$ & 37 & $5A_{1g}\oplus B_{2g}$ \\
$K_{14}=\mathrm{O}(2)\times Z^c_2$ & $-$ & $5U^{0gg}$ \\
$K_{15}=\mathcal{O}\times Z^c_2$ & 71 & $3A_{1g}$ \\
$K_{16}=\mathrm{O}(3)$ & $-$ & $2U^{0g}$ \\
\noalign{\smallskip}\hline
\end{tabular}
\end{table}

If $K_i=H$, then the space $\mathsf{V}$ is spanned by the spaces of the copies of all trivial representations of $K_i$ that belong to $\mathsf{S}^2(\mathsf{S}^2(g))$. This gives us a method for calculation of the dimension $\dim\mathsf{V}$ alternative to that in \cite{MR3165064}. Otherwise, it is spanned by the spaces of all irreducible representations of $K_i$ that contain at least one copy of the trivial representation of $H$. To determine such representations, we use Table~$n$.9.

\subsection{The basis of the space $\mathsf{V}^H$ for different groups}

We start from the basis for $K_2$. Gordienko \cite{MR1888117} proposed a basis $\{\,\mathbf{h}^m_{\ell}\colon -\ell\leq m\leq\ell\,\}$ in the space of the irreducible representation $U^{\ell}$ of the group~$\mathrm{SO}(3)$ in which all matrix entries of the representation's matrices become real-valued functions. Godunov and Cordienko \cite{MR2078714} found the coefficients $g^{m[m_1,m_2]}_{\ell[\ell_1,\ell_2]}$ of the expansion
\[
\mathbf{h}^{m_1}_{\ell_1}\otimes\mathbf{h}^{m_2}_{\ell_2}
=\sum_{\ell=|\ell_1-\ell_2|}^{\ell_1+\ell_2}\sum_{m=-\ell}^{\ell}
g^{m[m_1,m_2]}_{\ell[\ell_1,\ell_2]}\mathbf{h}^m_{\ell}.
\]
We call them the \emph{Godunov--Gordienko coefficients}. Malyarenko and Ostoja-Starzewski \cite{Malyarenko2016} calculated the tensors of the basis of the $21$-dimensional space $\mathsf{S}^2(\mathsf{S}^2(\mathbb{R}^3))$ for the group $K_2$ in terms of the above coefficients. Using MATLAB Symbolic Math Toolbox, we calculate the elements of the bases for the groups $K_1$, $K_3$--$K_{16}$ as linear combinations of the tensors of the basis for the group~$K_2$, see Table~\ref{tab:5}.

\begin{longtable}{|l|l|l|}
\caption{The basis of the space $\mathsf{V}^H$ for different groups}
\label{tab:5} \\
\hline $K_i$ & Tensor & Value  \\
\hline \endfirsthead \caption[]{continued} \\
\hline 1 & 2 & 3 \\
\hline \endhead \hline \multicolumn{3}{c}{\emph{Continued at next page}} \endfoot \hline \endlastfoot
$K_1$ & ${}_{Z^c_2}\mathsf{T}^{A_g,1,1}_{ijkl}$ & $\frac{1}{3}\mathsf{T}^{U^{0g},1,1}_{ijkl}+\frac{2}{3\sqrt{5}}\mathsf{T}^{U^{0g},2,1}_{ijkl}
+\frac{2}{3}\mathsf{T}^{U^{2g},1,0}_{ijkl}+\frac{2\sqrt{2}}{3\sqrt{7}}\mathsf{T}^{U^{2g},2,0}_{ijkl}
+\frac{2\sqrt{2}}{\sqrt{35}}\mathsf{T}^{U^{4g},1,0}_{ijkl}$ \\
$K_1$ & ${}_{Z^c_2}\mathsf{T}^{A_g,2,1}_{ijkl}$ & $\frac{1}{3}\mathsf{T}^{U^{0g},1,1}_{ijkl}+\frac{2}{3\sqrt{5}}\mathsf{T}^{U^{0g},2,1}_{ijkl}
-\frac{1}{3}\mathsf{T}^{U^{2g},1,0}_{ijkl}+\frac{1}{\sqrt{3}}\mathsf{T}^{U^{2g},1,2}_{ijkl}
-\frac{\sqrt{2}}{3\sqrt{7}}\mathsf{T}^{U^{2g},2,0}_{ijkl}$ \\
& & $+\frac{\sqrt{2}}{\sqrt{21}}\mathsf{T}^{U^{2g},2,2}_{ijkl}+\frac{3}{2\sqrt{70}}\mathsf{T}^{U^{4g},1,0}_{ijkl}
-\frac{1}{\sqrt{14}}\mathsf{T}^{U^{4g},1,2}_{ijkl}-\frac{1}{2\sqrt{2}}\mathsf{T}^{U^{4g},1,4}_{ijkl}$ \\
$K_1$ & ${}_{Z^c_2}\mathsf{T}^{A_g,3,1}_{ijkl}$ & $\frac{1}{3}\mathsf{T}^{U^{0g},1,1}_{ijkl}+\frac{2}{3\sqrt{5}}\mathsf{T}^{U^{0g},2,1}_{ijkl}
-\frac{1}{3}\mathsf{T}^{U^{2g},1,0}_{ijkl}-\frac{1}{\sqrt{3}}\mathsf{T}^{U^{2g},1,2}_{ijkl}
-\frac{\sqrt{2}}{3\sqrt{7}}\mathsf{T}^{U^{2g},2,0}_{ijkl}$ \\
& & $-\frac{\sqrt{2}}{\sqrt{21}}\mathsf{T}^{U^{2g},2,2}_{ijkl}-\frac{3}{2\sqrt{70}}\mathsf{T}^{U^{4g},1,0}_{ijkl}
+\frac{1}{\sqrt{14}}\mathsf{T}^{U^{4g},1,2}_{ijkl}-\frac{1}{2\sqrt{2}}\mathsf{T}^{U^{4g},1,4}_{ijkl}$ \\
$K_1$ & ${}_{Z^c_2}\mathsf{T}^{A_g,4,1}_{ijkl}$ & $\frac{1}{\sqrt{5}}\mathsf{T}^{U^{0g},2,1}_{ijkl}-\frac{\sqrt{2}}{\sqrt{7}}\mathsf{T}^{U^{2g},2,0}_{ijkl}
+\frac{1}{\sqrt{70}}\mathsf{T}^{U^{4g},1,0}_{ijkl}+\frac{1}{\sqrt{2}}\mathsf{T}^{U^{4g},1,4}_{ijkl}$ \\
$K_1$ & ${}_{Z^c_2}\mathsf{T}^{A_g,5,1}_{ijkl}$ & $\frac{1}{\sqrt{5}}\mathsf{T}^{U^{0g},2,1}_{ijkl}+\frac{\sqrt{2}}{\sqrt{7}}\mathsf{T}^{U^{2g},2,0}_{ijkl}
+\frac{\sqrt{3}}{\sqrt{7}}\mathsf{T}^{U^{2g},2,2}_{ijkl}-\frac{2\sqrt{2}}{\sqrt{35}}\mathsf{T}^{U^{4g},1,0}_{ijkl}
+\frac{\sqrt{2}}{\sqrt{7}}\mathsf{T}^{U^{4g},1,2}_{ijkl}$ \\
\hline
\end{longtable}

\subsection{The isotropy subgroups for the groups $K_i$}

Table~\ref{tab:4} shows the isotropy subgroups of the groups $K_i$. In this table, $Z^-_2$ is the order~2 group generated by the reflection through the $yz$-plane, and $\tilde{Z}_2$ is the group generated by a reflection leavind an edge of a cube invariant \cite{MR950168}. The group $H_0$ is always equal to $K_i$ and therefore is omitted.

\begin{table}[htbp]
\caption{The isotropy subgroups of the groups $K_i$}
\label{tab:4}
\begin{tabular}{lllllll}
\hline\noalign{\smallskip}
$K_i$ & $H_1$ & $H_2$ & $H_3$ & $H_4$ & $H_5$ & $H_6$ \\
\noalign{\smallskip}\hline\noalign{\smallskip}
$K_1$ & $I$ \\
$K_2$, $K_{16}$ & $\mathrm{O}(2)$ & \\
$K_3$ & $Z_2$ & $Z^-_2$ & $I$ \\
$K_4$, $K_{14}$ & $\mathrm{O}(2)$ & $Z^-_2\times Z^c_2$ & $Z^-_2$ \\
$K_5$ & $Z_2$ & $Z^-_2$ & $I$ \\
$K_6$, $K_{12}$ & $Z_4$ & $Z_2$ & $Z^-_2$ & $I$ \\
$K_7$, $K_{11}$ & $Z_3\times Z^c_2$ & $Z_2$ & $Z^-_2$ & $I$ \\
$K_8$ & $D_3$ & $D_2$ & $Z^c_2$ & $I$ \\
$K_9$, $K_{15}$ & $D_3$ & $D_4$ & $D_2$ & $\tilde{Z}_2$ & $Z_2$ & $I$ \\
$K_{10}$ & $Z_3$ & $Z^-_2$ & $Z_2$ & $I$ \\
$K_{13}$ & $Z_8$ & $Z_2$ & $Z^-_2$ & $I$ \\
\noalign{\smallskip}\hline
\end{tabular}
\end{table}

\subsection{The orbit type stratification}

The following formulae describe the orbit type stratification of the orbit space $\hat{\mathbb{R}}^3/ K_i$. The zeroth stratum is always equal to $\{\mathbf{0}\}$ and therefore is omitted.

$\hat{\mathbb{R}}^3/ K_1$:

\[
(\hat{\mathbb{R}}^3/ K_1)_1=\{p_3>0\}\cup\{p_2>0,p_3=0\}\cup\{\,(p_1,0,0)\colon p_1>0\,\}.
\]

$\hat{\mathbb{R}}^3/ K_2$, $\hat{\mathbb{R}}^3/ K_{16}$:

\[
(\hat{\mathbb{R}}^3/ K_2)_1=\{\,(0,0,p_3)\colon p_3>0\,\}.
\]

$\hat{\mathbb{R}}^3/ K_3$:

\[
\begin{aligned}
(\hat{\mathbb{R}}^3/ K_3)_1&=\{\,(0,0,p_3)\colon p_3>0\,\},\\
(\hat{\mathbb{R}}^3/ K_3)_2&=\{(p_1\neq 0,0,p_3>0)\},\\
(\hat{\mathbb{R}}^3/ K_3)_3&=\{\,(p_1,p_2>0,p_3>0)\colon p_3>0\,\}.
\end{aligned}
\]

$\hat{\mathbb{R}}^3/ K_4$, $\hat{\mathbb{R}}^3/ K_{14}$:

\[
\begin{aligned}
(\hat{\mathbb{R}}^3/ K_4)_1&=\{\,(p_1,0,0)\colon p_1>0\,\},\\
(\hat{\mathbb{R}}^3/ K_4)_2&=\{(0,0,p_3)\colon p_3>0\},\\
(\hat{\mathbb{R}}^3/ K_4)_3&=\{\,(p_1,0,p_3)\colon p_1>0,p_3>0\,\}.
\end{aligned}
\]

$\hat{\mathbb{R}}^3/ K_5$:

\begin{equation}\label{eq:6}
\begin{aligned}
(\hat{\mathbb{R}}^3/ K_5)_1&=\{\,(\lambda,0,0)\colon\lambda>0\,\},\\
(\hat{\mathbb{R}}^3/ K_5)_2&=\{\,(\lambda,\theta_p,0)\colon\lambda>0,0<\theta_p<\pi/2\,\},\\
(\hat{\mathbb{R}}^3/ K_5)_3&=\{\,(\lambda,\pi/2,\varphi_p)\colon\lambda>0,0<\varphi_p<\pi/m\,\},\\
(\hat{\mathbb{R}}^3/ K_5)_4&=\{\,(\lambda,\theta_p,\varphi_p)\colon\lambda>0,0<\theta_p<\pi/2,0<\varphi_p<\pi/m\,\}
\end{aligned}
\end{equation}
for $m=1$, where $(\lambda,\theta_p,\varphi_p)$ are the spherical coordinates in $\hat{\mathbb{R}}^3$.

$\hat{\mathbb{R}}^3/ K_6$, $\hat{\mathbb{R}}^3/ K_{12}$: \eqref{eq:6} with $m=2$.

$\hat{\mathbb{R}}^3/ K_7$, $\hat{\mathbb{R}}^3/ K_{11}$: \eqref{eq:6} with $m=3$.

$\hat{\mathbb{R}}^3/ K_8$:

\[
\begin{aligned}
(\hat{\mathbb{R}}^3/ K_8)_1&=\{\,(\lambda,\pi/4,0)\colon\lambda>0\,\},\\
(\hat{\mathbb{R}}^3/ K_8)_2&=\{\,(\lambda,0,0)\colon\lambda>0\,\},\\
(\hat{\mathbb{R}}^3/ K_8)_3&=\{\,(\lambda,\theta_p,0)\colon\lambda>0,0<\theta_p<\pi/4\,\},\\
(\hat{\mathbb{R}}^3/ K_8)_3&=\{\,(\lambda,\theta_p,\varphi)\colon\lambda>0,0<\varphi_p<\pi/2,
0<\theta_p<\cot^{-1}(\sqrt{2}\cos(\varphi_p-\pi/4))\,\}.
\end{aligned}
\]

$\hat{\mathbb{R}}^3/ K_9$, $\hat{\mathbb{R}}^3/ K_{15}$:

\[
\begin{aligned}
(\hat{\mathbb{R}}^3/ K_9)_1&=\{\,(p_1,p_2,p_3)\colon 0<p_1=p_2=p_3\,\},\\
(\hat{\mathbb{R}}^3/ K_9)_2&=\{\,(0,0,p_3)\colon p_3>0\,\},\\
(\hat{\mathbb{R}}^3/ K_9)_3&=\{\,(0,p_2,p_3)\colon 0<p_2=p_3\,\},\\
(\hat{\mathbb{R}}^3/ K_9)_4&=\{\,(0,p_2,p_3)\colon 0<p_2<p_3\,\},\\
(\hat{\mathbb{R}}^3/ K_9)_5&=\{\,(p_1,p_2,p_3)\colon 0<p_1=p_2<p_3\,\},\\
(\hat{\mathbb{R}}^3/ K_9)_6&=\{\,(p_1,p_2,p_3)\colon 0<p_1<p_2<p_3\,\}.
\end{aligned}
\]

$\hat{\mathbb{R}}^3/ K_{10}$:

\[
\begin{aligned}
(\hat{\mathbb{R}}^3/ K_{10})_1&=\{\,(\lambda,0,0)\colon\lambda>0\,\},\\
(\hat{\mathbb{R}}^3/ K_{10})_2&=\{\,(\lambda,\theta_p,0)\colon\lambda>0,0<\theta_p<\pi/2\,\},\\
(\hat{\mathbb{R}}^3/ K_{10})_3&=\{\,(\lambda,\pi/2,\varphi_p)\colon\lambda>0,0<\varphi_p<\pi/3\,\},\\
(\hat{\mathbb{R}}^3/ K_{10})_4&=\{\,(\lambda,\theta/2,\varphi_p)\colon\lambda>0,0<\theta_p<\pi/3,0<\varphi_p<\pi/3\,\}.
\end{aligned}
\]

$\hat{\mathbb{R}}^3/ K_{13}$: \eqref{eq:6} with $m=4$.

\section{The results}\label{sec:results}

In Theorem~$m$ below we denote by ${}_{K_m}\mathsf{T}_{ijkl}$ the tensors of the basis given in Table~\ref{tab:5} in the lines marked by $K_m$, $1\leq m\leq 16$. We say ``a triclinic (orthotropic, etc) random field'' instead of more rigourous ``a random field with triclinic (orthotropic, etc) symmetry''.

\subsection{The triclinic class}

\begin{theorem}[A triclinic random field in the triclinic class]\label{th:1}
The one-point correlation tensor of a homogeneous and $(Z^c_2,21A_g)$-isotropic random field $\mathsf{C}(\mathbf{x})$ is
\[
\langle\mathsf{C}(\mathbf{x})\rangle_{ijkl}=\sum_{m=1}^{21}C_m\,
{}_{Z^c_2}\mathsf{T}^{A_g,m,1}_{ijkl},
\]
where $C_m\in\mathbb{R}$. Its two-point correlation tensor has the form
\[
\langle\mathsf{C}(\mathbf{x}),\mathsf{C}(\mathbf{y})\rangle
=\int_{\hat{\mathbb{R}}^3/ Z^c_2}\cos(\mathbf{p},\mathbf{y}-\mathbf{x})
f(\mathbf{p})\,\mathrm{d}\Phi(\mathbf{p}),
\]
where $f(\mathbf{p})$ is a $\Phi$-equivalence class of measurable functions acting from $\hat{\mathbb{R}}^3/ Z^c_2$ to the set of nonnegative-definite symmetric linear operators on $\mathsf{V}^{Z^c_2}$ with unit trace, and $\Phi$ is a finite measure on $\hat{\mathbb{R}}^3/ Z^c_2$. The field has the form
\[
\begin{aligned}
\mathsf{C}_{ijkl}(\mathbf{x})&=\sum_{m=1}^{21}C_m\,
{}_{Z^c_2}\mathsf{T}^{A_g,m,1}_{ijkl}+\sum_{m=1}^{21}\int_{\hat{\mathbb{R}}^3/ Z^c_2}\cos(\mathbf{p},\mathbf{x})\,\mathrm{d}Z^{1}_m(\mathbf{p}){}_{Z^c_2}\mathsf{T}^{A_g,m,1}_{ijkl}\\
&\quad+\sum_{m=1}^{21}\int_{\hat{\mathbb{R}}^3/ Z^c_2}\sin(\mathbf{p},\mathbf{x})\,\mathrm{d}Z^{2}_m(\mathbf{p}){}_{Z^c_2}\mathsf{T}^{A_g,m,1}_{ijkl},
\end{aligned}
\]
where $(Z^{m}_1(\mathbf{p}),\dots,Z^{m}_{21}(\mathbf{p})^{\top}$ are two centred
uncorrelated $\mathsf{V}^{Z^c_2}$-valued random measures on $\hat{\mathbb{R}}^3/ Z^c_2$ with control measure $f(\mathbf{p})\,\mathrm{d}\Phi(\mathbf{p})$.
\end{theorem}

To formulate the next theorem, we need to introduce some notation. Let $f(\lambda)$, $\lambda\geq 0$ be a measurable function on $[0,\infty)$ taking values in the set of real symmetric nonnegative-definite matrices of size $21\times 21$ with unit trace. Assume that
\begin{equation}\label{eq:H1}
\begin{aligned}
f_{1,1}(\lambda)&=f_{1,3}(\lambda), & f_{1,2}(\lambda)&=f_{2,3}(\lambda), & f_{1,6}(\lambda)&=f_{3,4}(\lambda), \\
f_{1,7}(\lambda)&=f_{3,9}(\lambda), & f_{1,8}(\lambda)&=f_{3,8}(\lambda), & f_{2,4}(\lambda)&=f_{2,6}(\lambda), \\
f_{2,7}(\lambda)&=f_{2,9}(\lambda), & f_{4,4}(\lambda)&=f_{6,6}(\lambda), & f_{4,5}(\lambda)&=f_{5,6}(\lambda), \\
f_{4,9}(\lambda)&=f_{6,7}(\lambda), & f_{5,7}(\lambda)&=f_{5,9}(\lambda), & f_{7,7}(\lambda)&=f_{9,9}(\lambda), \\
f_{10,10}(\lambda)&=f_{14,14}(\lambda), & f_{10,11}(\lambda)&=f_{14,15}(\lambda), & f_{10,13}(\lambda)&=f_{14,17}(\lambda), \\
f_{11,11}(\lambda)&=f_{15,15}(\lambda), & f_{12,13}(\lambda)&=f_{16,17}(\lambda), & f_{13,13}(\lambda)&=f_{17,17}(\lambda), \\
f_{18,19}(\lambda)&=f_{18,20}(\lambda), & f_{19,19}(\lambda)&=f_{20,20}(\lambda), & f_{19,21}(\lambda)&=f_{20,21}(\lambda)
\end{aligned}
\end{equation}
and
\begin{equation}\label{eq:H2}
\begin{aligned}
f_{1,3}(\lambda)&=-f_{1,1}(\lambda)+8f_{5,5}(\lambda)-2f_{8,8}(\lambda)+4f_{1,8}(\lambda), \\
f_{1,4}(\lambda)&=f_{3,6}(\lambda)=f_{1,6}(\lambda)-4f_{18,19}(\lambda), \\
f_{1,5}(\lambda)&=f_{3,5}(\lambda)=\frac{1}{2}f_{1,1}(\lambda)-2f_{19,19}(\lambda)
-\frac{1}{2}f_{1,8}(\lambda), \\
f_{1,9}(\lambda)&=f_{3,7}(\lambda)=f_{1,7}(\lambda)-4f_{19,21}(\lambda), \\
f_{2,8}(\lambda)&=f_{1,2}(\lambda)-2f_{2,5}(\lambda), \\
f_{4,6}(\lambda)&=f_{4,4}(\lambda)-2f_{18,18}(\lambda), \\
f_{4,7}(\lambda)&=f_{6,9}(\lambda)=f_{4,9}(\lambda)-2f_{18,21}(\lambda), \\
f_{4,8}(\lambda)&=f_{6,8}(\lambda)=f_{1,6}(\lambda)-2f_{4,5}(\lambda)-2f_{18,19}(\lambda), \\
f_{5,8}(\lambda)&-\frac{1}{2}f_{1,1}(\lambda)+2f_{5,5}(\lambda)-f_{8,8}(\lambda)
+2f_{19,19}(\lambda)+\frac{3}{2}f_{1,8}(\lambda), \\
f_{7,8}(\lambda)&=f_{8,9}(\lambda)=f_{1,7}(\lambda)-2f_{5,7}(\lambda)-2f_{19,21}(\lambda), \\
f_{7,9}(\lambda)&=f_{7,7}(\lambda)-2f_{21,21}(\lambda), \\
f_{10,12}(\lambda)&=f_{14,16}(\lambda)=-\frac{1}{2}f_{11,11}(\lambda)
+\frac{1}{2}f_{12,12}(\lambda)-f_{10,11}(\lambda), \\
f_{11,12}(\lambda)&=f_{15,16}(\lambda)=-2f_{10,10}(\lambda)
+\frac{1}{2}f_{11,11}(\lambda)+\frac{1}{2}f_{12,12}(\lambda), \\
f_{11,13}(\lambda)&=f_{15,17}(\lambda)=f_{12,13}(\lambda)-2f_{10,13}(\lambda), \\
f_{19,20}(\lambda)&=\frac{1}{2}f_{1,1}(\lambda)-2f_{5,5}(\lambda)
+\frac{1}{2}f_{8,8}(\lambda)-f_{19,19}(\lambda)-f_{1,8}(\lambda).
\end{aligned}
\end{equation}
Assume also that all the entries of the matrix $f(\lambda)$ that lie over its main diagonal and were not mentioned previously, are equal to $0$.

Put
\begin{equation}\label{eq:uH}
\begin{aligned}
u_1(\lambda)&=2f_{1,1}(\lambda),\quad u_2(\lambda)=f_{2,2}(\lambda),\quad u_3(\lambda)=2f_{4,4}(\lambda),\\
u_4(\lambda)&=f_{5,5}(\lambda),\quad u_5(\lambda)=2f_{7,7}(\lambda),\quad u_6(\lambda)=f_{8,8}(\lambda),\\
u_7(\lambda)&=f_{1,2}(\lambda),\quad u_8(\lambda)=f_{1,6}(\lambda),\quad u_9(\lambda)=f_{1,7}(\lambda),\\
u_{10}(\lambda)&=f_{1,8}(\lambda),\quad u_{11}(\lambda)=f_{2,4}(\lambda),\quad u_{12}(\lambda)=f_{2,5}(\lambda),\\
u_{13}(\lambda)&=f_{2,7}(\lambda),\quad u_{14}(\lambda)=f_{4,5}(\lambda),\quad u_{15}(\lambda)=f_{4,9}(\lambda),\\
u_{16}(\lambda)&=f_{5,9}(\lambda),\quad u_{17}(\lambda)=2f_{10,10}(\lambda),\quad u_{18}(\lambda)=2f_{11,11}(\lambda),\\
u_{19}(\lambda)&=2f_{12,12}(\lambda),\quad u_{20}(\lambda)=2f_{13,13}(\lambda),\quad u_{21}(\lambda)=f_{10,11}(\lambda),\\
u_{22}(\lambda)&=f_{10,13}(\lambda),\quad u_{23}(\lambda)=f_{12,13}(\lambda),\quad u_{24}(\lambda)=f_{18,18}(\lambda),\\
u_{25}(\lambda)&=2f_{19,19}(\lambda),\quad u_{26}(\lambda)=f_{21,21}(\lambda),\quad u_{27}(\lambda)=f_{18,19}(\lambda),\\
u_{28}(\lambda)&=f_{18,21}(\lambda),\quad u_{29}(\lambda)=f_{19,21}(\lambda).
\end{aligned}
\end{equation}
and
\begin{equation}\label{eq:vi}
v_i(\lambda)=
\begin{cases}
  \frac{u_i(\lambda)}{u_1(\lambda)+\cdots+u_6(\lambda)}, & \mbox{if } 1\leq i\leq 5 \\
  \frac{u_{i+1}(\lambda)}{u_1(\lambda)+\cdots+u_6(\lambda)}, & \mbox{if } 6\leq i\leq 15 \\
  \frac{u_{i+1}(\lambda)}{u_{17}(\lambda)+\cdots+u_{20}(\lambda)}, & \mbox{if } 16\leq i\leq 18 \\
  \frac{u_{i+2}(\lambda)}{u_{17}(\lambda)+\cdots+u_{20}(\lambda)}, & \mbox{if } 19\leq i\leq 21 \\
  \frac{u_{i+2}(\lambda)}{u_{24}(\lambda)+u_{25}(\lambda)+u_{26}(\lambda)}, & \mbox{if } 22\leq i\leq 23 \\
  \frac{u_{i+3}(\lambda)}{u_{24}(\lambda)+u_{25}(\lambda)+u_{26}(\lambda)}, & \mbox{if } 24\leq i\leq 26.
\end{cases}
\end{equation}
The set $\mathcal{C}$ of the possible values of the function $f(\lambda)$ is a convex compact. The set of extreme points of $\mathcal{C}$ consists of three connected components. The functions $v_i(\lambda)$ with $1\leq i\leq 15$ (resp. $16\leq i\leq 21$, resp. $22\leq i\leq 26$) are coordinates in the closed convex hull of the first (resp. second, resp. third) component. The possible values for coordinates are determined by the following conditions: the principal minors of the matrix $f(\lambda)$ are non-negative.

Let $<$ be the lexicographic order on the sequences $tuijkl$, where $ijkl$ are indices that numerate the 21 component of the elasticity tensor, $t\geq 0$, and $-t\leq u\leq t$. Consider the infinite symmetric positive definite matrices given by
\[
\begin{aligned}
b^{t'u'i'j'k'l'}_{tuijkl}(m)&=\mathrm{i}^{t'-t}\sqrt{(2t+1)(2t'+1)}\sum^4_{n=0}
\frac{1}{4n+1}g^{0[0,0]}_{2n[t,t']}\\
&\quad\times\sum^{m_n}_{q=1}a_{nqm}\sum^{2n}_{v=-2n}
\mathsf{T}^{2n,q,v}_{i\cdots l'}g^{v[u,u']}_{2n[t,t']}
\end{aligned}
\]
with $1\leq m\leq 13$. Let $L(m)$ be the infinite lower triangular matrices of the Cholesky factorisation of the matrices $b^{t'u'i'j'k'l'}_{tuijkl}(m)$ constructed in \cite{MR2734774}. Let $Z'_{mtuijkl}$ be the sequence of centred scattered random measures with $\Phi_m$ as their control measures. Define
\[
Z_{mtuijkl}=\sum_{(t'u'i'j'k'l')\leq(tuijkl)}
L^{t'u'i'j'k'l'}_{tuijkl}(m)Z'_{mtuijkl}.
\]

\begin{theorem}[An isotropic random field in the triclinic class]\label{th:2}
The one-point correlation tensor of a homogeneous and $(\mathrm{O}(3),2U^{0g}\oplus 2U^{2g}\oplus U^{4g})$-isotropic random field $\mathsf{C}(\mathbf{x})$ is
\[
\langle\mathsf{C}(\mathbf{x})\rangle_{ijkl}=C_1\mathsf{T}^{U^{0g},1,1}_{ijkl}
+C_2\mathsf{T}^{U^{0g},2,1}_{ijkl},
\]
where $C_1$, $C_2\in\mathbb{R}$. Its two-point correlation tensor has the
spectral expansion
\begin{equation}\label{eq:2point}
\langle\mathsf{C}(\mathbf{x}),\mathsf{C}(\mathbf{y})\rangle_{ijkli'j'k'l'}
=\sum^3_{n=1}\int^{\infty}_0\sum^{29}_{q=1}N_{nq}(\lambda,\rho)L^q_{iikli'j'k'l'} (\mathbf{y}-\mathbf{x})\,\mathrm{d}\Phi_n(\lambda),
\end{equation}
where the functions $N_{nq}(\lambda,\rho)$ are given in Table~\emph{\ref{tab:11}} and the functions $L^q_{iikli'j'k'l'}$ are given in Table~\emph{\ref{tab:6}}. The measures $\Phi_n(\lambda)$ satisfy the condition
\begin{equation}\label{eq:phi2phi3}
\Phi_2(\{0\})=2\Phi_3(\{0\}).
\end{equation}
The spectral expansion of the field has the form
\[
\begin{aligned}
\mathsf{C}_{ijkl}(\rho,\theta,\varphi)&=C_1\mathsf{T}^{U^{0g},1,1}_{ijkl}
+C_2\mathsf{T}^{U^{0g},2,1}_{ijkl}\\
&\quad+2\sqrt{\pi}\sum^{13}_{m=1}\sum^{\infty}_{t=0}\sum^{t}_{u=-t}\int^{%
\infty}_0
j_t(\lambda\rho)\,\mathrm{d}Z_{mtuijkl}(\lambda)S^u_t(\theta,\varphi),
\end{aligned}
\]
where $S^u_t(\theta,\varphi)$ are real-valued spherical harmonics.
\end{theorem}

\begin{longtable}{|p{0.03\textwidth}|p{0.05\textwidth}|p{0.87\textwidth}|}
\caption{The functions $N_{nq}(\lambda,\rho)$.}\label{tab:11} \\
\hline $n$ & $q$ & $N_{nq}(\lambda,\rho)$ \\
\hline \endfirsthead \caption[]{continued} \\
\hline 1 & 2 & 3 \\
\hline \endhead \hline \multicolumn{3}{c}{\emph{Continued at next page}} \endfoot \hline \endlastfoot
1 & 1 & $[\frac{22}{405}+\frac{1}{81}v_1(\lambda)-\frac{127}{2835}v_2(\lambda)
-\frac{46}{2835}v_3(\lambda)-\frac{26}{81}v_4(\lambda)+\frac{34}{567}v_5(\lambda)$ \\
& & $+\frac{4}{35}v_6(\lambda)-\frac{32}{105}v_7(\lambda)
+\frac{64}{105}v_8(\lambda)-\frac{8}{35}v_9(\lambda)
+\frac{8}{105}v_{10}(\lambda)-\frac{12}{35}v_{11}(\lambda)$ \\
& & $+\frac{32}{105}v_{13}(\lambda)-\frac{5608}{19845}v_{14}(\lambda)
-\frac{32}{35}v_{15}(\lambda)]j_0(\lambda\rho)$ \\
& & $+[-\frac{1534}{1323}+\frac{198523}{218295}v_1(\lambda)+\frac{52226}{43659}v_2(\lambda)
+\frac{510788}{509355}v_3(\lambda)+\frac{3694}{1323}v_4(\lambda)+\frac{58846}{43659}v_5(\lambda)$ \\
& & $+\frac{8224}{43659}v_6(\lambda)-\frac{1696}{14553}v_7(\lambda)
+\frac{11800}{43659}v_8(\lambda)+\frac{21920}{27783}v_9(\lambda)
-\frac{1168}{14553}v_{10}(\lambda)-\frac{1760}{3969}v_{11}(\lambda)$ \\
& & $-\frac{3116}{43659}v_{12}(\lambda)+\frac{1712}{1323}v_{13}(\lambda)
-\frac{8224}{14553}v_{14}(\lambda)+\frac{664}{3969}v_{15}(\lambda)]j_2(\lambda\rho)$ \\
& & $+[-\frac{8}{35}-\frac{983}{5005}v_1(\lambda)+\frac{165622}{105105}v_2(\lambda)
+\frac{640}{1001}v_3(\lambda)+\frac{104}{35}v_4(\lambda)+\frac{2022}{5005}v_5(\lambda)$ \\
& & $+\frac{35866}{315315}v_6(\lambda)-\frac{34024}{105105}v_7(\lambda)
-\frac{206672}{315315}v_8(\lambda)-\frac{429718}{848925}v_9(\lambda)
-\frac{6056}{21021}v_{10}(\lambda)+\frac{42596}{121275}v_{11}(\lambda)$ \\
\hline
\end{longtable}

Let $\nu$ be a nonnegative integer. The \emph{Ogden tensor} \cite{MR0342001} $\mathsf{I}^{\nu}$ of rank $2\nu+2$ is determined inductively as
\[
\begin{aligned}
\mathsf{I}^0_{ij}&:=\delta_{ij},\qquad\mathsf{I}^1_{ijk\ell}:=
\frac{1}{2}(\delta_{ik}\delta_{j\ell}+\delta_{i\ell}\delta_{jk}),\\
\mathsf{I}^{\nu}_{i_1\cdots i_{2\nu+2}}&:=\nu^{-1}(\mathsf{I}^1_{i_1pi_3i_4}\mathsf{I}^{\nu-1}_{pi_2i_5\cdots i_{2\nu+2}}+\cdots+\mathsf{I}^1_{i_1pi_{2\nu+1}i_{2\nu+2}}\mathsf{I}^{\nu-1}_{pi_2\cdots i_{2\nu-1}i_{2\nu}}),
\end{aligned}
\]
where there is a summation over $p$. In what follows we will omit the upper index.

\begin{longtable}{|l|l|}
\caption{The functions $L^q_{ijkli'j'k'l'}(\mathbf{x})$.}\label{tab:6} \\
\hline \textbf{Function} & \textbf{Value} \\
\hline \endfirsthead \caption[]{continued} \\
\hline 1 & 2 \\
\hline \endhead \hline \multicolumn{2}{c}{\emph{Continued at next page}} \endfoot \hline \endlastfoot
$L^1_{i\cdots l'}$ & $\delta_{ij}\delta_{kl}\delta_{i'j'}\delta_{k'l'}$ \\
$L^2_{i\cdots l'}$ & $2(\delta_{ij}\delta_{kl}I_{i'j'k'l'}+\delta_{i'j'}\delta_{k'l'}I_{ijkl})$ \\
$L^3_{i\cdots l'}$ & $2(\delta_{ij}(\delta_{i'j'}I_{kl k'l'}+\delta_{k'l'}I_{kl i'j'})+\delta_{kl}(\delta_{i'j'}I_{ijk'l'}+\delta_{k'l'}I_{iji'j'}))$ \\
$L^4_{i\cdots l'}$ & $4I_{ijkl}I_{i'j'k'l'}$ \\
$L^5_{i\cdots l'}$ & $8(\delta_{ij}I_{kl i'j'k'l'}+\delta_{kl}I_{iji'j'k'l'}+\delta_{i'j'}I_{ijkl k'l'}+\delta_{k'l'}I_{ijkl i'j'})$ \\
$L^6_{i\cdots l'}$ & $4(I_{iji'j'}I_{kl k'l'}+I_{ijk'l'}I_{kl i'j'})$ \\
$L^7_{i\cdots l'}$ & $4(I_{iji'k'}I_{kl j'l'}+I_{iji'l'}I_{kl j'k'}+I_{ijj'k'}I_{kl i'l'}+I_{ijj'l'}I_{kl i'k'})$ \\
$\|\mathbf{x}\|^2L^8_{i\cdots l'}(\mathbf{x})$ & $\delta_{ij}\delta_{kl}(\delta_{i'j'}x_{k'}x_{l'}+\delta_{k'l'}x_{i'}x_{j'})
+\delta_{i'j'}\delta_{k'l'}(\delta_{ij}x_kx_{l}+\delta_{kl}x_ix_j)$ \\
$\|\mathbf{x}\|^2L^9_{i\cdots l'}(\mathbf{x})$ & $2(I_{ijkl}(\delta_{i'j'}x_{k'}x_{l'}+\delta_{k'l'}x_{i'}x_{j'})
+I_{i'j'k'l'}(\delta_{ij}x_kx_{l}+\delta_{kl}x_ix_j))$ \\
$\|\mathbf{x}\|^2L^{10}_{i\cdots l'}(\mathbf{x})$ & $\delta_{ij}\delta_{kl}(\delta_{i'k'}x_{j'}x_{l'}+\delta_{i'l'}x_{j'}x_{k'}
+\delta_{j'k'}x_{i'}x_{l'}+\delta_{j'l'}x_{i'}x_{k'})$ \\
& $+\delta_{i'j'}\delta_{k'l'}(\delta_{ik}x_jx_{l}+\delta_{il}x_jx_k
+\delta_{jk}x_ix_{l}+\delta_{jl}x_ix_k)$ \\
$\|\mathbf{x}\|^2L^{11}_{i\cdots l'}(\mathbf{x})$ & $\delta_{ij}\delta_{i'j'}(\delta_{kk'}x_{l}x_{l'}+\delta_{kl'}x_{l}x_{k'}
+\delta_{l k'}x_kx_{l'}+\delta_{ll'}x_kx_{k'})$ \\
& $+\delta_{ij}\delta_{k'l'}(\delta_{ki'}x_{l}x_{j'}+\delta_{kj'}x_{l}x_{j'}
+\delta_{l i'}x_kx_{j'}+\delta_{l j'}x_kx_{i'})$ \\
& $+\delta_{kl}\delta_{i'j'}(\delta_{ik'}x_jx_{l'}+\delta_{jk'}x_jx_{l'}
+\delta_{il'}x_jx_{k'}+\delta_{jl'}x_ix_{k'})$ \\
& $+\delta_{kl}\delta_{k'l'}(\delta_{ii'}x_jx_{j'}+\delta_{ij'}x_jx_{i'}
+\delta_{ji'}x_ix_{j'}+\delta_{jj'}x_ix_{i'})$ \\
$\|\mathbf{x}\|^2\mathbf{L^{12}_{i\cdots l'}(x)}$ & $2(I_{ijkl}(\delta_{i'k'}x_{j'}x_{l'}+\delta_{i'l'}x_{j'}x_{k'}
+\delta_{j'k'}x_{i'}x_{l'}+\delta_{j'l'}x_{i'}x_{k'})$ \\
& $+I_{i'j'k'l'}(\delta_{ik}x_jx_{l}+\delta_{il}x_jx_k
+\delta_{jk}x_ix_{l}+\delta_{jl}x_ix_k))$ \\
$\|\mathbf{x}\|^2L^{13}_{i\cdots l'}(\mathbf{x})$ & $2((\delta_{ij}I_{kl i'j'}+\delta_{kl}I_{iji'j'})x_{k'}x_{l'}+(\delta_{ij}I_{kl k'l'}+\delta_{kl}I_{ijk'l'})x_{i'}x_{j'}$ \\
& $+(\delta_{i'j'}I_{ijk'l'}+\delta_{k'l'}I_{iji'j'})x_kx_{l}
+(\delta_{i'j'}I_{kl k'l'}+\delta_{k'l'}I_{kl i'j'})x_ix_j)$ \\
$\|\mathbf{x}\|^2L^{14}_{i\cdots l'}(\mathbf{x})$ & $2((\delta_{ij}I_{kl i'k'}+\delta_{kl}I_{iji'k'})x_{j'}x_{l'}+(\delta_{ij}I_{kl i'l'}+\delta_{kl}I_{iji'l'})x_{j'}x_{k'}$ \\
& $+(\delta_{ij}I_{kl j'k'}+\delta_{kl}I_{ijj'k'})x_{i'}x_{l'}+(\delta_{ij}I_{kl j'l'}+\delta_{kl}I_{ijj'l'})x_{i'}x_{k'}$ \\
& $+(\delta_{i'j'}I_{ikk'l'}+\delta_{k'l'}I_{iki'j'})x_jx_{l}
+(\delta_{i'j'}I_{il k'l'}+\delta_{k'l'}I_{il i'j'})x_jx_k$ \\
& $+(\delta_{i'j'}I_{jkk'l'}+\delta_{k'l'}I_{jki'j'})x_ix_{l}
+(\delta_{i'j'}I_{jl k'l'}+\delta_{k'l'}I_{jl i'j'})x_ix_k)$ \\
$\|\mathbf{x}\|^2L^{15}_{i\cdots l'}(\mathbf{x})$ & $8(I_{ijkl i'j'}x_{k'}x_{l'}
+I_{ijkl k'l'}x_{i'}x_{j'}+I_{iji'j'k'l'}x_kx_{l}
+I_{kl i'j'k'l'}x_ix_j)$ \\
$\|\mathbf{x}\|^2L^{16}_{i\cdots l'}(\mathbf{x})$ & $8(I_{ijkl i'k'}x_{j'}x_{l'}
+I_{ijkl i'l'}x_{j'}x_{k'}+I_{ijkl j'k'}x_{i'}x_{l'}
+I_{ijkl j'l'}x_{i'}x_{k'}$ \\
& $+I_{iki'j'k'l'}x_jx_{l}+I_{il i'j'k'l'}x_jx_k+I_{jki'j'k'l'}x_ix_{l}+I_{jl i'j'k'l'}x_ix_k)$ \\
$\|\mathbf{x}\|^2L^{17}_{i\cdots l'}(\mathbf{x})$ & $2(I_{iji'j'}(\delta_{kk'}x_{l}x_{l'}+\delta_{kl'}x_{l}x_{k'}
+\delta_{l k'}x_kx_{l'}+\delta_{ll'}x_kx_{k'})$ \\
& $+I_{ijk'l'}(\delta_{ki'}x_{l}x_{j'}+\delta_{kj'}x_{l}x_{i'}
+\delta_{l i'}x_kx_{j'}+\delta_{l j'}x_kx_{i'}$ \\
& $+I_{kl i'j'}(\delta_{ik'}x_jx_{l'}+\delta_{il'}x_jx_{k'}
+\delta_{jk'}x_ix_{l'}+\delta_{jl'}x_ix_{k'})$ \\
& $+I_{kl k'l'}(\delta_{ii'}x_jx_{j'}+\delta_{ij'}x_jx_{i'}
+\delta_{ji'}x_ix_{j'}+\delta_{jj'}x_ix_{i'})$ \\
$\|\mathbf{x}\|^4L^{18}_{i\cdots l'}(\mathbf{x})$ & $\delta_{ij}\delta_{kl}x_{i'}x_{j'}x_{k'}x_{l'}
+\delta_{i'j'}\delta_{k'l'}x_ix_jx_kx_{l}$ \\
$\|\mathbf{x}\|^4L^{19}_{i\cdots l'}(\mathbf{x})$ & $(\delta_{ij}x_kx_{l}+\delta_{kl}x_ix_j)
(\delta_{i'j'}x_{k'}x_{l'}+\delta_{k'l'}x_{i'}x_{j'})$ \\
$\|\mathbf{x}\|^4L^{20}_{i\cdots l'}(\mathbf{x})$ & $2(I_{ijkl}x_{i'}x_{j'}x_{k'}x_{l'}+I_{i'j'k'l'}x_ix_jx_kx_{l})$ \\
$\|\mathbf{x}\|^4L^{21}_{i\cdots l'}(\mathbf{x})$ & $(\delta_{ij}x_kx_{l}+\delta_{kl}x_ix_j)(\delta_{i'k'}x_{j'}x_{l'}
+\delta_{i'l'}x_{j'}x_{k'}+\delta_{j'k'}x_{i'}x_{l'}+\delta_{j'l'}x_{i'}x_{k'})$ \\
& $+(\delta_{i'j'}x_{k'}x_{l'}+\delta_{k'l'}x_{i'}x_{j'})(\delta_{ik}x_jx_{l}
+\delta_{il}x_jx_k+\delta_{jk}x_ix_{l}+\delta_{jl}x_ix_k)$ \\
$\|\mathbf{x}\|^4L^{22}_{i\cdots l'}(\mathbf{x})$ & $\delta_{ij}(\delta_{ki'}x_{l}x_{j'}x_{k'}x_{l'}
+\delta_{kj'}x_{l}x_{i'}x_{k'}x_{l'}+\delta_{kk'}x_{l}x_{i'}x_{j'}x_{l'}
+\delta_{kl'}x_{l}x_{i'}x_{j'}x_{k'}$ \\
& $+\delta_{l i'}x_kx_{j'}x_{k'}x_{l'}+\delta_{l j'}x_kx_{i'}x_{k'}x_{l'}+\delta_{l k'}x_kx_{i'}x_{j'}x_{l'}+\delta_{ll'}x_kx_{i'}x_{j'}x_{k'})$ \\
& $+\delta_{kl}(\delta_{ii'}x_jx_{j'}x_{k'}x_{l'}
+\delta_{ij'}x_jx_{i'}x_{k'}x_{l'}+\delta_{ik'}x_jx_{i'}x_{j'}x_{l'}
+\delta_{il'}x_jx_{i'}x_{j'}x_{k'}$ \\
& $+\delta_{ji'}x_ix_{j'}x_{k'}x_{l'}+\delta_{jj'}x_ix_{i'}x_{k'}x_{l'}
+\delta_{jk'}x_ix_{i'}x_{j'}x_{l'}+\delta_{jl'}x_ix_{i'}x_{j'}x_{k'})$ \\
& $+\delta_{i'j'}(\delta_{ik'}x_jx_kx_{l}x_{l'}
+\delta_{jk'}x_ix_kx_{l}x_{l'}+\delta_{kk'}x_ix_jx_{l}x_{l'}
+\delta_{l k'}x_ix_jx_kx_{l'}$ \\
& $+\delta_{il'}x_kx_{l'}x_{j'}x_{k'}+\delta_{jl'}x_kx_{l'}x_{i'}x_{k'}
+\delta_{kl'}x_ix_jx_{k'}x_{l'}+\delta_{ll'}x_ix_jx_kx_{k'})$ \\
& $+\delta_{k'l'}(\delta_{ii'}x_jx_kx_{l}x_{j'}
+\delta_{ji'}x_ix_kx_{j'}x_{l'}+\delta_{ki'}x_jx_jx_{l}x_{j'}
+\delta_{l i'}x_ix_jx_kx_{j'}$ \\
& $+\delta_{ij'}x_jx_kx_{l}x_{i'}+\delta_{jj'}x_ix_kx_{l}x_{i'}
+\delta_{kj'}x_ix_jx_{l}x_{i'}+\delta_{l j'}x_ix_jx_kx_{i'})$ \\
$\|\mathbf{x}\|^4L^{23}_{i\cdots l'}(\mathbf{x})$ & $(\delta_{ik}x_jx_{l}+\delta_{il}x_jx_k+\delta_{jk}x_ix_{l}+\delta_{jl}x_ix_k)$ \\
& $\times(\delta_{i'k'}x_{j'}x_{l'}+\delta_{i'l'}x_{j'}x_{k'}
+\delta_{j'k'}x_{i'}x_{l'}+\delta_{j'l'}x_{i'}x_{k'})$ \\
$\|\mathbf{x}\|^4L^{24}_{i\cdots l'}(\mathbf{x})$ & $2(I_{iji'j'}x_kx_{l}x_{k'}x_{l'}+I_{ijk'l'}x_kx_{l}x_{i'}x_{j'}
+I_{kl i'j'}x_ix_jx_{k'}x_{l'}+I_{kl k'l'}x_ix_jx_{i'}x_{j'})$ \\
$\|\mathbf{x}\|^4L^{25}_{i\cdots l'}(\mathbf{x})$ & $2[(I_{iji'k'}x_{j'}x_{l'}+I_{iji'l'}x_{j'}x_{k'}+I_{ijj'k'}x_{i'}x_{l'}
+I_{ijj'l'}x_{i'}x_{k'})x_kx_{l}$ \\
& $+(I_{iki'j'}x_{k'}x_{l'}+I_{ikk'l'}x_{i'}x_{j'})x_jx_{l}
+(I_{il i'j'}x_{k'}x_{l'}+I_{il k'l'}x_{i'}x_{j'})x_jx_k$ \\
& $+(I_{jki'j'}x_{k'}x_{l'}+I_{jkk'l'}x_{i'}x_{j'})x_ix_{l}
+(I_{jl i'j'}x_{k'}x_{l'}+I_{jl k'l'}x_{i'}x_{j'})x_jx_k$ \\
& $+(I_{kl i'k'}x_{j'}x_{l'}+I_{kl i'l'}x_{j'}x_{k'}+I_{kl j'k'}x_{i'}x_{l'}+I_{kl j'l'}x_{i'}x_{k'})x_ix_j]$ \\
$\|\mathbf{x}\|^6L^{26}_{i\cdots l'}(\mathbf{x})$ & $(\delta_{ij}x_kx_{l}+\delta_{kl}x_ix_j)x_{i'}x_{j'}x_{k'}x_{l'}
+(\delta_{i'j'}x_{k'}x_{l'}+\delta_{k'l'}x_{i'}x_{j'})x_ix_jx_kx_{l}$ \\
$\|\mathbf{x}\|^6L^{27}_{i\cdots l'}(\mathbf{x})$ & $(\delta_{ik}x_jx_{l}
+\delta_{il}x_jx_k+\delta_{jk}x_ix_{l}+\delta_{jl}x_ix_k)
x_{i'}x_{j'}x_{k'}x_{l'}$ \\
& $+(\delta_{i'k'}x_{j'}x_{l'}+\delta_{i'l'}x_{j'}x_{k'}
+\delta_{j'k'}x_{i'}x_{l'}+\delta_{j'l'}x_{i'}x_{k'})
x_ix_jx_kx_{l}$ \\
$\|\mathbf{x}\|^6L^{28}_{i\cdots l'}(\mathbf{x})$ & $(\delta_{ii'}x_{j'}x_{k'}x_{l'}+\delta_{ij'}x_{i'}x_{k'}x_{l'}
+\delta_{ik'}x_{i'}x_{j'}x_{l'}+\delta_{il'}x_{i'}x_{j'}x_{k'})x_jx_kx_{l}$ \\
& $+(\delta_{ji'}x_{j'}x_{k'}x_{l'}+\delta_{jj'}x_{i'}x_{k'}x_{l'}
+\delta_{jk'}x_{i'}x_{j'}x_{l'}+\delta_{jl'}x_{i'}x_{j'}x_{k'})x_ix_kx_{l}$ \\
& $+(\delta_{ki'}x_{j'}x_{k'}x_{l'}+\delta_{kj'}x_{i'}x_{k'}x_{l'}
+\delta_{kk'}x_{i'}x_{j'}x_{l'}+\delta_{kl'}x_{i'}x_{j'}x_{k'})x_ix_jx_{l}$ \\
& $+(\delta_{l i'}x_{j'}x_{k'}x_{l'}+\delta_{l j'}x_{i'}x_{k'}x_{l'}+\delta_{l k'}x_{i'}x_{j'}x_{l'}+\delta_{ll'}x_{i'}x_{j'}x_{k'})x_ix_jx_k$ \\
$\mathbf{\|x\|^8L^{29}_{i\cdots l'}(x)}$ & $x_ix_jx_kx_{l}x_{i'}x_{j'}x_{k'}x_{l'}$ \\
\hline
\end{longtable}

\subsection{The monoclinic class}

\begin{theorem}[A monoclinic random field in the monoclinic class]\label{th:3}
The one-point correlation tensor of a homogeneous and $(Z_2\times Z^c_2,13A_g)$-isotropic random field $\mathsf{C}(\mathbf{x})$ is
\[
\langle\mathsf{C}(\mathbf{x})\rangle_{ijkl}=\sum_{m=1}^{13}C_m\,
{}_{Z_2\times Z^c_2}\mathsf{T}^{A_g,m,1}_{ijkl},
\]
where $C_m\in\mathbb{R}$. Its two-point correlation tensor has the form
\[
\langle\mathsf{C}(\mathbf{x}),\mathsf{C}(\mathbf{y})\rangle =
\frac{1}{2}\int_{\hat{\mathbb{R}}^3/ Z_2\times Z^c_2}\cos(p_1(y_1-x_1)+p_2(y_2-x_2))\cos(p_3(y_3-x_3))
f(\mathbf{p})\,\mathrm{d}\Phi(\mathbf{p}),
\end{equation*}
where $f(\mathbf{p})$ is a $\Phi$-equivalence class of measurable functions acting from $\hat{\mathbb{R}}^3/Z_2\times Z^c_2$ to the set of nonnegative-definite symmetric linear operators on $\mathsf{V}^{Z_2\times Z^c_2}$ with unit trace, and $\Phi$ is a finite Radon measure on $\hat{\mathbb{R}}^3/Z_2\times Z^c_2$. The field has the
form
\[
\begin{aligned}
\mathsf{C}(\mathbf{x})_{ijkl}&=\sum_{m=1}^{13}C_m\,
{}_{Z_2\times Z^c_2}\mathsf{T}^{A_g,m,1}_{ijkl}\\
&\quad+\frac{1}{\sqrt{2}}\sum_{m=1}^{13}\int_{\hat{\mathbb{R}}^3/Z_2\times Z^c_2}\cos(p_1x+p_2y)\cos(p_3z)
\,\mathrm{d}\mathbf{Z}^{1}_m(\mathbf{p}){}_{Z_2\times
Z^c_2}\mathsf{T}^{A_g,m,1}_{ijkl}\\
&\quad+\frac{1}{\sqrt{2}}\sum_{m=1}^{13}\int_{\hat{\mathbb{R}}^3/Z_2\times Z^c_2}\sin(p_1x+p_2y)\sin(p_3z) \,\mathrm{d}Z^{2}_m(\mathbf{p}){}_{Z_2\times
Z^c_2}\mathsf{T}^{A_g,m,1}_{ijkl}\\
&\quad+\frac{1}{\sqrt{2}}\sum_{m=1}^{13}\int_{\hat{\mathbb{R}}^3/ Z_2\times Z^c_2}\cos(p_1x+p_2y)\sin(p_3z) \,\mathrm{d}Z^{3}_m(\mathbf{p}){}_{Z_2\times
Z^c_2}\mathsf{T}^{A_g,m,1}_{ijkl}\\
&\quad+\frac{1}{\sqrt{2}}\sum_{m=1}^{13}\int_{\hat{\mathbb{R}}^3/ Z_2\times Z^c_2}\sin(p_1x+p_2y)\cos(p_3z) \,\mathrm{d}Z^{4}_m(\mathbf{p}){}_{Z_2\times
Z^c_2}\mathsf{T}^{A_g,m,1}_{ijkl},
\end{aligned}
\]
where $(Z^{n}_1(\mathbf{p}),\dots,Z^n_{13}(\mathbf{p}))^{\top}$ are four
centred uncorrelated $\mathsf{V}^{Z_2\times Z^c_2}$-valued random measures on $\hat{\mathbb{R}}^3/ Z_2\times Z^c_2$ with control measure $f(\mathbf{p})\,\mathrm{d}\Phi(\mathbf{p})$.
\end{theorem}

\begin{theorem}[A transverse isotropic random field in the monoclinic class]\label{th:4}
The one-point correlation tensor of a homogeneous and $(\mathrm{O}(2)\times Z^c_2,5U^{0gg}\oplus 3U^{2g}\oplus U^{4g})$-isotropic random field $\mathsf{C}(\mathbf{x})$ is
\[
\langle\mathsf{C}(\mathbf{x})\rangle_{ijkl}=\sum_{m=1}^{5}C_m\, {}_{\mathrm{O}(2)\times Z^c_2}\mathsf{T}^{U^{0gg},m,1}_{ijkl},
\]
where $C_m\in\mathbb{R}$. Its two-point correlation tensor has the form
\[
\langle\mathsf{C}(\mathbf{x}),\mathsf{C}(\mathbf{y})\rangle
=\int_{\hat{\mathbb{R}}^3/\mathrm{O}(2)\times Z^c_2}J_0\left(\sqrt{(p^2_1+p^2_2)(z^2_1+z^2_2)}\right)\cos(p_3z_3)
f(\mathbf{p}) \,\mathrm{d}\Phi(\mathbf{p}),
\]
where $\Phi$ is a measure on $\hat{\mathbb{R}}^3/\mathrm{O}(2)\times Z^c_2$, and $f(\mathbf{p})$ is a $\Phi$-equivalence class of measurable functions on $\hat{\mathbb{R}}^3/\mathrm{O}(2)\times Z^c_2$ with values in the compact set of all nonnegative-definite linear operators in the space $\mathsf{V}^{Z_2\times Z^c_2}$ with unit trace of the form
\[
\begin{pmatrix}
  A & 0 & 0 & 0 & 0 \\
  0 & B_1 & B_2 & B_3 & 0 \\
  0 & B_2 & B_4 & B_5 & 0 \\
  0 & B_3 & B_5 & B_6 & 0 \\
  0 & 0 & 0 & 0 & B_7
\end{pmatrix}
,
\]
where $A$ is a nonnegative-definite $5\times 5$ matrix, and $B_m$, $1\leq m\leq 7$ are $2\times 2$ matrices proportional to the identity matrix. The field has the form
\[
\begin{aligned}
\mathsf{C}(\mathbf{x})&=\sum_{m=1}^{5}C_m\mathsf{T}^m_{ijkl}\\
&\quad+\sum_{m=1}^{13}\int_{\hat{\mathbb{R}}^3/\mathrm{O}(2)\times Z^c_2}J_0\left(\sqrt{(p^2_1+p^2_2)(z^2_1+z^2_2)}\right)\\
&\quad\times(\cos(p_3z)\mathrm{d}Z^{01}_m(\mathbf{p})\mathsf{T}^m_{ijkl}
+\sin(p_3z)\mathrm{d}Z^{02}_m(\mathbf{p})\mathsf{T}^m_{ijkl})\\
&\quad+\sqrt{2}\sum_{\ell=1}^{\infty}\sum_{m=1}^{13}\int_{\hat{\mathbb{R}}^3
/\mathrm{O}(2)\times Z^c_2}J_{\ell}\left(\sqrt{(p^2_1+p^2_2)(z^2_1+z^2_2)}\right)\\
&\quad\times(\cos(p_3z)\cos(\ell\varphi_p)\mathrm{d}Z^{\ell
1}_m(\mathbf{p})\mathsf{T}^m_{ijkl}+\cos(p_3z)\sin(\ell\varphi_p)\mathrm{d}Z^{\ell
2m}(\mathbf{p})\mathsf{T}^m_{ijkl}\\
&\quad+\sin(p_3z)\cos(\ell\varphi_p)\mathrm{d}Z^{\ell
3}_m(\mathbf{p})\mathsf{T}^m_{ijkl}+\sin(p_3z)\sin(\ell\varphi_p)\mathrm{d}Z^{\ell
4}_m(\mathbf{p})\mathsf{T}^m_{ijkl}),
\end{aligned}
\]
where $(Z^{\ell i}_1(\mathbf{p}),\dots,Z^{\ell i}_{13}(\mathbf{p}))^{\top}$ are centred uncorrelated $\mathsf{V}^{Z_2\times Z^c_2}$-valued random measures on $\hat{\mathbb{R}}^3/\mathrm{O}(2)\times Z^c_2$ with control measure $f(\mathbf{p})\,\mathrm{d}\Phi(\mathbf{p})$, $J_{\ell}$ are the Bessel functions, and
\[
\mathsf{T}^m_{ijkl}=
\begin{cases}
  {}_{\mathrm{O}(2)\times Z^c_2}\mathsf{T}^{U^{0gg},m,1}, & \mbox{if } 1\leq m\leq 5, \\
  {}_{\mathrm{O}(2)\times Z^c_2}\mathsf{T}^{U^{2g},\lfloor m/2\rfloor-2,m\bmod 2+1}, & \mbox{if } 6\leq m\leq 11,\\
  {}_{\mathrm{O}(2)\times Z^c_2}\mathsf{T}^{U^{4g},1,m-11}, & \mbox{if } 12\leq m\leq 13.
\end{cases}
\]
\end{theorem}

\subsection{The orthotropic class}

\begin{theorem}[An orthotropic random field in the orthotropic class]\label{th:5}
The one-point correlation tensor of a homogeneous and $(D_2\times Z^c_2,9A_g)$-isotropic random field $\mathsf{C}(\mathbf{x})$ is
\[
\langle\mathsf{C}(\mathbf{x})\rangle_{ijkl}=\sum_{m=1}^9C_m\,
{}_{D_2\times Z^c_2}\mathsf{T}^{A_g,m,1}_{ijkl},
\]
where $C_m\in\mathbb{R}$. Its two-point correlation tensor has the form
\[
\langle\mathsf{C}(\mathbf{x}),\mathsf{C}(\mathbf{y})\rangle
=\int_{\hat{\mathbb{R}}^3/ D_2\times Z^c_2}\cos(p_1(y_1-x_1))\cos(p_2(y_2-x_2))\cos(p_3(y_3-x_3))f(\mathbf{p})
\,\mathrm{d}\Phi(\mathbf{p}),
\]
where $f(\mathbf{p})$ is a $\Phi$-equivalence class of measurable functions
acting from $\hat{\mathbb{R}}^3/ D_2\times Z^c_2$ to the set of nonnegative-definite symmetric linear operators on $\mathsf{V}^{D_2\times Z^c_2}$ with unit trace, and $\Phi$ is a finite measure on $\hat{\mathbb{R}}^3/ D_2\times Z^c_2$. The field has the form
\[
\mathsf{C}(\mathbf{x})_{ijkl}=\sum_{m=1}^9C_m\, {}_{D_2\times Z^c_2}
\mathsf{T}^{A_g,m,1}_{ijkl}+\sum_{m=1}^{9}\sum_{n=1}^{8}\int_{\hat{\mathbb{R}}^3/ D_2\times Z^c_2}u_n(\mathbf{p},\mathbf{x})\,\mathrm{d}Z^{n}_m(\mathbf{p}){}_{D_2\times
Z^c_2}\mathsf{T}^{A_g,m,1}_{ijkl},
\]
where $(Z^{n}_1(\mathbf{p}),\dots,Z^{n}_9(\mathbf{p}))^{\top}$ are eight centred uncorrelated $\mathsf{V}^{D_2\times Z^c_2}$-valued random measures on $\hat{\mathbb{R}}^3/ D_2\times Z^c_2$ with control measure $f(\mathbf{p})\,\mathrm{d}\Phi(\mathbf{p})$, and where $u_n(\mathbf{p},\mathbf{x})$ are eight different product of sines and cosines of $p_rx_r$.
\end{theorem}

Consider a $9\times 9$ symmetric nonnegative-definite matrix with the unit trace of the following structure:
\begin{equation}\label{eq:18}
\begin{pmatrix}
  A & B \\
  B^{\top} & C
\end{pmatrix}
,
\end{equation}
where $A$ is a $6\times 6$ matrix. Introduce the following notation:
\[
\begin{aligned}
j_1(\mathbf{p},\mathbf{z})&=\cos(p_1z_1)\cos(p_2z_2)\cos(p_3z_3),\\
j_2(\mathbf{p},\mathbf{z})&=\cos(p_1z_2)\cos(p_2z_1)\cos(p_3z_3).
\end{aligned}
\]
Let $\Phi$ be a finite measure on $\hat{\mathbb{R}}^3/ D_4\times
Z^c_2$. Let $f^0(\mathbf{p})$ be a $\Phi$-equivalence class of measurable functions acting from $(\hat{\mathbb{R}}^3/ D_4\times
Z^c_2)_m$, $0\leq m\leq 1$ to the set of nonnegative-definite symmetric matrices with unit trace satisfying $B=0$. Let $f^+(\mathbf{p})$ be a $\Phi$-equivalence class of measurable functions acting from $(\hat{\mathbb{R}}^3/ D_4\times
Z^c_2)_m$, $2\leq m\leq 4$ to the set of nonnegative-definite symmetric linear operators on $\mathsf{V}^{D_2\times Z^c_2}$ with unit trace, and let $f^-(\mathbf{p})$ is obtained from $f^+(\mathbf{p})$ by multiplying $B$ and $B^{\top}$ by $-1$.

\begin{theorem}[A tetragonal random field in the orthotropic class]\label{th:6}
The one-point correlation tensor of a homogeneous and $(D_4\times Z^c_2,6A_{1g}\oplus 3B_{1g})$-isotropic random field $\mathsf{C}(\mathbf{x})$ is
\[
\langle\mathsf{C}(\mathbf{x})\rangle_{ijkl}=\sum_{m=1}^6C_m\,
{}_{D_4\times Z^c_2}\mathsf{T}^{A_{g1},m,1}_{ijkl},
\]
where $C_m\in\mathbb{R}$. Its two-point correlation tensor has the form
\begin{equation}\label{eq:19}
\begin{aligned}
\langle\mathsf{C}(\mathbf{x}),\mathsf{C}(\mathbf{y})\rangle
&=\frac{1}{2}\sum_{m=0}^{1}\int_{(\hat{\mathbb{R}}^3/ D_4\times
Z^c_2)_m}[j_1(\mathbf{p},\mathbf{y}-\mathbf{x})+j_2(\mathbf{p},\mathbf{y}-\mathbf{x})]
f^0(\mathbf{p})\,\mathrm{d}\Phi(\mathbf{p})\\
&\quad+\frac{1}{2}\sum_{m=2}^{4}\int_{(\hat{\mathbb{R}}^3/ D_4\times
Z^c_2)_m}[j_1(\mathbf{p},\mathbf{y}-\mathbf{x})f^+(\mathbf{p})
+j_2(\mathbf{p},\mathbf{y}-\mathbf{x})f^-(\mathbf{p})]\,\mathrm{d}\Phi(\mathbf{p}).
\end{aligned}
\end{equation}
The field has the form
\[
\begin{aligned}
\mathsf{C}(\mathbf{x})_{ijkl}&=\sum_{m=1}^6C_m\,{}_{D_4\times Z^c_2}\mathsf{T}^{A_{1g},m,1}_{ijkl}\\
&\quad+\frac{1}{\sqrt{2}}\sum_{q=1}^{9}\sum_{n=1}^{16}\sum_{m=0}^{1}
\int_{(\hat{\mathbb{R}}^3/ D_4\times
Z^c_2)_m}u_n(\mathbf{p},\mathbf{x})\,\mathrm{d}Z^{n0}_q(\mathbf{p}){}_{D_4\times
Z^c_2}\mathsf{T}^{q}_{ijkl}\\
&\quad+\frac{1}{\sqrt{2}}\sum_{q=1}^{9}\sum_{n=1}^{8}\sum_{m=2}^{4}
\int_{(\hat{\mathbb{R}}^3/ D_4\times
Z^c_2)_m}u_n(\mathbf{p},\mathbf{x})\,\mathrm{d}Z^{n+}_q(\mathbf{p}){}_{D_4\times
Z^c_2}\mathsf{T}^{q}_{ijkl}\\
&\quad+\frac{1}{\sqrt{2}}\sum_{q=1}^{9}\sum_{n=9}^{16}\sum_{m=2}^{4}
\int_{(\hat{\mathbb{R}}^3/ D_4\times
Z^c_2)_m}u_n(\mathbf{p},\mathbf{x})\,\mathrm{d}Z^{n-}_q(\mathbf{p}){}_{D_4\times
Z^c_2}\mathsf{T}^{q}_{ijkl},
\end{aligned}
\]
where $(Z^{n0}_1(\mathbf{p}),\dots,Z^{n0}_9(\mathbf{p}))^{\top}$ \emph{(}resp. $(Z^{n+}_1(\mathbf{p}),\dots,Z^{n+}_9(\mathbf{p}))^{\top}$, resp. $(Z^{n-}_1(\mathbf{p}),\dots,Z^{n-}_9(\mathbf{p}))^{\top}$\emph{)} are centred uncorrelated $\mathsf{V}^{D_2\times Z^c_2}$-valued random measures on the spaces $(\hat{\mathbb{R}}^3/ D_4\times
Z^c_2)_m$, $0\leq m\leq 1$ \emph{(}resp. $2\leq m\leq 4$\emph{)} with control measure $f^0(\mathbf{p})\,\mathrm{d}\Phi(\mathbf{p})$ \emph{(}resp. $f^+(\mathbf{p})\,\mathrm{d}\Phi(\mathbf{p})$, resp. $f^-(\mathbf{p})\,\mathrm{d}\Phi(\mathbf{p})$\emph{)}, $u_n(\mathbf{p},\mathbf{x})$ are different product of sines and cosines of $p_rx_r$ for $1\leq n\leq 8$ and eight different product of sines and cosines of $p_1x_2$, $p_2x_1$, and $p_3x_3$ for $9\leq n\leq 16$, and
\[
\mathsf{T}^{q}_{ijkl}=
\begin{cases}
  \mathsf{T}^{A_{1g},q,1}_{ijkl}, & \text{if } 1\leq q\leq 6, \\
  \mathsf{T}^{B_{1g},q-6,1}_{ijkl}, & \text{otherwise}.
\end{cases}
\]
\end{theorem}

Consider a $9\times 9$ symmetric nonnegative-definite matrix with unit trace of the following structure
\[
\begin{pmatrix}
  * & * & * & * & * & \mathbf{c}^{\top}_1 & \mathbf{c}^{\top}_2 \\
  * & * & * & * & * & \mathbf{c}^{\top}_3 & \mathbf{c}^{\top}_4 \\
  * & * & * & * & * & \mathbf{c}^{\top}_5 & \mathbf{c}^{\top}_6 \\
  * & * & * & * & * & \mathbf{c}^{\top}_7 & \mathbf{c}^{\top}_8 \\
  * & * & * & * & * & \mathbf{c}^{\top}_9 & \mathbf{c}^{\top}_{10} \\
  \mathbf{c}_1 & \mathbf{c}_3 & \mathbf{c}_5 & \mathbf{c}_7 & \mathbf{c}_9 & A_1 & A_2 \\
  \mathbf{c}_2 & \mathbf{c}_4 & \mathbf{c}_6 & \mathbf{c}_8 & \mathbf{c}_{10} & A_2 & A_3
\end{pmatrix}
,
\]
where stars are arbitrary numbers, $\mathbf{c}_i$ are vectors with two components, and $A_i$ are $2\times 2$ matrices of the form
\begin{equation}\label{eq:13new}
A_j=
\begin{pmatrix}
  a-b & c+d \\
  c-d & a+b
\end{pmatrix}
.
\end{equation}
Let $\Phi$ be a finite measure on $\hat{\mathbb{R}}^3/ D_6\times Z^c_2$. Let $f^0(\mathbf{p})$ be a $\Phi$-equivalence class of measurable functions acting from $(\hat{\mathbb{R}}^3/ D_6\times Z^c_2)_m$, $0\leq m\leq 1$ to the set of nonnegative-definite symmetric matrices with unit trace such that $\mathbf{c}_i=\mathbf{0}$ and $A_i$ are proportional to the identity matrix. Let $f^-(\mathbf{p})$ be a $\Phi$-equivalence class of measurable functions acting from $(\hat{\mathbb{R}}^3/ D_6\times Z^c_2)_2$ to the set of nonnegative-definite symmetric matrices with unit trace such that $A_i$ are symmetric. Let $f^+(\mathbf{p})$ be a $\Phi$-equivalence class of measurable functions acting from $(\hat{\mathbb{R}}^3/ D_6\times Z^c_2)_m$, $3\leq m\leq 4$ to the set of nonnegative-definite symmetric matrices with unit trace. Consider matrices and functions of Table~\ref{tab:3}.

\begin{table}[htbp]
\caption{The matrices $g_n$ and the functions $j_n(\mathbf{p},\mathbf{z})$ for the group $D_6\times Z^c_2$}\label{tab:3}
\begin{tabular}{lll}
\hline\noalign{\smallskip}
$n$ & $g_n$ & $j_n(\mathbf{p},\mathbf{z})$ \\
\noalign{\smallskip}\hline\noalign{\smallskip}
$1$ & $\left(\begin{smallmatrix}
      1 & 0 \\
      0 & 1
    \end{smallmatrix}\right)$ & $\cos(p_3z_3)\cos(p_1z_1+p_2z_2)$ \\
$2$ & $\left(\begin{smallmatrix}
         -1 & 0 \\
         0 & 1
       \end{smallmatrix}\right)$ & $\cos(p_3z_3)\cos(p_1z_1-p_2z_2)$ \\
$3$ & $\frac{1}{2}\left(\begin{smallmatrix}
                    1 & -\sqrt{3} \\
                    \sqrt{3} & 1
                  \end{smallmatrix}\right)$ & $\cos(p_3z_3)\cos[(p_1+\sqrt{3}p_2)z_1+(-\sqrt{3}p_1+p_2)z_2]/2$ \\
$4$ & $\frac{1}{2}\left(\begin{smallmatrix}
                    -1 & \sqrt{3} \\
                    \sqrt{3} & -1
                  \end{smallmatrix}\right)$ & $\cos(p_3z_3)\cos[(p_1-\sqrt{3}p_2)z_1+(\sqrt{3}p_1+p_2)z_2]/2$ \\
$5$ & $\frac{1}{2}\left(\begin{smallmatrix}
                    1 & \sqrt{3} \\
                    \sqrt{3} & -1
                  \end{smallmatrix}\right)$ & $\cos(p_3z_3)\cos[(-p_1+\sqrt{3}p_2)z_1+(\sqrt{3}p_1+p_2)z_2]/2$ \\
$6$ & $\frac{1}{2}\left(\begin{smallmatrix}
                    1 & -\sqrt{3} \\
                    -\sqrt{3} & -1
                  \end{smallmatrix}\right)$ & $\cos(p_3z_3)\cos[(p_1+\sqrt{3}p_2)z_1+(\sqrt{3}p_1-p_2)z_2]/2$ \\
\noalign{\smallskip}\hline
\end{tabular}
\end{table}

Let $f^{-i}(\mathbf{p})$ is obtained from $f^-(\mathbf{p})$ by replacing all $\mathbf{c}_j$ with $g_i\mathbf{c}_j$ and the vectors $(b,c)^{\top}$ in all $A_j$ with $g_i(b,c)^{\top}$. Let $f^{+i}(\mathbf{p})$ is obtained from $f^+(\mathbf{p})$ by replacing all $\mathbf{c}_j$ with $g_i\mathbf{c}_j$ and all $A_j$ with $g_iA_jg^{-1}_i$.

\begin{theorem}[A hexagonal random field in the orthotropic class]\label{th:7}
The one-point correlation tensor of a homogeneous and $(D_6\times Z^c_2,5A_{1g}\oplus 2E_{2g})$-isotropic random field $\mathsf{C}(\mathbf{x})$ is
\[
\langle\mathsf{C}(\mathbf{x})\rangle_{ijkl}=\sum_{m=1}^5C_m\,
{}_{D_6\times Z^c_2}\mathsf{T}^{A_{1g},m,1}_{ijkl},
\]
where $C_m\in\mathbb{R}$. Its two-point correlation tensor has the form
\[
\begin{aligned}
\langle\mathsf{C}(\mathbf{x}),\mathsf{C}(\mathbf{y})\rangle
&=\frac{1}{6}\left(\sum_{m=0}^{1}\int_{(\hat{\mathbb{R}}^3/ D_6\times Z^c_2)_m}\sum_{n=1}^{6}j_n(\mathbf{p},\mathbf{y}-\mathbf{x})f^0(\mathbf{p})\,
\mathrm{d}\Phi(\mathbf{p})\right.\\
&\quad+\int_{(\hat{\mathbb{R}}^3/ D_6\times Z^c_2)_2}\sum_{n=1}^{6}j_n(\mathbf{p},\mathbf{y}-\mathbf{x})f^{-n}(\mathbf{p})\,
\mathrm{d}\Phi(\mathbf{p})\\
&\quad+\left.\sum_{m=3}^{4}\int_{(\hat{\mathbb{R}}^3/ D_6\times Z^c_2)_m}\sum_{n=1}^{6}j_n(\mathbf{p},\mathbf{y}-\mathbf{x})f^{+n}(\mathbf{p})\,
\mathrm{d}\Phi(\mathbf{p})\right).
\end{aligned}
\]
The field has the form
\[
\begin{aligned}
\mathsf{C}(\mathbf{x})_{ijkl}&=\sum_{m=1}^5C_m\,
{}_{D_6\times Z^c_2}\mathsf{T}^{A_{1g},m,1}_{ijkl}
+\frac{1}{\sqrt{6}}\sum_{q=1}^9\sum_{n=1}^{24}\sum_{m=0}^{1}\int_{(\hat{\mathbb{R}}^3/ D_6\times Z^c_2)_m}u_n(\mathbf{p},\mathbf{x})\,\mathrm{d}Z^{0n}_q(\mathbf{p}){}_{D_6\times
Z^c_2}T^{q}_{ijkl}\\
&\quad+\frac{1}{\sqrt{6}}\sum_{q=1}^9\sum_{s=1}^{6}\sum_{n=4s-3}^{4s}
\int_{(\hat{\mathbb{R}}^3/ D_6\times Z^c_2)_2}u_n(\mathbf{p},\mathbf{x})\,\mathrm{d}Z^{-ns}_q(\mathbf{p}){}_{D_6\times
Z^c_2}T^{q}_{ijkl}\\
&\quad+\frac{1}{\sqrt{6}}\sum_{q=1}^9\sum_{s=1}^{6}\sum_{n=4s-3}^{4s}\sum_{m=3}^{4}
\int_{(\hat{\mathbb{R}}^3/ D_6\times Z^c_2)_m}u_n(\mathbf{p},\mathbf{x})\,\mathrm{d}Z^{+ns}_q(\mathbf{p}){}_{D_6\times
Z^c_2}T^{q}_{ijkl},
\end{aligned}
\]
where $(Z^{0n}_1(\mathbf{p}),\dots,Z^{0n}_9(\mathbf{p}))^{\top}$ \emph{(}resp. $(Z^{-ns}_1(\mathbf{p}),\dots,Z^{-ns}_9(\mathbf{p}))^{\top}$, resp. $(Z^{+ns}_1(\mathbf{p}),\dots,Z^{+ns}_9(\mathbf{p}))^{\top}$\emph{)} are centred uncorrelated $\mathsf{V}^{D_2\times Z^c_2}$-valued random measures on $(\hat{\mathbb{R}}^3/ D_6\times Z^c_2)_m$, $0\leq m\leq 1$ \emph{(}resp. on $(\hat{\mathbb{R}}^3/ D_6\times Z^c_2)_2$, resp. on $(\hat{\mathbb{R}}^3/ D_6\times Z^c_2)_2$, $3\leq m\leq 4$\emph{)} with control measure $f^0(\mathbf{p})\,\mathrm{d}\Phi(\mathbf{p})$ \emph{(}resp. $f^{-s}(\mathbf{p})\,\mathrm{d}\Phi(\mathbf{p})$, resp. $f^{+s}(\mathbf{p})\,\mathrm{d}\Phi(\mathbf{p})$\emph{)}, $u_n(\mathbf{p},\mathbf{x})$, $1\leq n\leq 8$ are different product of sines and cosines of angles in Table~\emph{\ref{tab:3}}, and where
\[
\mathsf{T}^q_{ijkl}=
\begin{cases}
  {}_{D_6\times
Z^c_2}\mathsf{T}^{A_g,q,1}_{ijkl}, & \mbox{if } 1\leq q\leq 5 \\
  {}_{D_6\times
Z^c_2}\mathsf{T}^{E^{2g},\lfloor q/2\rfloor-1,q\bmod 2+1}_{ijkl}, & \mbox{otherwise}.
\end{cases}
\]
\end{theorem}

Consider a $9\times 9$ symmetric nonnegative-definite matrix with unit trace of the following structure
\[
\begin{pmatrix}
  * & * & * & \mathbf{c}^{\top}_1 & \mathbf{c}^{\top}_2 & \mathbf{c}^{\top}_3 \\
  * & * & * & \mathbf{c}^{\top}_4 & \mathbf{c}^{\top}_5 & \mathbf{c}^{\top}_6 \\
  * & * & * & \mathbf{c}^{\top}_7 & \mathbf{c}^{\top}_8 & \mathbf{c}^{\top}_9 \\
  \mathbf{c}_1 & \mathbf{c}_4 & \mathbf{c}_7 & A_1 & A_2 & A_3 \\
  \mathbf{c}_2 & \mathbf{c}_5 & \mathbf{c}_8 & A_2 & A_4 & A_5 \\
  \mathbf{c}_3 & \mathbf{c}_6 & \mathbf{c}_9 & A_3 & A_5 & A_6
\end{pmatrix}
,
\]
where stars are arbitrary numbers, $\mathbf{c}_i$ are vectors with two components, and $A_i$ are $2\times 2$ matrices. Let $\Phi$ be a finite measure on $\hat{\mathbb{R}}^3/\mathcal{T}\times Z^c_2$. Let $f^0(\mathbf{p})$ be a $\Phi$-equivalence class of measurable functions acting from $(\hat{\mathbb{R}}^3/\mathcal{T}\times Z^c_2)_m$, $0\leq m\leq 1$ to the set of nonnegative-definite symmetric linear operators on $\mathsf{V}^{D_2\times Z^c_2}$ with unit trace such that $\mathbf{c}_i=\mathbf{0}$ and $A_i$ are proportional to the identity matrix. Let $f^1(\mathbf{p})$ be a $\Phi$-equivalence class of measurable functions acting from $(\hat{\mathbb{R}}^3/\mathcal{T}\times Z^c_2)_m$, $2\leq m\leq 4$ to the set of nonnegative-definite symmetric linear operators on $\mathsf{V}^{D_2\times Z^c_2}$ with unit trace. Denote
\begin{equation}\label{eq:20}
g=\frac{1}{2}
\begin{pmatrix}
  -1 & \sqrt{3} \\
  -\sqrt{3} & -1
\end{pmatrix}
.
\end{equation}
Let $f^+(\mathbf{p})$ (resp. $f^-(\mathbf{p})$) is obtained from $f^1(\mathbf{p})$ by replacing all $\mathbf{c}_i$ with $g\mathbf{c}_1$ (resp. with $g^{-1}\mathbf{c}_i$) and all $A_i$ with $gA_ig^{-1}$ (resp. $g^{-1}A_ig$). Finally, let $j_m(\mathbf{p},\mathbf{z})$ be functions from Table~\ref{tab:1old}.

\begin{table}[htbp]
\caption{The functions $j_n(\mathbf{p},\mathbf{z})$ for the tetrahedral
group}\label{tab:1old}
\begin{tabular}{ll}
\hline\noalign{\smallskip}
$n$ & $j_n(\mathbf{p},\mathbf{z})$ \\
\noalign{\smallskip}\hline\noalign{\smallskip}
1 & $[\cos(p_1z_1+p_2z_2)+\cos(p_1z_2+p_2z_1)]\cos(p_3z_3)$ \\
2 & $\cos[\frac{1}{2}((p_1+\sqrt{2}p_3)z_1+p_2z_2+\sqrt{2}p_1z_3)]
\cos[\frac{1}{2}(p_2z_1+(-p_1+\sqrt{2}p_3)z_2-\sqrt{2}p_2z_3)]$ \\
& $+\cos[\frac{1}{2}((z_1(p_1-\sqrt{2}p_3)-p_2z_2-\sqrt{2}p_1z_3)
]\cos[\frac{1}{2}(p_2z_1+(-p_1-\sqrt{2}p_3)z_2+\sqrt{2}p_2z_3)]$ \\
3 & $\cos[\frac{1}{2}(-p_1z_1+(p_2+\sqrt{2}p_3)z_2+\sqrt{2}p_2z_3)%
]\cos[\frac{1}{2}((p_2-\sqrt{2}p_3)z_1-p_1z_2+\sqrt{2}p_1z_3)]$ \\
& $+\cos[\frac{1}{2}(-p_1z_1+(p_2-\sqrt{2}p_3)z_2-\sqrt{2}p_2z_3)%
]\cos[\frac{1}{2}((p_2+\sqrt{2}p_3)z_1-p_1z_2-\sqrt{2}p_1z_3)]$ \\
\noalign{\smallskip}\hline
\end{tabular}
\end{table}

\begin{theorem}[A cubic random field in the orthotropic class]\label{th:8}
The one-point correlation tensor of a homogeneous and $(\mathcal{T}\times Z^c_2,3A_g\oplus 3({}^1E_g\oplus{}^2E_g))$-isotropic random field $\mathsf{C}(\mathbf{x})$ is
\[
\langle\mathsf{C}(\mathbf{x})\rangle_{ijkl}=\sum_{m=1}^3C_m\, {}_{\mathcal{T}\times Z^c_2}\mathsf{T}^{A_g,m,1}_{ijkl},
\]
where $C_m\in\mathbb{R}$. Its two-point correlation tensor has the form
\[
\begin{aligned}
\langle\mathsf{C}(\mathbf{x}),\mathsf{C}(\mathbf{y})\rangle&=\frac{1}{6}\sum_{m=0}^{1}
\int_{(\hat{\mathbb{R}}^3/\mathcal{T}\times Z^c_2)_m}\sum_{n=1}^{3}
j_n(\mathbf{p},\mathbf{y}-\mathbf{x})f^0(\mathbf{p})\,\mathrm{d}\Phi(\mathbf{p})\\
&\quad+\frac{1}{6}\sum_{m=2}^{4}\int_{(\hat{\mathbb{R}}^3/\mathcal{T}\times Z^c_2)_m}[j_1(\mathbf{p},\mathbf{y}-\mathbf{x})f^1(\mathbf{p})+j_2(\mathbf{p},\mathbf{y}-\mathbf{x})f^+(\mathbf{p})\\
&\quad+j_3(\mathbf{p},\mathbf{y}-\mathbf{x})f^-(\mathbf{p})]\,\mathrm{d}\Phi(\mathbf{p}).
\end{aligned}
\]
The field has the form
\[
\begin{aligned}
\mathsf{C}(\mathbf{x})&=\sum_{m=1}^3C_m\,
{}_{\mathcal{T}\times
Z^c_2}\mathsf{T}^{A_g,m,1}_{ijkl}+\frac{1}{\sqrt{6}}\left(\sum_{q=1}^{9}\sum_{n=1}^{24}\sum_{m=0}^{1}
\int_{(\hat{\mathbb{R}}^3/\mathcal{T}\times Z^c_2)_m}
u_n(\mathbf{p},\mathbf{x})\,\mathrm{d}Z^{n0}_q(\mathbf{p})\mathsf{T}^q_{ijkl}\right.\\
&\quad+\sum_{q=1}^{9}\sum_{n=1}^{8}\sum_{m=2}^{4}
\int_{(\hat{\mathbb{R}}^3/\mathcal{T}\times Z^c_2)_m}
u_n(\mathbf{p},\mathbf{x})\,\mathrm{d}Z^{n1}_q(\mathbf{p})\mathsf{T}^q_{ijkl}\\
&\quad+\sum_{q=1}^{9}\sum_{n=9}^{16}\sum_{m=2}^{4}
\int_{(\hat{\mathbb{R}}^3/\mathcal{T}\times Z^c_2)_m}
u_n(\mathbf{p},\mathbf{x})\,\mathrm{d}Z^{n+}_q(\mathbf{p})\mathsf{T}^q_{ijkl}\\
&\quad+\left.\sum_{q=1}^{9}\sum_{n=17}^{24}\sum_{m=2}^{4}
\int_{(\hat{\mathbb{R}}^3/\mathcal{T}\times Z^c_2)_m}
u_n(\mathbf{p},\mathbf{x})\,\mathrm{d}Z^{n-}_q(\mathbf{p})\mathsf{T}^q_{ijkl}\right),
\end{aligned}
\]
where $u_n(\mathbf{p},\mathbf{x})$ are various products of sines and cosines of angles from Table~\emph{\ref{tab:1old}},
\[
\mathsf{T}^q_{ijkl}=
\begin{cases}
  {}_{\mathcal{T}\times
Z^c_2}\mathsf{T}^{A_g,q,1}_{ijkl}, & \mbox{if } 1\leq q\leq 3 \\
  {}_{\mathcal{T}\times
Z^c_2}\mathsf{T}^{E^{2g},\lfloor q/2\rfloor-1,q\bmod 2+1}_{ijkl}, & \mbox{otherwise},
\end{cases}
\]
and where $(Z^{n0}_1(\mathbf{p}),\dots,Z^{n0}_9(\mathbf{p}))^{\top}$ \emph{(}resp. $(Z^{n1}_1(\mathbf{p}),\dots,Z^{n1}_9(\mathbf{p}))^{\top}$, resp. $(Z^{n+}_1(\mathbf{p}),\dots,Z^{n+}_9(\mathbf{p}))^{\top}$ resp. $(Z^{n-}_1(\mathbf{p}),\dots,Z^{n-}_9(\mathbf{p}))^{\top}$\emph{)} are centred uncorrelated $\mathsf{V}^{D_2\times Z^c_2}$-valued random measures on $(\hat{\mathbb{R}}^3/\mathcal{T}\times Z^c_2)_m$ for $0\leq m\leq 1$ \emph{(}resp. $2\leq m\leq 4$\emph{)} with control measure $f^0(\mathbf{p})\,\mathrm{d}\Phi(\mathbf{p})$ \emph{(}resp. $f^1(\mathbf{p})\,\mathrm{d}\Phi(\mathbf{p})$, resp. $f^+(\mathbf{p})\,\mathrm{d}\Phi(\mathbf{p})$, resp. $f^-(\mathbf{p})\,\mathrm{d}\Phi(\mathbf{p})$\emph{)}.
\end{theorem}

Consider a $9\times 9$ symmetric nonnegative-definite matrix with unit trace of the following structure
\[
\begin{pmatrix}
  * & * & * & \mathbf{c}^{\top}_1 & \mathbf{c}^{\top}_2 & \mathbf{c}^{\top}_3 \\
  * & * & * & \mathbf{c}^{\top}_4 & \mathbf{c}^{\top}_5 & \mathbf{c}^{\top}_6 \\
  * & * & * & \mathbf{c}^{\top}_7 & \mathbf{c}^{\top}_8 & \mathbf{c}^{\top}_9 \\
  \mathbf{c}_1 & \mathbf{c}_4 & \mathbf{c}_7 & A_1 & A_2 & A_3 \\
  \mathbf{c}_2 & \mathbf{c}_5 & \mathbf{c}_8 & A_2 & A_4 & A_5 \\
  \mathbf{c}_3 & \mathbf{c}_6 & \mathbf{c}_9 & A_3 & A_5 & A_6
\end{pmatrix},
\]
where stars are arbitrary numbers, $\mathbf{c}_i$ are vectors with two components, and $A_i$ are $2\times 2$ matrices of the form \eqref{eq:13new}. Let $\Phi$ be a finite measure on $\hat{\mathbb{R}}^3/\mathcal{O}\times Z^c_2$. Let $f^0(\mathbf{p})$ be a $\Phi$-equivalence class of measurable functions acting from $(\hat{\mathbb{R}}^3/\mathcal{O}\times Z^c_2)_m$, $0\leq m\leq 1$ to the set of nonnegative-definite symmetric matrices with unit trace such that $\mathbf{c}_i=\mathbf{0}$ and $A_i$ are proportional to the identity matrix. Let $f^-(\mathbf{p})$ be a $\Phi$-equivalence class of measurable functions acting from $(\hat{\mathbb{R}}^3/\mathcal{O}\times Z^c_2)_2$ to the set of nonnegative-definite symmetric matrices with unit trace such that $A_i$ are symmetric. Let $f^+(\mathbf{p})$ be a $\Phi$-equivalence class of measurable functions acting from $(\hat{\mathbb{R}}^3/\mathcal{O}\times Z^c_2)_m$, $3\leq m\leq 6$ to the set of nonnegative-definite symmetric matrices with unit trace. Consider matrices and functions of Table~\ref{tab:2old}.

Let $f^{-i}(\mathbf{p})$ is obtained from $f^-(\mathbf{p})$ by replacing all $\mathbf{c}_j$ with $g_i\mathbf{c}_i$ and the vectors $(b,c)^{\top}$ in all $A_j$ with $g_i(b,c)^{\top}$. Let $f^{+i}(\mathbf{p})$ is obtained from $f^+(\mathbf{p})$ by replacing all $\mathbf{c}_j$ with $g_i\mathbf{c}_i$ and all $A_j$ with $g_iA_jg^{-1}_i$.

\begin{theorem}[A cubic random field in the orthotropic class]\label{th:9}
The one-point correlation tensor of a homogeneous and $(\mathcal{O}\times Z^c_2,3A_{1g}\oplus 3E_g)$-isotropic random field $\mathsf{C}(\mathbf{x})$ is
\[
\langle\mathsf{C}(\mathbf{x})\rangle_{ijkl}=\sum_{m=1}^3C_m\,
{}_{\mathcal{O}\times Z^c_2}\mathsf{T}^{A_{1g},m,1}_{ijkl},
\]
where $C_m\in\mathbb{R}$. Its two-point correlation tensor has the form
\[
\begin{aligned}
\langle\mathsf{C}(\mathbf{x}),\mathsf{C}(\mathbf{y})\rangle
&=\frac{1}{12}\left(\sum_{m=0}^{1}\int_{(\hat{\mathbb{R}}^3/ \mathcal{O}\times Z^c_2)_m}\sum_{n=1}^{6}j_n(\mathbf{p},\mathbf{y}-\mathbf{x})f^0(\mathbf{p})\,
\mathrm{d}\Phi(\mathbf{p})\right.\\
&\quad+\int_{(\hat{\mathbb{R}}^3/\mathcal{O}\times Z^c_2)_2}\sum_{n=1}^{6}j_n(\mathbf{p},\mathbf{y}-\mathbf{x})f^{-n}(\mathbf{p})\,
\mathrm{d}\Phi(\mathbf{p})\\
&\quad+\left.\sum_{m=3}^{6}\int_{(\hat{\mathbb{R}}^3/\mathcal{O}\times Z^c_2)_m}\sum_{n=1}^{6}j_n(\mathbf{p},\mathbf{y}-\mathbf{x})f^{+n}(\mathbf{p})\,
\mathrm{d}\Phi(\mathbf{p})\right).
\end{aligned}
\]
The field has the form
\[
\begin{aligned}
\mathsf{C}(\mathbf{x})_{ijkl}&=\sum_{m=1}^3C_m\,
{}_{\mathcal{O}\times Z^c_2}\mathsf{T}^{A_{1g},m,1}_{ijkl}
+\frac{1}{\sqrt{12}}\sum_{q=1}^6\sum_{n=1}^{48}\sum_{m=0}^{1}\int_{(\hat{\mathbb{R}}^3/ \mathcal{O}\times Z^c_2)_m}u_n(\mathbf{p},\mathbf{x})\,\mathrm{d}Z^{0n}_q(\mathbf{p}){}_{\mathcal{O}\times
Z^c_2}T^{q}_{ijkl}\\
&\quad+\frac{1}{\sqrt{12}}\sum_{q=1}^6\sum_{s=1}^{6}\sum_{n=8s-7}^{8s}
\int_{(\hat{\mathbb{R}}^3/\mathcal{O}\times Z^c_2)_2}u_n(\mathbf{p},\mathbf{x})\,\mathrm{d}Z^{-ns}_q(\mathbf{p}){}_{\mathcal{O}\times
Z^c_2}T^{q}_{ijkl}\\
&\quad+\frac{1}{\sqrt{12}}\sum_{q=1}^6\sum_{s=1}^{6}\sum_{n=8s-7}^{8s}\sum_{m=3}^{6}
\int_{(\hat{\mathbb{R}}^3/\mathcal{O}\times Z^c_2)_m}u_n(\mathbf{p},\mathbf{x})\,\mathrm{d}Z^{+ns}_q(\mathbf{p}){}_{\mathcal{O}\times
Z^c_2}T^{q}_{ijkl},
\end{aligned}
\]
where $u_n(\mathbf{p},\mathbf{x})$, $1\leq n\leq 8$ are different products of sines and cosines of angles in Table~\emph{\ref{tab:2old}}, $(Z^{0n}_1(\mathbf{p}),\dots,Z^{0n}_9(\mathbf{p}))^{\top}$ \emph{(}resp. $(Z^{-ns}_1(\mathbf{p}),\dots,Z^{-ns}_9(\mathbf{p}))^{\top}$, resp. $(Z^{+ns}_1(\mathbf{p}),\dots,Z^{+ns}_9(\mathbf{p}))^{\top}$\emph{)} are centred uncorrelated $\mathsf{V}^{D_2\times Z^c_2}$-valued random measures on $(\hat{\mathbb{R}}^3/\mathcal{O}\times Z^c_2)_m$, $0\leq m\leq 1$ \emph{(}resp. on $(\hat{\mathbb{R}}^3/\mathcal{O}\times Z^c_2)_2$, resp. on $(\hat{\mathbb{R}}^3/\mathcal{O}\times Z^c_2)_2$, $3\leq m\leq 6$\emph{)} with control measure $f^0(\mathbf{p})\,\mathrm{d}\Phi(\mathbf{p})$ \emph{(}resp. $f^{-s}(\mathbf{p})\,\mathrm{d}\Phi(\mathbf{p})$, resp. $f^{+s}(\mathbf{p})\,\mathrm{d}\Phi(\mathbf{p})$\emph{)}, and where
\[
\mathsf{T}^m_{ijkl}=
\begin{cases}
  {}_{\mathcal{O}\times
Z^c_2}\mathsf{T}^{A_g,m,1}_{ijkl}, & \mbox{if } 1\leq m\leq 3 \\
  {}_{\mathcal{O}\times
Z^c_2}\mathsf{T}^{E^{2g},\lfloor m/2\rfloor-1,m\bmod 2+1}_{ijkl}, & \mbox{if } 4\leq m\leq 9.
\end{cases}
\]
\end{theorem}

\begin{table}[htbp]
\caption{The matrices $g_n$ and the functions $j_n(\mathbf{p},\mathbf{z})$ for the group $\mathcal{O}\times Z^c_2$}\label{tab:2old}
\begin{tabular}{lll}
\hline\noalign{\smallskip}
$n$ & $g_n$ & $j_n(\mathbf{p},\mathbf{z})$ \\
\noalign{\smallskip}\hline\noalign{\smallskip}
$1$ & $\left(\begin{smallmatrix}
      1 & 0 \\
      0 & 1
    \end{smallmatrix}\right)$ & $\cos(p_3z_3)[\cos(p_1z_1+p_2z_2)+\cos(p_1z_2+p_2z_1)]$ \\
$2$ & $\left(\begin{smallmatrix}
         -1 & 0 \\
         0 & 1
       \end{smallmatrix}\right)$ & $\cos(p_3z_3)[\cos(-p_1z_2+p_2z_1)+\cos(p_2z_2-p_1z_1)]$ \\
$3$ & $\frac{1}{2}\left(\begin{smallmatrix}
                    1 & -\sqrt{3} \\
                    \sqrt{3} & 1
                  \end{smallmatrix}\right)$ & $2\cos[\sqrt{2}p_3(z_2-z_1)/2]\cos[(p_2-p_1)(z_1+z_2)/2]\cos[\sqrt{2}(p_1+p_2)z_3/2]$ \\
$4$ & $\frac{1}{2}\left(\begin{smallmatrix}
                    -1 & \sqrt{3} \\
                    \sqrt{3} & -1
                  \end{smallmatrix}\right)$ & $\cos[\sqrt{2}(p_1-p_2)z_3/2]\{\cos[(-p_1-p_2+\sqrt{2}p_3)z_1/2+(p_1+p_2+\sqrt{2}p_3)z_2/2]$  \\
& & $+\cos[(-p_1-p_2-\sqrt{2}p_3)z_1/2+(p_1+p_2-\sqrt{2}p_3)z_2/2]\}$ \\
$5$ & $\frac{1}{2}\left(\begin{smallmatrix}
                    1 & \sqrt{3} \\
                    \sqrt{3} & -1
                  \end{smallmatrix}\right)$ & $2\cos[\sqrt{2}p_3(z_1+z_2)/2]\cos[(p_2-p_1)(z_2-z_1)/2]\cos[\sqrt{2}(p_1+p_2)z_3/2]$ \\
$6$ & $\frac{1}{2}\left(\begin{smallmatrix}
                    1 & -\sqrt{3} \\
                    -\sqrt{3} & -1
                  \end{smallmatrix}\right)$ & $\cos[\sqrt{2}(p_1-p_2)z_3/2]\{\cos[(p_1+p_2-\sqrt{2}p_3)z_1/2+(p_1+p_2+\sqrt{2}p_3)z_2/2]$  \\
& & $+\cos[(p_1+p_2+\sqrt{2}p_3)z_1/2+(p_1+p_2-\sqrt{2}p_3)z_2/2]\}$ \\
\noalign{\smallskip}\hline
\end{tabular}
\end{table}

\subsection{The trigonal class}

Introduce the following notation:
\[
\begin{aligned}
j_{10}(\mathbf{p},\mathbf{z})&=\cos(p_1z_1+p_3z_3)\cos(p_2z_2)\\
&\quad+\cos\left[\frac{1}{2}(p_1+\sqrt{3}p_2)z_1\right]\cos\left[\frac{1}{2}
(\sqrt{3}p_1-p_2)z_2+p_3z_3\right]\\
&\quad+\cos\left[\frac{1}{2}(p_1-\sqrt{3}p_2)z_1\right]\cos\left[%
\frac{1}{2}(-\sqrt{3}p_1+p_2)z_2+p_3z_3\right].
\end{aligned}
\]

\begin{theorem}[A trigonal random field in the trigonal class]\label{th:10}
The one-point correlation tensor of a homogeneous and $(D_3\times Z_2^c,6A_{1g})$-isotropic random field $\mathsf{C}(\mathbf{x})$ is
\[
\langle\mathsf{C}(\mathbf{x})\rangle_{ijkl}=\sum_{m=1}^{6}C_m\,{}_{D_3\times Z_2^c}T_{ijkl}^{A_{1g},m,1},
\]
where $C_m\in \mathbb{R}$. Its two-point correlation tensor has the form
\[
\langle\mathsf{C}(\mathbf{x}),\mathsf{C}(\mathbf{y})\rangle
=\frac{1}{3}\int_{\hat{\mathbb{R}}^3/ D_3\times
Z^c_2}j_{10}(\mathbf{p},\mathbf{y}-\mathbf{x})f(\mathbf{p})\,
\mathrm{d}\Phi(\mathbf{p}),
\]
where $f(\mathbf{p})$ is the $\Phi $-equivalence class of measurable functions acting from $\hat{\mathbb{R}}^3/ D_3\times Z^c_2$ to the set of nonnegative-definite symmetric linear operators on $\mathsf{V}^{D_3\times Z_2^c}$ with unit trace, and $\Phi$ is a finite measure on $\hat{\mathbb{R}}^3/ D_3\times Z^c_2$. The field has the form
\[
\mathsf{C}(\mathbf{x})_{ijkl}=\sum_{m=1}^{6}C_{m}\,{}_{D_3\times
Z_2^c}T_{ijkl}^{A_{1g},m,1}+\frac{1}{\sqrt{3}}\sum_{m=1}^6
\sum_{n=1}^{12}\int_{\hat{\mathbb{R}}^3/ D_3\times
Z^c_2}u_{n}(\mathbf{p},\mathbf{x})\,\mathrm{d}Z^{mn}(\mathbf{p}){}_{D_3\times
Z_2^c}T_{ijkl}^{A_{1g},m,1},
\]
where $(Z^{1n}(\mathbf{p}),\dots ,Z^{6n}(\mathbf{p}))^{\top}$ are $12$ centred uncorrelated $\mathsf{V}^{D_3\times Z_2^c}$-valued random measures on $\hat{\mathbb{R}}^3/ D_3\times Z^c_2$ with control measure $f(\mathbf{p})\,\mathrm{d}\Phi(\mathbf{p})$, and where $u_{n}(\mathbf{p},\mathbf{x})$, $1\leq n\leq 4$ are four different products of sines and cosines of $p_{1}x_{1}+p_{3}x_{3}$ and $p_{2}x_{2}$, $u_{n}(\mathbf{p},\mathbf{x})$, $5\leq n\leq 8$ are four different product of sines and cosines of $\frac{1}{2}(p_{1}+\sqrt{3}p_{2})x_{1}$ and $\frac{1}{2}(\sqrt{3}p_{1}-p_{2})x_{2}+p_{3}x_{3}$, $u_{n}(\mathbf{p},\mathbf{x})$, $9\leq n\leq 12$ are four different product of sines and cosines of $\frac{1}{2}(p_{1}-\sqrt{3}p_{2})x_{1}$ and $\frac{1}{2}(-\sqrt{3}p_{1}+p_{2})x_{2}+p_{3}x_{3}$.
\end{theorem}

Consider a $6\times 6$ symmetric nonnegative-definite matrix with unit trace of the following structure
\begin{equation}\label{eq:14new}
\begin{pmatrix}
  * & * & * & * & * & c_1 \\
  * & * & * & * & * & c_2 \\
  * & * & * & * & * & c_3 \\
  * & * & * & * & * & c_4 \\
  * & * & * & * & * & c_5 \\
  c_1 & c_2 & c_3 & c_4 & c_ 5 & * \\
  \end{pmatrix}
  ,
\end{equation}
where stars and $c_i$ are arbitrary numbers. Let $\Phi$ be a finite measure on $\hat{\mathbb{R}}^3/ D_6\times Z^c_2$. Let $f^0(\mathbf{p})$ be a $\Phi$-equivalence class of measurable functions acting from $(\hat{\mathbb{R}}^3/ D_6\times Z^c_2)_m$, $0\leq m\leq 2$ to the set of nonnegative-definite symmetric matrices with unit trace such that $c_i=0$. Let $f^+(\mathbf{p})$ be a $\Phi$-equivalence class of measurable functions acting from $(\hat{\mathbb{R}}^3/ D_6\times Z^c_2)_m$, $3\leq m\leq 4$ to the set of nonnegative-definite symmetric matrices with unit trace, and let $f^-(\mathbf{p})$ be a $\Phi$-equivalence class of measurable functions acting from $(\hat{\mathbb{R}}^3/ D_6\times Z^c_2)_m$, $3\leq m\leq 4$ to the set of nonnegative-definite symmetric matrices with unit trace such that all $c_i$s are multiplied by $-1$.

\begin{theorem}[A hexagonal random field in the trigonal class]\label{th:11}
The one-point correlation tensor of a homogeneous and $(D_6\times Z^c_2,5A_{1g}\oplus B_{1g}$-isotropic random field $\mathsf{C}(\mathbf{x})$ is
\[
\langle\mathsf{C}(\mathbf{x})\rangle_{ijkl}=\sum_{m=1}^5C_m\,
{}_{D_6\times Z^c_2}T^{A_{1g},1,1}_{ijkl},
\]
where $C_m\in\mathbb{R}$. Its two-point correlation tensor has the form
\[
\begin{aligned}
\langle\mathsf{C}(\mathbf{x}),\mathsf{C}(\mathbf{y})\rangle
&=\frac{1}{6}\left(\sum_{m=0}^{2}\int_{(\hat{\mathbb{R}}^3/ D_6\times Z^c_2)_m}\sum_{n=1}^{6}j_n(\mathbf{p},\mathbf{y}-\mathbf{x})f^0(\mathbf{p})\,
\mathrm{d}\Phi(\mathbf{p})\right.\\
&\quad+\sum_{m=3}^{4}\int_{(\hat{\mathbb{R}}^3/ D_6\times Z^c_2)_m}\sum_{n=1}^{3}j_n(\mathbf{p},\mathbf{y}-\mathbf{x})f^{+}(\mathbf{p})\,
\mathrm{d}\Phi(\mathbf{p})\\
&\quad+\left.\sum_{m=3}^{4}\int_{(\hat{\mathbb{R}}^3/ D_6\times Z^c_2)_m}\sum_{n=4}^{6}j_n(\mathbf{p},\mathbf{y}-\mathbf{x})f^{-}(\mathbf{p})\,
\mathrm{d}\Phi(\mathbf{p})\right).
\end{aligned}
\]
The field has the form
\[
\begin{aligned}
\mathsf{C}(\mathbf{x})_{ijkl}&=\sum_{m=1}^5C_m\,
{}_{D_6\times Z^c_2}\mathsf{T}^{A_{1g},m,1}_{ijkl}
+\frac{1}{\sqrt{6}}\sum_{q=1}^6\sum_{n=1}^{24}\sum_{m=0}^{2}\int_{(\hat{\mathbb{R}}^3/ D_6\times Z^c_2)_m}u_n(\mathbf{p},\mathbf{x})\,\mathrm{d}Z^{0n}_q(\mathbf{p}){}_{D_6\times
Z^c_2}T^{q}_{ijkl}\\
&\quad+\frac{1}{\sqrt{6}}\sum_{q=1}^6\sum_{s=1}^{6}\sum_{n=4s-3}^{4s}\sum_{m=3}^{4}
\int_{(\hat{\mathbb{R}}^3/ D_6\times Z^c_2)_m}u_n(\mathbf{p},\mathbf{x})\,\mathrm{d}Z^{+ns}_q(\mathbf{p}){}_{D_6\times
Z^c_2}T^{q}_{ijkl}\\
&\quad+\frac{1}{\sqrt{6}}\sum_{q=1}^9\sum_{s=1}^{6}\sum_{n=4s-3}^{4s}\sum_{m=3}^{4}
\int_{(\hat{\mathbb{R}}^3/ D_6\times Z^c_2)_m}u_n(\mathbf{p},\mathbf{x})\,\mathrm{d}Z^{-ns}_q(\mathbf{p}){}_{D_6\times
Z^c_2}T^{q}_{ijkl},
\end{aligned}
\]
where $(Z^{0n}_1(\mathbf{p}),\dots,Z^{0n}_6(\mathbf{p}))^{\top}$ \emph{(}resp. $(Z^{+ns}_1(\mathbf{p}),\dots,Z^{+ns}_6(\mathbf{p}))^{\top}$, resp. $(Z^{-ns}_1(\mathbf{p}),\dots,Z^{-ns}_6(\mathbf{p}))^{\top}$\emph{)} are centred uncorrelated $\mathsf{V}^{D_3\times Z^c_2}$-valued random measures on $(\hat{\mathbb{R}}^3/ D_6\times Z^c_2)_m$, $0\leq m\leq 2$ \emph{(}resp. on $(\hat{\mathbb{R}}^3/ D_6\times Z^c_2)_m$, $3\leq m\leq 4$\emph{)} with control measure $f^0(\mathbf{p})\,\mathrm{d}\Phi(\mathbf{p})$ \emph{(}resp. $f^{+}(\mathbf{p})\,\mathrm{d}\Phi(\mathbf{p})$, resp. $f^{-}(\mathbf{p})\,\mathrm{d}\Phi(\mathbf{p})$\emph{)}, $u_n(\mathbf{p},\mathbf{x})$, $1\leq n\leq 8$ are different product of sines and cosines of angles in Table~\emph{\ref{tab:3}}, and where
\[
T^{q}_{ijkl}=
\begin{cases}
  {}_{D_6\times Z^c_2}\mathsf{T}^{A_{1g},q,1}_{ijkl}, & \mbox{if } 1\leq q\leq 5, \\
  {}_{D_6\times Z^c_2}\mathsf{T}^{B_{1g},m,1}_{ijkl}, & \mbox{otherwise}.
\end{cases}
\]
\end{theorem}

\subsection{The tetragonal class}

\begin{theorem}[A tetragonal random field in the tetragonal class]\label{th:12}
The one-point correlation tensor of a homogeneous and $(D_4\times Z^c_2,6A_{1g})$-isotropic random field $\mathsf{C}(\mathbf{x})$
is
\[
\langle\mathsf{C}(\mathbf{x})\rangle_{ijkl}=\sum_{m=1}^6C_m\,
{}_{D_4\times Z^c_2}\mathsf{T}^{A_{1g},m,1}_{ijkl},
\]
where $C_m\in\mathbb{R}$. Its two-point correlation tensor has the form
\[
\begin{aligned}
\langle\mathsf{C}(\mathbf{x}),\mathsf{C}(\mathbf{y})\rangle&=
\frac{1}{2}\int_{\hat{\mathbb{R}}^3/ D_4\times
Z^c_2}[\cos(p_1(x_1-y_1))\cos(p_2(x_2-y_2))\\
&\quad+\cos(p_1(x_2-y_2))\cos(p_2(x_1-y_1))]\cos(p_3(x_3-y_3)) f(\mathbf{p})\,\mathrm{d}\Phi(\mathbf{p}),
\end{aligned}
\]
where $f(\mathbf{p})$ is a $\Phi$-equivalence class of measurable functions
acting from $\hat{\mathbb{R}}^3/ D_4\times Z^c_2$ to the set of nonnegative-definite symmetric linear operators on $\mathsf{V}^{D_4\times Z^c_2}$ with unit trace, and $\Phi$ is a finite measure on $\hat{\mathbb{R}}^3/ D_4\times Z^c_2$. The field has the form
\[
\mathsf{C}(\mathbf{x})_{ijkl}=\sum_{m=1}^6C_m\, {}_{D_4\times Z^c_2}
\mathsf{T}^{A_{1g},m,1}_{ijkl}+\frac{1}{\sqrt{2}}\sum_{m=1}^{6}
\sum_{n=1}^{16}\int_{\hat{\mathbb{R}}^3/ D_4\times
Z^c_2}u_n(\mathbf{p},\mathbf{x})\,\mathrm{d}Z^{mn}(\mathbf{p}){}_{D_4\times Z^c_2}\mathsf{T}^{A_{1g},m,1}_{ijkl},
\]
where $(Z^{1n}(\mathbf{p}),\dots,Z^{6n}(\mathbf{p}))^{\top}$ are $16$ centred uncorrelated $\mathsf{V}^{D_4\times Z^c_2}$-valued random measures on $\hat{\mathbb{R}}^3/ D_4\times
Z^c_2$ with control measure $f(\mathbf{p})\,\mathrm{d}\Phi(\mathbf{p})$, and where $u_n(\mathbf{p},\mathbf{x})$ are eight different product of sines and cosines of $p_rx_r$ for $1\leq n\leq 8$ and eight different product of sines and cosines of $p_1x_2$, $p_2x_1$, and $p_3x_3$ for $9\leq n\leq 16$.
\end{theorem}

Consider a $6\times 6$ symmetric nonnegative-definite matrix with unit trace of the structure \eqref{eq:14new}. Let $\Phi$ be a finite measure on $\hat{\mathbb{R}}^3/ D_8\times Z^c_2$. Let $f^0(\mathbf{p})$ be a $\Phi$-equivalence class of measurable functions acting from $(\hat{\mathbb{R}}^3/ D_8\times Z^c_2)_m$, $0\leq m\leq 1$ to the set of nonnegative-definite symmetric matrices with unit trace such that $c_i=0$. Let $f^+(\mathbf{p})$ be a $\Phi$-equivalence class of measurable functions acting from $(\hat{\mathbb{R}}^3/ D_8\times Z^c_2)_m$, $2\leq m\leq 4$ to the set of nonnegative-definite symmetric matrices with unit trace, and let $f^-(\mathbf{p})$ be a $\Phi$-equivalence class of measurable functions acting from $(\hat{\mathbb{R}}^3/ D_8\times Z^c_2)_m$, $2\leq m\leq 4$ to the set of nonnegative-definite symmetric matrices with unit trace such that all $c_i$s are multiplied by $-1$.

Introduce the following notation.
\[
\begin{aligned}
j^+_{13}(\mathbf{p},\mathbf{z})&=2\cos(p_3z_3)[\cos(p_1z_1+p_2z_2)+\cos(p_2z_1-p_1z_2)
+\cos((p_1+p_2)(z_1+z_2)/\sqrt{2})\\
&\quad+\cos((p_2z_2-p_1z_1)/\sqrt{2}-p_3z_3)\cos((p_1z_2+p_2z_1)/\sqrt{2})],\\
j^-_{13}(\mathbf{p},\mathbf{z})&=\cos(p_3z_3)[2\cos(p_1z_1-p_2z_2)+2\cos(p_2z_1+p_1z_2)\\
&\quad+\cos((p_1z_1+p_2z_2)/\sqrt{2})\cos((p_2z_1-p_1z_2)/\sqrt{2})].
\end{aligned}
\]

\begin{theorem}[An octagonal random field in the tetragonal class]\label{th:13}
The one-point correlation tensor of a homogeneous and $(D_8\times Z^c_2,5A_{1g}\oplus B_{1g}$-isotropic random field $\mathsf{C}(\mathbf{x})$ is
\[
\langle\mathsf{C}(\mathbf{x})\rangle_{ijkl}=\sum_{m=1}^5C_m\,
{}_{D_8\times Z^c_2}T^{A_{1g},1,1}_{ijkl},
\]
where $C_m\in\mathbb{R}$. Its two-point correlation tensor has the form
\[
\begin{aligned}
\langle\mathsf{C}(\mathbf{x}),\mathsf{C}(\mathbf{y})\rangle
&=\frac{1}{4}\left(\sum_{m=0}^{1}\int_{(\hat{\mathbb{R}}^3/ D_8\times Z^c_2)_m}(j^+_{13}(\mathbf{p},\mathbf{y}-\mathbf{x})+j^-_{13}(\mathbf{p},\mathbf{y}-\mathbf{x}))f^0(\mathbf{p})\,
\mathrm{d}\Phi(\mathbf{p})\right.\\
&\quad+\sum_{m=2}^{4}\int_{(\hat{\mathbb{R}}^3/ D_8\times Z^c_2)_m}j^+_{13}(\mathbf{p},\mathbf{y}-\mathbf{x})f^{+}(\mathbf{p})\,
\mathrm{d}\Phi(\mathbf{p})\\
&\quad+\left.\sum_{m=2}^{4}\int_{(\hat{\mathbb{R}}^3/ D_8\times Z^c_2)_m}j^-_{13}(\mathbf{p},\mathbf{y}-\mathbf{x})f^{-}(\mathbf{p})\,
\mathrm{d}\Phi(\mathbf{p})\right).
\end{aligned}
\]
The field has the form
\[
\begin{aligned}
\mathsf{C}(\mathbf{x})_{ijkl}&=\sum_{m=1}^5C_m\,
{}_{D_8\times Z^c_2}\mathsf{T}^{A_{1g},m,1}_{ijkl}
+\frac{1}{2}\sum_{q=1}^6\sum_{n=1}^{32}\sum_{m=0}^{1}\int_{(\hat{\mathbb{R}}^3/ D_8\times Z^c_2)_m}u_n(\mathbf{p},\mathbf{x})\,\mathrm{d}Z^{0n}_q(\mathbf{p}){}_{D_8\times
Z^c_2}T^{q}_{ijkl}\\
&\quad+\frac{1}{2}\sum_{q=1}^6\sum_{n=1}^{16}\sum_{m=2}^{4}\int_{(\hat{\mathbb{R}}^3/ D_8\times Z^c_2)_m}u_n(\mathbf{p},\mathbf{x})\,\mathrm{d}Z^{+n}_q(\mathbf{p}){}_{D_8\times
Z^c_2}T^{q}_{ijkl}\\
&\quad+\frac{1}{2}\sum_{q=1}^6\sum_{n=17}^{32}\sum_{m=2}^{4}\int_{(\hat{\mathbb{R}}^3/ D_8\times Z^c_2)_m}u_n(\mathbf{p},\mathbf{x})\,\mathrm{d}Z^{-n}_q(\mathbf{p}){}_{D_8\times
Z^c_2}T^{q}_{ijkl},
\end{aligned}
\]
where $(Z^{0n}_1(\mathbf{p}),\dots,Z^{0n}_6(\mathbf{p}))^{\top}$ \emph{(}resp. $(Z^{+n}_1(\mathbf{p}),\dots,Z^{+n}_6(\mathbf{p}))^{\top}$, resp. $(Z^{-n}_1(\mathbf{p}),\dots,Z^{-n}_6(\mathbf{p}))^{\top}$\emph{)} are centred uncorrelated $\mathsf{V}^{D_4\times Z^c_2}$-valued random measures on $(\hat{\mathbb{R}}^3/ D_8\times Z^c_2)_m$, $0\leq m\leq 1$ \emph{(}resp. on $(\hat{\mathbb{R}}^3/ D_8\times Z^c_2)_m$, $2\leq m\leq 4$\emph{)} with control measure $f^0(\mathbf{p})\,\mathrm{d}\Phi(\mathbf{p})$ \emph{(}resp. $f^{+}(\mathbf{p})\,\mathrm{d}\Phi(\mathbf{p})$, resp. $f^{-}(\mathbf{p})\,\mathrm{d}\Phi(\mathbf{p})$\emph{)}, $u_n(\mathbf{p},\mathbf{x})$, $1\leq n\leq 8$ are different product of sines and cosines of angles in Table~\emph{\ref{tab:3}}, and where
\[
T^{q}_{ijkl}=
\begin{cases}
  {}_{D_8\times Z^c_2}\mathsf{T}^{A_{1g},q,1}_{ijkl}, & \mbox{if } 1\leq q\leq 5, \\
  {}_{D_8\times Z^c_2}\mathsf{T}^{B_{1g},m,1}_{ijkl}, & \mbox{otherwise}.
\end{cases}
\]
\end{theorem}

\subsection{The transverse isotropic class}

\begin{theorem}[A transverse isotropic random field in the transverse isotropic class]\label{th:14}
The one-point correlation tensor of the homogeneous and $(\mathrm{O}(2)\times Z^c_2,5U^{0gg})$-isotropic mean-square continuous random field $\mathsf{C}(\mathbf{x})$ has the form
\[
\langle\mathsf{C}(\mathbf{x})\rangle=\sum_{m=1}^{5}C_m\,{}_{\mathrm{O}(2)\times Z^c_2}T^{U^{0gg},m,1}_{ijkl},
\]
where $C_m\in\mathbb{R}$. Its two-point correlation tensor has the form
\[
\begin{aligned}
\langle\mathsf{C}(\mathbf{x}),\mathsf{C}(\mathbf{y})\rangle&=
\int_{\hat{\mathbb{R}}^3/\mathrm{O}(2)\times Z^c_2}J_0\left(\sqrt{(p^2_1+p^2_2)((y_1-x_1)^2+(y_2-x_2)^2)}\right)\\
&\quad\times\cos(p_3(y_3-x_3))f(\mathbf{p}) \,\mathrm{d}\Phi(\mathbf{p}),
\end{aligned}
\]
where $\Phi$ is a measure on $\hat{\mathbb{R}}^3/\mathrm{O}(2)\times Z^c_2$, and $f(\mathbf{p})$ is a $\Phi$-equivalence class of measurable functions on $\hat{\mathbb{R}}^3/\mathrm{O}(2)\times Z^c_2$ with values in the compact set of all nonnegative-definite linear operators in the space $\mathsf{V}^{\mathrm{O}(2)\times Z^c_2}$ with unit trace. The field has the form
\[
\begin{aligned}
\mathsf{C}(\mathbf{x})&=\sum_{m=1}^{5}C_m\,{}_{\mathrm{O}(2)\times
Z^c_2}T^{0\otimes A,m,1}_{ijkl}\\
&\quad+\sum_{m=1}^{5}\int_{\hat{\mathbb{R}}^3/\mathrm{O}(2)\times Z^c_2}J_0\left(\sqrt{(p^2_1+p^2_2)(x^2_1+x^2_2)}\right) \\
&\quad\times(\cos(p_3x_3)\mathrm{d}Z^{01m}(\mathbf{p}){}_{\mathrm{O}(2)\times
Z^c_2}T^{U^{0gg},m,1}_{ijkl}
+\sin(p_3x_3)\mathrm{d}Z^{02m}(\mathbf{p}){}_{\mathrm{O}(2)\times
Z^c_2}T^{U^{0gg},m,1}_{ijkl})\\
&\quad+\sqrt{2}\sum_{\ell=1}^{\infty}\sum_{m=1}^{5}\int_{\hat{\mathbb{R}}^3/\mathrm{O}(2)\times Z^c_2}J_{\ell}\left(\sqrt{(p^2_1+p^2_2)(x^2_1+x^2_2)}\right)\\
&\quad\times(\cos(p_3x_3)\cos(\ell\varphi_p)\mathrm{d}Z^{\ell
1m}(\mathbf{p}){}_{\mathrm{O}(2)\times Z^c_2}T^{U^{0gg},m,1}_{ijkl}\\
&\quad+\cos(p_3x_3)\sin(\ell\varphi_p)\mathrm{d}Z^{\ell
2m}(\mathbf{p}){}_{\mathrm{O}(2)\times Z^c_2}T^{U^{0gg},m,1}_{ijkl}\\
&\quad+\sin(p_3x_3)\cos(\ell\varphi_p)\mathrm{d}Z^{\ell
3m}(\mathbf{p}){}_{\mathrm{O}(2)\times Z^c_2}T^{U^{0gg},m,1}_{ijkl}\\
&\quad+\sin(p_3x_3)\sin(\ell\varphi_p)\mathrm{d}Z^{\ell
4m}(\mathbf{p}){}_{\mathrm{O}(2)\times Z^c_2}T^{U^{0gg},m,1}_{ijkl}),
\end{aligned}
\]
where $(Z^{\ell i1}(\mathbf{p}),\dots,Z^{\ell i5}(\mathbf{p}))^{\top}$ are centred uncorrelated $\mathsf{V}^{\mathrm{O}(2)\times Z^c_2}$-valued random
measures on $\hat{\mathbb{R}}^3/\mathrm{O}(2)\times Z^c_2$ with control measure $f(\mathbf{p})\,\mathrm{d}\Phi(\mathbf{p})$, and where $J_{\ell}$ are the Bessel functions.
\end{theorem}

\subsection{The cubic class}

\begin{theorem}[A cubic random field in the cubic class]\label{th:15}
The one-point correlation tensor of the homogeneous and $(\mathcal{O}\times Z^c_2,3A_{1g})$-isotropic mean-square continuous random field $\mathsf{C}(\mathbf{x})$ has the form
\[
\langle\mathsf{C}(\mathbf{x})\rangle=\sum_{m=1}^{3}C_m\,{}_{\mathcal{O}\times Z^c_2}\mathsf{T}^{A_{1g},m,1}_{ijkl},
\]
where $C_m\in\mathbb{R}$. Its two-point correlation tensor has the form
\[
\langle\mathsf{C}(\mathbf{x}),\mathsf{C}(\mathbf{y})\rangle=
\int_{\hat{\mathbb{R}}^3/\mathcal{O}\times Z^c_2}\sum_{m=0}^{8}j_m(\mathbf{x}-\mathbf{y},\mathbf{p})f(\mathbf{p}) \,\mathrm{d}\Phi(\mathbf{p}),
\]
where the functions $j_m(\mathbf{z},\mathbf{p})$ are shown in Table~\emph{\ref{tab:2old}}, $\Phi$ is a measure on $\hat{\mathbb{R}}^3/\mathcal{O}\times Z^c_2$, and $f(\mathbf{p}) $ is a $\Phi$-equivalence class of measurable functions on $\hat{\mathbb{R}}^3/\mathcal{O}\times Z^c_2$ with values in the compact set of all nonnegative-definite linear operators in the space $\mathsf{V}^{\mathcal{O}\times Z^c_2}$ with unit trace. The field has the form
\[
\mathsf{C}(\mathbf{x})=\sum_{m=1}^{3}C_m\,{}_{\mathcal{O}\times Z^c_2}
\mathsf{T}^{A_{1g},m,1}_{ijkl}+\sum_{m=1}^{3}\sum_{n=1}^{48}
\int_{\hat{\mathbb{R}}^3/\mathcal{O}\times Z^c_2}u_n(\mathbf{x},\mathbf{p})\,\mathrm{d}Z^{mn}(\mathbf{p}){}_{\mathcal{O}\times Z^c_2}\mathsf{T}^{A_{1g},m,1}_{ijkl},
\]
where $(Z^{1n}(\mathbf{p}),\dots,Z^{3n}(\mathbf{p}))^{\top}$ are $48$ centred uncorrelated $\mathsf{V}^{\mathcal{O}\times Z^c_2}$-valued random measures on $\hat{\mathbb{R}}^3/\mathcal{O}\times Z^c_2$ with control measure $f(\mathbf{p})\,\mathrm{d}\mu(\mathbf{p})$, and where $u_n(\mathbf{x},\mathbf{p})$ are different products of sines and cosines of angles from Table~\emph{\ref{tab:2old}}.
\end{theorem}

\subsection{The isotropic class}

\begin{theorem}[An isotropic random field in the isotropic class]\label{th:16}
The one-point correlation tensor of the homogeneous and $(\mathrm{O}(3),2U^{0g})$-isotropic mean-square continuous random field $\mathsf{C}(\mathbf{x})$ has the form
\[
\langle\mathsf{C}(\mathbf{x})\rangle=C_1\delta_{ij}\delta_{kl}
+C_2(\delta_{ik}\delta_{jl}+\delta_{il}\delta_{jk}),\qquad C_m\in \mathbb{R}.
\]
Its two-point correlation tensor has the form
\[
\langle\mathsf{C}(\mathbf{x}),\mathsf{C}(\mathbf{y})\rangle=\int_{0}^{\infty} \frac{\sin(\lambda\|\mathbf{y}-\mathbf{x}\|)}{\lambda\|\mathbf{y}-\mathbf{x}\|} f(\lambda)\,\mathrm{d}\Phi(\lambda),
\]
where $\Phi(\lambda)$ is a finite measure on $[0,\infty)$,
\[
f(\lambda)=
\begin{pmatrix}
v_1(\lambda) & v_2(\lambda) \\
v_2(\lambda) & 1-v_1(\lambda)
\end{pmatrix}
,
\]
and where $\mathbf{v}(\lambda)=(v_1(\lambda),v_2(\lambda))^{\top}$ is a $\Phi$-equivalence class of measurable functions on $[0,\infty)$ taking
values in the closed disk $(v_1(\lambda)-1/2)^2+v^2_2(\lambda)\leq 1/4$. The
field itself has the form
\[
\begin{aligned}
\mathsf{C}_{ijkl}(\rho,\theta,\varphi)&=C_1\delta_{ij}\delta_{kl}
+C_2(\delta_{ik}\delta_{jl}+\delta_{il}\delta_{jk})
+2\sqrt{\pi}\sum_{\ell=0}^{\infty}\sum_{m=\ell}^{\ell}
S^m_{\ell}(\theta,\varphi)\int_{0}^{\infty}j_{\ell}(\lambda\rho)\\
&\quad\times({}_{\mathrm{O}(3)}\mathsf{T}^{0,1,0}_{ijkl}\,\mathrm{d}Z^m_{\ell 1}(\lambda)+{}_{\mathrm{O}(3)}\mathsf{T}^{0,2,0}_{ijkl}\,\mathrm{d}Z^m_{\ell
2}(\lambda)),
\end{aligned}
\]
where $(Z^m_{\ell 1},Z^m_{\ell 2})^{\top}$ is the set of
mutually uncorrelated $\mathsf{V}^{\mathrm{O}(3)}$-valued random measures with $f(\lambda)\,\mathrm{d}\Phi(\lambda)$ as their common control measure.
\end{theorem}

\section{A sketch of proofs of Theorems~\ref{th:1}--\ref{th:16}}

The first display formulae in Theorems~\ref{th:1}--\ref{th:16} follow directly from Theorem~\ref{th:0}.

Now we need to prove that \eqref{eq:12new} is equivalent to the second display formulae in each theorem. The easiest cases arise when $K=H$, i.e., in Theorems~\ref{th:1}, \ref{th:3}, \ref{th:5}, \ref{th:10}, \ref{th:12}, \ref{th:14}--\ref{th:16}. Then the representation $U$ is the direct sum of the $\dim\mathsf{V}$ copies of the trivial representation of the group $K$, the matrix $f(\mathbf{p})$ is nonnegative-definite with unit trace and no further restrictions appear. In Theorems~\ref{th:14} and \ref{th:16}, the group $K$ is infinite, and the integral in \eqref{eq:17} is calculated directly. Otherwise, the group $K$ is discrete. The sets $(\mathbb{R}^3/K)_{M-1}\subset\mathbb{R}^3$ and $(\hat{\mathbb{R}}^3/K)_{M-1}\subset\hat{\mathbb{R}^3}$ have nonempty interior. The coordinate $\bm{\rho}_{M-1}\in(\mathbb{R}^3/K)_{M-1}$ may be identified with the coordinate $\mathbf{x}\in\mathbb{R}^3$, similarly for $\bm{\lambda}_{M-1}\in(\hat{\mathbb{R}}^3/K)_{M-1}$ and $\mathbf{p}\in\hat{\mathbb{R}}^3$. The representation $U^{\ell}$ is trivial. Equation \eqref{eq:17} takes the form
\[
j(\mathbf{p},\mathbf{x})=\frac{1}{|G|}\sum_{g\in G}
\mathrm{e}^{\mathrm{i}(g\mathbf{p},\mathbf{x})},
\]
where $|G|$ is the number of elements in $G$. The matrix entries $g'_{ij}$ of the matrix $g\in K$ in the Wigner basis may be found in \cite[Table~N.7]{altmann1994point}. To calculate the entries $g_{ij}$ in the Gordienko basis, use the following result obtained in \cite{MR2078714}:
\[
g_{ij}=\sum_{k,l=1}^{3}u_{ik}g'_{kl}\overline{u_{jl}},
\]
where $u_{ik}$ are the matrix entries of the unitary matrix
\[
U=\frac{1}{\sqrt{2}}
\begin{pmatrix}
  -1 & 0 & \mathrm{i} \\
  0 & -\sqrt{2}\mathrm{i} & 0 \\
  -1 & 0 & -\mathrm{i}
\end{pmatrix}
.
\]

In Theorem~\ref{th:2} we proceed as follows. By Theorem~\ref{th:0},
\begin{equation}\label{eq:2pcg}
\langle\mathsf{C}(\mathbf{x}),\mathsf{C}(\mathbf{y})
=\int_{\hat{E}}\mathrm{e}^{\mathrm{i}(\mathbf{p},\mathbf{y}-\mathbf{x})}
f(\mathbf{p})\,\mathrm{d}\nu(\lambda).
\end{equation}
The basis of the $21$-dimensional space $\mathsf{V}$ is formed by the basis tensors of the group $K_2$ shown in Table~\ref{tab:3}. We are interested in the tensors of the uncoupled basis of the space $\mathsf{S}^2(\mathsf{V})$ that lie in the spaces of the irreducible components $U^{2t,g}$. They are shown in Table~\ref{tab:7}.

\begin{longtable}{|l|l|}
\caption{The tensors of the uncoupled basis of the space $\mathsf{S}^2(\mathsf{V})$ that lie in the spaces of the irreducible components $U^{2t,g}$.}\label{tab:7} \\
\hline \textbf{Tensor} & \textbf{Value} \\
\hline \endfirsthead \caption[]{continued} \\
\hline 1 & 2 \\
\hline \endhead \hline \multicolumn{2}{c}{\emph{Continued at next page}} \endfoot \hline \endlastfoot
$\mathsf{T}^{0,1,0}_{i\cdots l'}$ & $\mathsf{T}^{0,1}_{ijkl}\mathsf{T}^{0,1}_{i'j'k'l'}$ \\
$\mathsf{T}^{0,2}_{i\cdots l'}$ & $\frac{1}{\sqrt{2}}(\mathsf{T}^{0,1}_{ijkl}\mathsf{T}^{0,2}_{i'j'k'l'}+\mathsf{T}^{0,1}_{i'j'k'l'}
\mathsf{T}^{0,2}_{ijkl})$ \\
$\mathsf{T}^{0,3,0}_{i\cdots l'}$ & $\displaystyle\sum_{q,q'=-2}^2g^{0[q,q']}_{0[2,2]} \mathsf{T}^{2,1,q}_{ijkl}\mathsf{T}^{2,1,q'}_{i'j'k'l'}$ \\
$\mathsf{T}^{0,4,0}_{i\cdots l'}$ & $\mathsf{T}^{0,2}_{ijkl}\mathsf{T}^{0,2}_{i'j'k'l'}$ \\
$\mathsf{T}^{0,5,0}_{i\cdots l'}$ & $\frac{1}{\sqrt{2}}(\displaystyle\sum_{q,q'=-2}^2g^{0[q,q']}_{0[2,2]} \mathsf{T}^{2,1,q}_{ijkl}\mathsf{T}^{2,2,q'}_{i'j'k'l'}+\displaystyle\sum_{q,q'=-2}^2g^{0[q',q]}_{0[2,2]} \mathsf{T}^{2,1,q'}_{i'j'k'l'}\mathsf{T}^{2,2,q}_{ijkl})$ \\
$\mathsf{T}^{0,6,0}_{i\cdots l'}$ & $\displaystyle\sum_{q,q'=-2}^2g^{0[q,q']}_{0[2,2]} \mathsf{T}^{2,2,q}_{ijkl}\mathsf{T}^{2,2,q'}_{i'j'k'l'}$ \\
$\mathsf{T}^{0,7,0}_{i\cdots l'}$ & $\displaystyle\sum_{q,q'=-4}^4g^{0[q,q']}_{0[4,4]} \mathsf{T}^{4,1,q}_{ijkl}\mathsf{T}^{4,1,q'}_{i'j'k'l'}$ \\
$\mathsf{T}^{2,1,u}_{i\cdots l'}$ & $\frac{1}{\sqrt{2}}(\mathsf{T}^{0,1}_{ijkl}\mathsf{T}^{2,1,u}_{i'j'k'l'}
+\mathsf{T}^{0,1}_{i'j'k'l'}\mathsf{T}^{2,1,u}_{ijkl})$ \\
$\mathsf{T}^{2,2,u}_{i\cdots l'}$ & $\frac{1}{\sqrt{2}}(\mathsf{T}^{0,2}_{ijkl}\mathsf{T}^{2,1,u}_{i'j'k'l'}
+\mathsf{T}^{0,2}_{i'j'k'l'}\mathsf{T}^{2,1,u}_{ijkl})$ \\
$\mathsf{T}^{2,3,u}_{i\cdots l'}$ & $\frac{1}{\sqrt{2}}(\mathsf{T}^{0,1}_{ijkl}\mathsf{T}^{2,2,u}_{i'j'k'l'}
+\mathsf{T}^{0,1}_{i'j'k'l'}\mathsf{T}^{2,2,u}_{ijkl})$  \\
$\mathsf{T}^{2,4,u}_{i\cdots l'}$ & $\displaystyle\sum^2_{q,q'=-2}g^{u[q,q']}_{2[2,2]}
\mathsf{T}^{2,1,q}_{ijkl}\mathsf{T}^{2,1,q'}_{i'j'k'l'}$ \\
$\mathsf{T}^{2,5,u}_{i\cdots l'}$ & $\frac{1}{\sqrt{2}}(\mathsf{T}^{0,2}_{ijkl}\mathsf{T}^{2,2,u}_{i'j'k'l'}
+\mathsf{T}^{0,2}_{i'j'k'l'}\mathsf{T}^{2,2,u}_{ijkl})$ \\
$\mathsf{T}^{2,6,u}_{i\cdots l'}$ & $\frac{1}{\sqrt{2}}(\displaystyle\sum^2_{q=-2}\displaystyle\sum^4_{q'=-4} g^{u[q,q']}_{2[2,4]}\mathsf{T}^{2,1,q}_{ijkl}\mathsf{T}^{4,1,q'}_{i'j'k'l'}
+\displaystyle\sum^2_{q'=-2}\displaystyle\sum^4_{q=-4} g^{u[q',q]}_{2[2,4]}\mathsf{T}^{2,1,q'}_{i'j'k'l'}\mathsf{T}^{4,1,q}_{ijkl})$ \\
$\mathsf{T}^{2,7,u}_{i\cdots l'}$ & $\frac{1}{\sqrt{2}}(\displaystyle\sum^2_{q,q'=-2} g^{u[q,q']}_{2[2,2]}\mathsf{T}^{2,2,q}_{ijkl}\mathsf{T}^{2,1,q'}_{i'j'k'l'}
+\displaystyle\sum^2_{q',q=-2}g^{u[q',q]}_{2[2,2]}\mathsf{T}^{2,2,q'}_{i'j'k'l'}
\mathsf{T}^{2,1,q}_{ijkl})$ \\
$\mathsf{T}^{2,8,u}_{i\cdots l'}$ & $\displaystyle\sum^2_{q,q'=-2}g^{u[q,q']}_{2[2,2]}
\mathsf{T}^{2,2,q}_{ijkl}\mathsf{T}^{2,2,q'}_{i'j'k'l'}$ \\
$\mathsf{T}^{2,9,u}_{i\cdots l'}$ & $\frac{1}{\sqrt{2}}(\displaystyle\sum^2_{q=-2}\displaystyle\sum^4_{q'=-4} g^{u[q,q']}_{2[2,4]}\mathsf{T}^{2,2,q}_{ijkl}\mathsf{T}^{4,1,q'}_{i'j'k'l'}
+\displaystyle\sum^2_{q'=-2}\displaystyle\sum^4_{q=-4} g^{u[q',q]}_{2[2,4]}\mathsf{T}^{2,2,q'}_{i'j'k'l'}\mathsf{T}^{4,1,q}_{ijkl})$ \\
$\mathsf{T}^{2,10,u}_{i\cdots l'}$ & $\displaystyle\sum^4_{q,q'=-4}g^{u[q,q']}_{2[4,4]}
\mathsf{T}^{4,1,q}_{ijkl}\mathsf{T}^{4,1,q'}_{i'j'k'l'}$ \\
$\mathsf{T}^{4,1,u}_{i\cdots l'}$ & $\frac{1}{\sqrt{2}}(\mathsf{T}^{0,1}_{ijkl}\mathsf{T}^{4,1,u}_{i'j'k'l'}
+\mathsf{T}^{0,1}_{i'j'k'l'}\mathsf{T}^{4,1,u}_{ijkl})$ \\
$\mathsf{T}^{4,2,u}_{i\cdots l'}$ & $\displaystyle\sum^4_{q,q'=-4}g^{u[q,q']}_{4[2,2]}
\mathsf{T}^{2,1,q}_{ijkl}\mathsf{T}^{2,1,q'}_{i'j'k'l'}$ \\
$\mathsf{T}^{4,3,u}_{i\cdots l'}$ & $\frac{1}{\sqrt{2}}(\mathsf{T}^{0,2}_{ijkl}\mathsf{T}^{4,1,u}_{i'j'k'l'}
+\mathsf{T}^{0,2}_{i'j'k'l'}\mathsf{T}^{4,1,u}_{ijkl})$ \\
$\mathsf{T}^{4,4,u}_{i\cdots l'}$ & $\frac{1}{\sqrt{2}}(\displaystyle\sum^4_{q,q'=-4} g^{u[q,q']}_{4[2,2]}\mathsf{T}^{2,2,q}_{ijkl}\mathsf{T}^{2,1,q'}_{i'j'k'l'}
+\displaystyle\sum^4_{q',q=-4}g^{u[q',q]}_{4[2,2]}\mathsf{T}^{2,2,q'}_{i'j'k'l'}
\mathsf{T}^{2,1,q}_{ijkl})$ \\
$\mathsf{T}^{4,5,u}_{i\cdots l'}$ & $\frac{1}{\sqrt{2}}(\displaystyle\sum^2_{q=-2} \displaystyle\sum^4_{q'=-4}g^{u[q,q']}_{4[2,4]}\mathsf{T}^{2,1,q}_{ijkl}\mathsf{T}^{4,1,q'}_{i'j'k'l'}
+\displaystyle\sum^2_{q'=-2} \displaystyle\sum^4_{q=-4}g^{u[q',q]}_{4[2,4]}\mathsf{T}^{2,1,q'}_{i'j'k'l'}\mathsf{T}^{4,1,q}_{ijkl})$ \\
$\mathsf{T}^{4,6,u}_{i\cdots l'}$ & $\displaystyle\sum^4_{q,q'=-4}g^{u[q,q']}_{4[2,2]}
\mathsf{T}^{2,2,q}_{ijkl}\mathsf{T}^{2,2,q'}_{i'j'k'l'}$ \\
$\mathsf{T}^{4,7,u}_{i\cdots l'}$ & $\frac{1}{\sqrt{2}}(\displaystyle\sum^2_{q=-2} \displaystyle\sum^4_{q'=-4}g^{u[q,q']}_{4[2,4]}\mathsf{T}^{2,2,q}_{ijkl}\mathsf{T}^{4,1,q'}_{i'j'k'l'}
+\displaystyle\sum^2_{q'=-2} \displaystyle\sum^4_{q=-4}g^{u[q',q]}_{4[2,4]}\mathsf{T}^{2,2,q'}_{i'j'k'l'}\mathsf{T}^{4,1,q}_{ijkl})$ \\
$\mathsf{T}^{4,8,u}_{i\cdots l'}$ & $\displaystyle\sum^4_{q,q'=-4}g^{u[q,q']}_{4[4,4]}
\mathsf{T}^{4,1,q}_{ijkl}\mathsf{T}^{4,1,q'}_{i'j'k'l'}$ \\
$\mathsf{T}^{6,1,u}_{i\cdots l'}$ & $\frac{1}{\sqrt{2}}(\displaystyle\sum^2_{q=-2} \displaystyle\sum^4_{q'=-4}g^{u[q,q']}_{6[2,4]}\mathsf{T}^{2,1,q}_{ijkl}\mathsf{T}^{4,1,q'}_{i'j'k'l'}
+\displaystyle\sum^2_{q'=-2} \displaystyle\sum^4_{q=-4}g^{u[q',q]}_{6[2,4]}\mathsf{T}^{2,1,q'}_{i'j'k'l'}\mathsf{T}^{4,1,q}_{ijkl})$ \\
$\mathsf{T}^{6,2,u}_{i\cdots l'}$ & $\frac{1}{\sqrt{2}}(\displaystyle\sum^2_{q=-2} \displaystyle\sum^4_{q'=-4}g^{u[q,q']}_{6[2,4]}\mathsf{T}^{2,2,q}_{ijkl}\mathsf{T}^{4,1,q'}_{i'j'k'l'}
+\displaystyle\sum^2_{q'=-2} \displaystyle\sum^4_{q=-4}g^{u[q',q]}_{6[2,4]}\mathsf{T}^{2,2,q'}_{i'j'k'l'}\mathsf{T}^{4,1,q}_{ijkl})$ \\
$\mathsf{T}^{6,3,u}_{i\cdots l'}$ & $\displaystyle\sum^4_{q,q'=-4}g^{u[q,q']}_{6[4,4]}
\mathsf{T}^{4,1,q}_{ijkl}\mathsf{T}^{4,1,q'}_{i'j'k'l'}$ \\
$\mathsf{T}^{8,1,u}_{i\cdots l'}$ & $\displaystyle\sum^4_{q,q'=-4}g^{u[q,q']}_{8[4,4]}
\mathsf{T}^{4,1,q}_{ijkl}\mathsf{T}^{4,1,q'}_{i'j'k'l'}$ \\
\hline
\end{longtable}

The function $f(\lambda)$ takes the form
\begin{equation}\label{eq:Hp}
f_{i\cdots l'}(\lambda)=\sum^4_{t=0}\sum^{m_{2t}}_{v=1}f_{2t,v}(\lambda)
\mathsf{T}^{2t,v,0}_{i\cdots l'}
\end{equation}
with $f_{2t,v}(0)=0$ for $t\geq 1$, where $m_0=7$, $m_2=10$, $m_4=8$, $m_6=3$, and $m_8=1$. When $\lambda=0$, we obtain
\[
f_{i\cdots l'}(0)=\sum^7_{v=1}f_{0,v}(\lambda)\mathsf{T}^{0,v,0}_{i\cdots l'}.
\]
To simplify this expression, we note that $\mathsf{T}^{0,v,0}_{i\cdots l'}=M^{0,v}_{i\cdots l'}$. Using MATLAB Symbolic Math Toolbox, we express the functions $M^{n,m}(\mathbf{p})$ as linear combinations of the functions $L^m_{i\cdots l'}(\mathbf{p})$ given in Table~\ref{tab:6}. The results are given in Table~\ref{tab:8}.

\begin{longtable}{|l|l|}
\caption{The functions $M^{n,m}(\mathbf{p})$ expressed as linear combinations of the functions $L^m_{i\cdots l'}(\mathbf{p})$}\label{tab:8}  \\
\hline $\mathbf{M^{n,m}_{i\cdots l'}(p)}$ & \textbf{Linear combination} \\
\hline \endfirsthead \caption[]{continued} \\
\hline 1 & 2 \\
\hline \endhead \hline \multicolumn{2}{c}{\emph{Continued at next page}} \endfoot \hline \endlastfoot
$M^{0,1}_{i\cdots l'}$ & $\frac{1}{9}L^{0,1}_{i\cdots l'}$ \\
$M^{0,2}_{i\cdots l'}$ & $-\frac{\sqrt{2}}{9\sqrt{5}}L^1_{i\cdots l'}
+\frac{1}{6\sqrt{10}}L^2_{i\cdots l'}$ \\
$M^{0,3}_{i\cdots l'}$ & $-\frac{2}{9\sqrt{5}}L^1_{i\cdots l'}
+\frac{1}{12\sqrt{5}}L^3_{i\cdots l'}$ \\
$M^{0,4}_{i\cdots l'}$ & $\frac{1}{45}L^1_{i\cdots l'}
-\frac{1}{30}L^2_{i\cdots l'}+\frac{1}{20}L^4_{i\cdots l'}$ \\
$M^{0,5}_{i\cdots l'}$ & $\frac{8}{9\sqrt{35}}L^1_{i\cdots l'}
-\frac{1}{3\sqrt{35}}L^2_{i\cdots l'}-\frac{1}{3\sqrt{35}}L^3_{i\cdots l'}
+\frac{1}{8\sqrt{35}}L^5_{i\cdots l'}$ \\
$M^{0,6}_{i\cdots l'}$ & $\frac{11}{63\sqrt{5}}L^1_{i\cdots l'}
-\frac{1}{42\sqrt{5}}L^2_{i\cdots l'}-\frac{\sqrt{5}}{42}L^3_{i\cdots l'}
-\frac{1}{28\sqrt{5}}L^4_{i\cdots l'}+\frac{1}{28\sqrt{5}}L^5_{i\cdots l'}$ \\
& $+\frac{3}{28\sqrt{5}}L^6_{i\cdots l'}-\frac{3}{56\sqrt{5}}L^7_{i\cdots l'}$ \\
$M^{0,7}_{i\cdots l'}$ & $-\frac{1}{35}L^1_{i\cdots l'}
+\frac{2}{105}L^2_{i\cdots l'}+\frac{1}{84}L^3_{i\cdots l'}
-\frac{1}{210}L^4_{i\cdots l'}-\frac{1}{84}L^5_{i\cdots l'}$ \\
\hline
\end{longtable}

We enumerate the 21 indexes $ijk\ell$ in the following order: $-1-1-1-1$, $0000$, $1111$, $0101$, $-11-11$, $-10-10$, $-1-100$, $-1-111$, $0011$, $-11-10$, $-1-101$, $1101$, $0001$, $01-11$, $11-10$, $-1-1-10$, $00-10$, $01-10$, $-1-1-11$, $11-11$, $00-11$. With this order, the matrix $f_{II'}(\lambda)$ becomes block-diagonal. We chose $29$ linearly independent elements of the above matrix according to Table~\ref{tab:9}.

\begin{longtable}{|p{0.05\textwidth}|p{0.05\textwidth}|p{0.85\textwidth}|}
\caption{The elements of the matrix $f_{II'}(\lambda)$.}\label{tab:9}\\
\hline $I$ & $I'$ & $f_{II'}(\lambda)$ \\
\hline \endfirsthead \caption[]{continued} \\
\hline 1 & 2 & 3 \\
\hline \endhead \hline \multicolumn{3}{c}{\emph{Continued at next page}} \endfoot \hline \endlastfoot
$1$ & $1$ &
$\frac{1}{9}f^{0,1}(\lambda)+\frac{4}{45}f^{0,4}(\lambda)
+\frac{8\sqrt{5}}{315}f^{0,6}(\lambda)+\frac{4\sqrt{5}}{45}f^{0,3}(\lambda)
+\frac{8}{105}f^{0,7}(\lambda)$ \\ & & $+\frac{2\sqrt{10}}{45}f^{0,2}(\lambda)+\frac{8\sqrt{35}}{315}f^{0,5}(\lambda)
-\frac{4\sqrt{14}}{441}f^{2,8}(\lambda)-\frac{2\sqrt{14}}{63}f^{2,4}(\lambda)$ \\ & & $-\frac{8\sqrt{77}}{1617}f^{2,10}(\lambda)-\frac{4\sqrt{10}}{105}f^{2,6}(\lambda)
-\frac{8\sqrt{35}}{735}f^{2,9}(\lambda)-\frac{\sqrt{2}}{9}f^{2,1}(\lambda)$ \\ & & $-\frac{2\sqrt{7}}{63}f^{2,3}(\lambda)-\frac{4\sqrt{35}}{315}f^{2,5}(\lambda)
-\frac{4\sqrt{2}}{63}f^{2,7}(\lambda)-\frac{2\sqrt{10}}{45}f^{2,2}(\lambda)$ \\ & & $+\frac{\sqrt{70}}{245}f^{4,6}(\lambda)+\frac{\sqrt{70}}{70}f^{4,2}(\lambda)
+\frac{27\sqrt{2002}}{35035}f^{4,8}(\lambda)+\frac{2\sqrt{11}}{77}f^{4,5}(\lambda)$ \\ & & $+\frac{2\sqrt{154}}{539}f^{4,7}(\lambda)+\frac{\sqrt{7}}{35}f^{4,3}(\lambda)
+\frac{\sqrt{35}}{70}f^{4,1}(\lambda)+\frac{\sqrt{10}}{35}f^{4,4}(\lambda)$ \\ & & $-\frac{\sqrt{55}}{231}f^{6,3}(\lambda)-\frac{5\sqrt{77}}{462}f^{6,1}(\lambda)
-\frac{5\sqrt{22}}{462}f^{6,2}(\lambda)+\frac{7\sqrt{1430}}{6864}f^{8,1}(\lambda)$\\
$1$ & $2$ &
$\frac{1}{9}f^{0,1}(\lambda)+\frac{4}{45}f^{0,4}(\lambda)
-\frac{4\sqrt{5}}{315}f^{0,6}(\lambda)-\frac{2\sqrt{5}}{45}f^{0,3}(\lambda)
+\frac{1}{35}f^{0,7}(\lambda)$ \\
\hline
\end{longtable}

The remaining non-zero entries of the matrix $f(\lambda)$ are defined by \eqref{eq:H1} and \eqref{eq:H2}. Let $u_i(\lambda)$, $1\leq i\leq 29$, be the functions \eqref{eq:uH}. Define the functions $v_i(\lambda)$, $1\leq i\leq 26$, by \eqref{eq:vi}.

We see that the set of extreme points of the set $\mathcal{C}$ consists of three connected components. The first one is the $14$-dimensional boundary of the $15$-dimensional set of all $9\times 9$ symmetric nonnegative-definite matrices with unit trace with coordinates $v_1(\lambda)$, \dots, $v_{15}(\lambda)$. The second one is the $5$-dimensional boundary of the $6$-dimensional set of all $4\times 4$ symmetric nonnegative-definite matrices with unit trace with coordinates $v_{16}(\lambda)$, \dots, $v_{21}(\lambda)$. Finally, the third one is the $4$-dimensional boundary of the $5$-dimensional set of all $4\times 4$ symmetric nonnegative-definite matrices with unit trace with coordinates $v_{22}(\lambda)$, \dots, $v_{26}(\lambda)$.

The functions $f^{2t,v}(\lambda)$ are expressed in terms of $u_i(\lambda)$ according to Table~\ref{tab:10}.

\begin{longtable}{|p{0.1\textwidth}|p{0.85\textwidth}|}
\caption{The functions $f^{2t,v}(\lambda)$.}\label{tab:10}\\
\hline Function & Value \\

\hline \endfirsthead \caption[]{continued} \\
\hline 1 & 2 \\
\hline \endhead \hline \multicolumn{2}{c}{\emph{Continued at next page}} \endfoot \hline \endlastfoot
$f^{0,1}(\lambda)$ & $\frac{1}{9}u_2(\lambda)+\frac{16}{9}u_4(\lambda)
+\frac{8}{9}u_5(\lambda)+\frac{8}{9}u_7(\lambda)+\frac{32}{9}u_9(\lambda)
+\frac{16}{9}u_{10}(\lambda)$ \\
& $-\frac{8}{9}u_{12}(\lambda)+\frac{8}{9}u_{13}(\lambda)-\frac{32}{9}u_{16}(\lambda)
-\frac{16}{9}u_{26}(\lambda)-\frac{64}{9}u_{29}(\lambda)$ \\
$f^{0,2}(\lambda)$ & $\frac{2\sqrt{2}}{9\sqrt{5}}u_2(\lambda)
+\frac{56\sqrt{2}}{9\sqrt{5}}u_4(\lambda)-\frac{8\sqrt{2}}{9\sqrt{5}}u_5(\lambda)
-\frac{8\sqrt{2}}{3\sqrt{5}}u_6(\lambda)+\frac{2\sqrt{10}}{9}u_7(\lambda)$ \\
& $+\frac{16\sqrt{2}}{3\sqrt{5}}u_8(\lambda)-\frac{8\sqrt{2}}{9\sqrt{5}}u_9(\lambda)
+\frac{32\sqrt{2}}{9\sqrt{5}}u_{10}(\lambda)+\frac{4\sqrt{2}}{3\sqrt{5}}u_{11}(\lambda)
+\frac{2\sqrt{2}}{9\sqrt{5}}u_{12}(\lambda)$ \\
& $+\frac{4\sqrt{2}}{9\sqrt{5}}u_{13}(\lambda)-\frac{16\sqrt{2}}{3\sqrt{5}}u_{14}(\lambda)
+\frac{16\sqrt{2}}{3\sqrt{5}}u_{15}(\lambda)+\frac{56\sqrt{2}}{9\sqrt{5}}u_{16}(\lambda)
+\frac{16\sqrt{2}}{9\sqrt{5}}u_{26}(\lambda)$ \\
& $-\frac{32\sqrt{2}}{3\sqrt{5}}u_{27}(\lambda)-\frac{16\sqrt{2}}{3\sqrt{5}}u_{28}(\lambda)
+\frac{16\sqrt{2}}{9\sqrt{5}}u_{29}(\lambda)$ \\
$f^{0,3}(\lambda)$ & $\frac{4}{3\sqrt{5}}u_1(\lambda)+\frac{4}{9\sqrt{5}}u_2(\lambda)
-\frac{16\sqrt{5}}{9}u_4(\lambda)+\frac{2}{9\sqrt{5}}u_5(\lambda)
+\frac{8}{3\sqrt{5}}u_6(\lambda)$ \\
& $-\frac{16}{9\sqrt{5}}u_7(\lambda)-\frac{16}{9\sqrt{5}}u_9(\lambda)
-\frac{32}{9\sqrt{5}}u_{10}(\lambda)+\frac{16}{9\sqrt{5}}u_{12}(\lambda)
+\frac{8}{9\sqrt{5}}u_{13}(\lambda)$ \\
\hline
\end{longtable}

Substitute these values to \eqref{eq:Hp}. We obtain the matrix entries $f_{i\cdots\ell'}(\mathbf{p})$ expresses in terms of $u_i(\lambda)$ and $M^{n,m}(\mathbf{p})$. Using Table~\ref{tab:8}, we express $f_{i\cdots\ell'}(\mathbf{p})$ in terms of $u_i(\lambda)$ and $L^q_{i\cdots\ell'}(\mathbf{p})$. Substitute the obtained expression into \eqref{eq:2pcg} and use the \emph{Rayleigh expansion}
\begin{equation}\label{eq:Rayleigh}
\mathrm{e}^{\mathrm{i}(\mathbf{p},\mathbf{y})}=4\pi\sum^{\infty}_{\ell=0}
\sum^{\ell}_{m=-\ell}\mathrm{i}^{\ell}j_{\ell}(\lambda\rho)
S^m_{\ell}(\theta_{\mathbf{p}},\varphi_{\mathbf{p}})
S^m_{\ell}(\theta_{\mathbf{y}},\varphi_{\mathbf{y}}),
\end{equation}
where $j_{\ell}$ are the spherical Bessel functions, $S^m_{\ell}$ are real-valued spherical harmonics, $(\rho,\theta_{\mathbf{y}},\varphi_{\mathbf{y}})$ are the spherical coordinates of the point $\mathbf{y}\in E$, and $(\lambda,\theta_{\mathbf{p}},\varphi_{\mathbf{p}})$ are those of the point $\mathbf{p}\in\hat{E}$. We obtain \eqref{eq:2point} with
\[
\begin{aligned}
\mathrm{d}\Phi_1(\lambda)&=(u_1(\lambda)+\cdots+u_6(\lambda))\,\mathrm{d}\nu(\lambda),\\
\mathrm{d}\Phi_2(\lambda)&=(u_{17}(\lambda)+\cdots+u_{20}(\lambda))\,\mathrm{d}\nu(\lambda),\\
\mathrm{d}\Phi_3(\lambda)&=(u_{24}(\lambda)+\cdots+u_{26}(\lambda))\,\mathrm{d}\nu(\lambda).
\end{aligned}
\]

Using Table~\ref{tab:9}, we obtain
\[
\begin{aligned}
u_{17}(0)+\cdots+u_{20}(0)&=\frac{1}{2\sqrt{5}}f^{0,3}(0)+\frac{11}{28\sqrt{5}}
f^{0,6}(0)+\frac{2}{7}f^{0,7}(0),\\
u_{24}(0)+\cdots+u_{26}(0)&=\frac{1}{4\sqrt{5}}f^{0,3}(0)+\frac{11}{56\sqrt{5}}
f^{0,6}(0)+\frac{1}{7}f^{0,7}(0),
\end{aligned}
\]
which proves \eqref{eq:phi2phi3}.

In the case of Theorem~\ref{th:6}, the restrictions of the representation $B_{1g}$ to the subgroups $H_0=D_4\times Z^c_2$ and $H_1=Z_4=\{E,C^+_4,C^-_4,C_2\}$ do not contain the trivial representations of these groups, therefore the matrix $B$ in \eqref{eq:18} is $0$. The restrictions of the representation $B_{1g}$ to the subgroups $H_2=Z_2=\{E,C_2\}$, $H_3=Z^-_2=\{E,\sigma_h\}$, and $H_4=\{E\}$ contain the trivial representations of these groups. Choose the set $(D_4\times Z^c_2)/Z_4$ as
\[
(D_4\times Z^c_2)/Z_4=\{Z_4,C'_{21}Z_4,iZ_4,\sigma_{v1}Z_4\}.
\]
By \cite[Table~33.10]{altmann1994point}, the representation $g\mapsto g$ of the group~$D_4\times Z^c_2$ has the form $A_{2u}\oplus E_u$. We calculate the matrix entries of the above representation using \cite[Table~33.7]{altmann1994point}. The sum
\[
\frac{1}{8}\sum\exp[\mathrm{i}((A_{2u}\oplus E_u)(g_j)\mathbf{p},\mathbf{z})]
\]
over $g_j\in Z_4\cup iZ_4$ is $j_1(\mathbf{p},\mathbf{z})$, the similar sum over $g_j\in C'_{21}Z_4\cup\sigma_{v1}Z_4$ is $j_2(\mathbf{p},\mathbf{z})$. The representation $B_{1g}$ takes value $-1$ on $C'_{21}Z_4\cup\sigma_{v1}Z_4$, hence $f^-(\mathbf{p})$ is obtained from $f^+(\mathbf{p})$ by multiplying $B$ and $B^{\top}$ by $-1$. Equation~\ref{eq:19} follows.

In the case of Theorem~\ref{th:7}, the symmetric part of the tensor square of the representation $E_{2g}$ is $A_{1g}\oplus E_{2g}$, where $A_{1g}$ acts in the linear space of matrices \eqref{eq:13new} with $b=c=d=0$, and $E_{2g}$ acts in the space with $a=d=0$. The skew-symmetric part of the above tensor square is $A_{2g}$ and acts in the space with $a=b=c=0$. The restrictions of the representations $E_{2g}$ and $A_{2g}$ to the subgroups $H_0=D_6\times Z^c_2$ and $H_1=D_3$ do not contain the trivial representation of these groups, therefore $\mathbf{c}_i=\mathbf{0}$ and $A_i$ satisfy $b=c=d=0$ on $(\hat{\mathbb{R}}^3/D_6\times Z^c_2)_m$, $0\leq m\leq 1$. The restriction of the representation $E_{2g}$ to the subgroup $H_2=D_2$ contains the trivial representation of $H_2$, while that of the representation $A_{2g}$ does not contain the trivial representation of $H_2$. Therefore $A_i$ are symmetric in $f^-(\mathbf{p})$. Finally, the restrictions of the representations $E_{2g}$ and $A_{2g}$ to the subgroups $H_3=Z^c_2$ and $H_4=\{E\}$ contain the trivial representation of these groups. By \cite[Table~35.10]{altmann1994point}, the representation $g\mapsto g$ of the group~$D_6\times Z^c_2$ is $A_{2u}\oplus E_{1u}$. The group $D_6\times Z^c_2$ is the union of the sets $G_n$, $1\leq n\leq 6$ as follows: $G_1=\{E,C_2,i,\sigma_h\}$, $G_2=\{C'_{21},C''_{21},\sigma_{d1},\sigma_{v1}\}$, $G_3=\{C^+_6,C^-_3,S^-_3,S^+_6\}$, $G_4=\{C^-_6,C^+_3,S^+_3,S^-_6\}$, $G_5=\{C'_{22},C''_{22},\sigma_{d2},\sigma_{v2}\}$, and $G_6=\{C'_{23},C''_{23},\sigma_{d3},\sigma_{v3}\}$. The representation $E_{2g}$ maps the elements of the set~$G_n$ to the matrix $g_n$, and the sum
\[
\frac{1}{4}\sum_{g\in G_n}\exp[\mathrm{i}((A_{2u}\oplus E_{1u})(g)\mathbf{p},\mathbf{z})]
\]
is equal to $j_n(\mathbf{p},\mathbf{z})$. Under the action of the representation $E_{2g}$ all $\mathbf{c}_j$ become $g_i\mathbf{c}_j$, the vectors $(b,c)^{\top}$ become $g_i(b,c)^{\top}$, all $A_j$ become $g_iA_jg^{-1}_i$.

In the case of Theorem~\ref{th:8}, the restrictions of the representation ${}^1E_g\oplus{}^2E_g$ to the subgroups $H_0=\mathcal{T}\times Z^c_2$ and $H_1=Z_3\times Z^c_2$ do not contain the trivial representations of these groups, therefore $\mathbf{c}_i=\mathbf{0}$ in $f^0(\mathbf{p})$. The symmetric tensor square of ${}^1E_g\oplus{}^2E_g$ is $A_g\oplus E$, where $A_g$ acts in the one-dimensional space generated by the identity matrix, therefore $A_i$ must be proportional to the identity matrix. The restrictions of the representation ${}^1E_g\oplus{}^2E_g$ to the subgroups $H_2=Z_2$, and $H_3=Z^-_2$, and $H_4=\{E\}$ contain the trivial representations of these groups. By \cite[Table~72.10]{altmann1994point}, the representation $g\mapsto g$ of the group~$\mathcal{T}\times Z^c_2$ is $T_u$. The representation ${}^1E_g\oplus{}^2E_g$ maps the elements $E$, $C_{2x}$, $C_{2y}$, $C_{2z}$ of the group $\mathcal{T}\times Z^c_2$ to the $2\times 2$ identity matrix, and the elements $i$, $\sigma_x$, $\sigma_y$, $\sigma_z$ to the $2\times 2$ identity matrix times $-1$. The sum
\begin{equation}\label{eq:21}
\frac{1}{8}\sum\exp[\mathrm{i}(T_u(g_j)\mathbf{p},\mathbf{z})]
\end{equation}
over the above elements is $j_1(\mathbf{p},\mathbf{z})$. Similarly, the elements $C^+_{31}$, $C^+_{32}$, $C^+_{33}$, and $C^+_{34}$ are mapping to the matrix $g$ of equation~\eqref{eq:20}, and the elements $S^-_{61}$, $S^-_{62}$, $S^-_{63}$, and $S^-_{64}$ are mapping to $-g$. The sum~\eqref{eq:21} over the above elements is $j_2(\mathbf{p},\mathbf{z})$. Finally, the elements $C^-_{31}$, $C^-_{32}$, $C^-_{33}$, and $C^-_{34}$ are mapping to $g^{-1}$, and the elements $S^+_{61}$, $S^+_{62}$, $S^+_{63}$, and $S^+_{64}$ to $-g^{-1}$. The sum~\eqref{eq:21} over the above elements is $j_3(\mathbf{p},\mathbf{z})$. Under the action of the representation  ${}^1E_g\oplus{}^2E_g$ the vectors $\mathbf{c}_i$ and the matrices $A_i$ are transformed according to the explanations before the text of Theorem~\ref{th:8}.

In the case of Theorem~\ref{th:9}, the symmetric part of the tensor square of the representation $E_g$ is $A_{1g}\oplus E_g$, where $A_{1g}$ acts in the linear space of matrices \eqref{eq:13new} with $b=c=d=0$, and $E_g$ acts in the space with $a=d=0$. The skew-symmetric part of the above tensor square is $A_{2g}$ and acts in the space with $a=b=c=0$. The restrictions of the representations $E_g$ and $A_{2g}$ to the subgroups $H_0=\mathcal{O}\times Z^c_2$ and $H_1=D_3$ do not contain the trivial representation of these groups, therefore $\mathbf{c}_i=\mathbf{0}$ and $A_i$ satisfy $b=c=d=0$ on $(\hat{\mathbb{R}}^3/\mathcal{O}\times Z^c_2)_m$, $0\leq m\leq 1$. The restriction of the representation $E_g$ to the subgroup $H_2=D_4$ contains the trivial representation of $H_2$, while that of the representation $A_{2g}$ does not contain the trivial representation of $H_2$. Therefore $A_i$ are symmetric in $f^-(\mathbf{p})$. Finally, the restrictions of the representations $E_g$ and $A_{2g}$ to the subgroups $H_3=D_2$, $H_4=\tilde{Z}_2$, $H_5=Z_2$, and $H_6=\{E\}$ contain the trivial representation of these groups. By \cite[Table~35.10]{altmann1994point}, the representation $g\mapsto g$ of the group~$\mathcal{O}\times Z^c_2$ is $T_{1u}$. The group $\mathcal{O}\times Z^c_2$ is the union of the sets $G_n$, $1\leq n\leq 6$ as follows:
\[
\begin{aligned}
G_1&=\{E,C_{2x},C_{2y},C_{2z},i,\sigma_x,\sigma_y,\sigma_z\},\\ G_2&=\{C^+_{4z},C^-_{4z},C'_{2a},C'_{2b},S^-_{4z},S^+_{4z},\sigma_{d1},\sigma_{d2}\},\\
G_3&=\{C^+_{31},C^+_{32},C^+_{33},C^+_{34},S^-_{61},S^-_{62},S^-_{63},S^-_{64}\},\\ G_4&=\{C^+_{31},C^+_{32},C^+_{33},C^+_{34},S^+_{61},S^+_{62},S^+_{63},S^+_{64}\},\\ G_5&=\{C^+_{4x},C^-_{4x},C'_{2d},C'_{2f},S^-_{4x},S^+_{4x},\sigma_{d4},\sigma_{d6}\},\\
G_6&=\{C^+_{4y},C^-_{4y},C'_{2c},C'_{2e},S^-_{4y},S^+_{4y},\sigma_{d3},\sigma_{d5}\}.
\end{aligned}
\]
The representation $E_g$ maps the elements of the set~$G_n$ to the matrix $g_n$, and the sum
\[
\frac{1}{8}\sum_{g\in G_n}\exp[\mathrm{i}((A_{2u}\oplus E_{1u})(g)\mathbf{p},\mathbf{z})]
\]
is equal to $j_n(\mathbf{p},\mathbf{z})$. Under the action of the representation $E_g$ all $\mathbf{c}_j$ become $g_i\mathbf{c}_j$, the vectors $(b,c)^{\top}$ become $g_i(b,c)^{\top}$, all $A_j$ become $g_iA_jg^{-1}_i$.

To prove the last part of each theorem, we first observe that any homogeneous random field $\mathsf{C}(\mathbf{x})$ may be written as
\[
\mathsf{C}(\mathbf{x})=\langle\mathsf{C}(\mathbf{x})\rangle
+[\mathsf{C}(\mathbf{x})-\langle\mathsf{C}(\mathbf{x})\rangle].
\]
The first term in the right hand side is the same as that in the spectral expansions in Theorems~\ref{th:1}--\ref{th:16}. The second term is centred and has the same two-point correlation tensor as $\mathsf{C}(\mathbf{x})$ has. Assume that the above tensor has the form
\begin{equation}\label{eq:9}
\langle\mathsf{C}(\mathbf{x}),\mathsf{C}(\mathbf{y})\rangle
=\int_{\Lambda}\overline{u(\mathbf{x},\lambda)}u(\mathbf{y},\lambda)
\,\mathrm{d}F(\lambda),
\end{equation}
where $\Lambda$ is a set, and where $F$ is a measure on a $\sigma$-field $\mathfrak{L}$ of subsets of $\Lambda$ taking values in the set of Hermitian nonnegative-definite operators on $\mathsf{V}_{\mathbb{C}}$. Let $\Phi$ be the following measure:
\[
\Phi(A):=\tr F(A),\qquad A\in\mathfrak{L}.
\]
Assume that the set $\{\,u(\mathbf{x},\lambda)\colon\mathbf{x}\in\mathbb{R}^3\,\}$ is \emph{total} in the Hilbert space $L^2(\Lambda,\Phi)$ of the measurable complex-valued functions on $\Lambda$ that are square-integrable with respect to the measure $\Phi$, that is, the set of finite linear combinations $\sum c_nu(\mathbf{x}_n,\lambda)$ is dense in the above space. By Karhunen's theorem \cite{MR0023013}, the field $\mathsf{C}(\mathbf{x})$ has the following spectral expansion:
\[
\mathsf{C}(\mathbf{x})=\int_{\Lambda}u(\mathbf{x},\lambda)
\,\mathrm{d}\mathsf{Z}(\lambda),
\]
where $\mathsf{Z}$ is a measure on the measurable space $(\Lambda,\mathfrak{L})$ taking values in the Hilbert space of random tensors $\mathsf{Z}\colon\Omega\to\mathsf{V}_{\mathbb{C}}$ with $\mathsf{E}[\mathsf{Z}]=\mathsf{0}$ and $\mathsf{E}[\|\mathsf{Z}\|^2]<\infty$. The measure $F$ is the \emph{control measure} of the measure $\mathsf{Z}$, i.e.,
\[
\mathsf{E}[J\mathsf{Z}(A)\mathsf{Z}^{\top}(B)]=\Phi(A\cap B),\qquad
A,B\in\mathfrak{L},
\]
where $J$ is the \emph{real structure} in the space $\mathsf{V}_{\mathbb{C}}$: $J(\mathsf{v}+\mathrm{i}\mathsf{w})=\mathsf{v}-\mathrm{i}\mathsf{w}$.

We illustrate the use of Kahrunen's theorem in Theorem~\ref{th:1}. The two-point correlation tensor of the random field $\mathsf{C}(\mathbf{x})$ has the form \eqref{eq:9}, where $\Lambda$ is the union $Z^c_2\backslash\hat{E}_1\cup Z^c_2\backslash\hat{E}_2$ of two copies of the space $Z^c_2\backslash\hat{E}$ and
\[
u(\mathbf{x},\mathbf{p})=
\begin{cases}
  \cos(\mathbf{p},\mathbf{x}), & \mbox{if } \mathbf{p}\in Z^c_2\backslash\hat{E}_1, \\
  \sin(\mathbf{p},\mathbf{x}), & \mbox{if } \mathbf{p}\in Z^c_2\backslash\hat{E}_2.
\end{cases}
\]
This follows from the elementary formula $\cos(\mathbf{p},\mathbf{y}-\mathbf{x})=\cos(\mathbf{p},\mathbf{x})\cos(\mathbf{p},\mathbf{y})
+\sin(\mathbf{p},\mathbf{x})\sin(\mathbf{p},\mathbf{y})$. Similar considerations are applicable in Theorems~\ref{th:3}, \ref{th:5}--\ref{th:13}, and \ref{th:15}, where the group $K$ is discrete.

In Theorems~\ref{th:2} and \ref{th:16}, where $K=\mathrm{O}(3)$, we represent the \emph{plane wave} $\mathrm{e}^{\mathrm{i}(\mathbf{p},\mathbf{y}-\mathbf{x})}$ in the form $u(\mathbf{x},\lambda)\overline{u(\mathbf{y},\lambda)}$, using the real version of the Rayleigh expansion \eqref{eq:Rayleigh}. We see that $\Lambda$ is the union of countably many copies of the half-line $[0,\infty)$ enumerated by the pairs of integers $(\ell,m)$ with $\ell\geq 0$ and $-\ell\leq m\leq\ell$, and the function $u(\mathbf{y},\lambda)$ has the form
\[
u(\mathbf{y},\lambda^m_{\ell})=2\sqrt{\pi}\mathrm{i}^{\ell}j_{\ell}(\lambda^m_{\ell}\rho)
S^m_{\ell}(\theta_{\mathbf{y}},\varphi_{\mathbf{y}}).
\]

Similarly, in Theorems~\ref{th:4} and \ref{th:14}, where $K=\mathrm{O}(2)\times Z^c_2$, we use the \emph{Jacobi--Anger expansion}
\[
\mathrm{e}^{\mathrm{i}(\mathbf{p},\mathbf{y})}=J_0(\lambda\rho)
+2\sum^{\infty}_{\ell=1}\mathrm{i}^{\ell}J_{\ell}(\lambda\rho)
(\cos(\ell\varphi_{\mathbf{p}})\cos(\ell\varphi_{\mathbf{y}})
+\sin(\ell\varphi_{\mathbf{p}})\sin(\ell\varphi_{\mathbf{y}})),
\]
where $J_{\ell}$ are the Bessel functions of the first kind, $(\rho,\varphi_{\mathbf{y}})$ are the polar coordinates of the point $\mathbf{y}\in\mathbb{R}^2$, and $(\lambda,\varphi_{\mathbf{p}})$ are those of the point $\mathbf{p}\in\hat{\mathbb{R}}^2$. The set $\Lambda$ becomes the union of countably many copies of the half-line $[0,\infty)$ enumerated by integers.

Note that the random fields $\mathsf{C}(\mathbf{x})$ and $\mathrm{e}^{\mathrm{i}\varphi}\mathsf{C}(\mathbf{x})$ have the same two-point correlation tensor. Using this freedom, we can always force the random measure $\mathsf{Z}$ to become $\mathsf{V}$-valued rather than $\mathsf{V}^{\mathbb{C}}$-valued.

\section{Conclusions}

Hooke's law describes the physical phenomenon of \emph{elasticity} and belongs to the family of \emph{linear constitutive laws}, see \cite{Olive2013}. A physical quantity is a tensor of rank $p$ over $V$, that is, an element of the space $V^{\otimes p}$. Usually, physical quantities have symmetries. To describe symmetries mathematically, consider a subgroup $\Sigma$ of the symmetric group $\Sigma_p$ on $p$ symbols. Let $\tau$ be linear operator acting from $V^p$ to $V^{\otimes p}$ by
\[
\tau(\mathbf{x}_1,\dots,\mathbf{x}_p)=\mathbf{x}_1\otimes\cdots\otimes\mathbf{x}_p.
\]
The group $\Sigma$ acts linearly on $\tau(V^p)$ by permuting the positions of the factors in the tensor product:
\[
\sigma\cdot(\mathbf{x}_1\otimes\cdots\otimes\mathbf{x}_p)
=\mathbf{x}_{\sigma^{-1}(1)}\otimes\cdots\otimes\mathbf{x}_{\sigma^{-1}(p)}.
\]
This action can be extended by linearity to $V^{\otimes p}$. Define the linear operator $P_{\Sigma}\colon V^{\otimes p}\to V^{\otimes p}$ by
\[
P_{\Sigma}\mathsf{T}=\frac{1}{|\Sigma|}\sum_{\sigma\in\Sigma}\sigma\cdot\mathsf{T},
\]
where $|\Sigma|$ is the number of elements in $\Sigma$. The range of the operator $P_{\Sigma}$ is called the \emph{state tensor space}. A \emph{linear constitutive law} $C$ is a linear map between two state tensor spaces, say $\mathsf{V}_1$ and $\mathsf{V}_2$. It may be identified with an element of the tensor product $\mathsf{V}_1\otimes\mathsf{V}_2$, because the state tensor spaces inherit the Euclidean metric from $V$.

A linear constitutive law $C$ describes \emph{proper physics} or a single physical phenomenon if $\mathsf{V}_1=\mathsf{V}_2$ and $C$ is symmetric. Otherwise, $C$ describes \emph{coupled physics}, or a coupling between two different physics.

For example, Hooke's law corresponds to the case when $\mathsf{V}_1=\mathsf{V}_2=P_{\Sigma_2}V^{\otimes 2}$ and $C$ is symmetric. It describes the single physical phenomenon, elasticity. On the other hand, the \emph{photoelasticity tensor} is a general linear map $C\colon P_{\Sigma_2}V^{\otimes 2}\to P_{\Sigma_2}V^{\otimes 2}$. It couples two different physics and maps the space of strain tensors  to the space of the increments of dielectric tensors, see \cite{Forte19971317}. The \emph{piezoelectricity tensor} maps the space $P_{\Sigma_2}V^{\otimes 2}$ of strain tensors to the space $V$ of electric displacement vectors and couples two different physics, see \cite{Geymonat2002847}.

In general, a linear constitutive law is an element of a subspace of the tensor product $V^{\otimes(p+q)}$, where $p$ (resp. $q$) is the rank of tensors in the first (resp. second) state tensor space. Denote by $U$ the restriction of the representation $g\mapsto g^{\otimes(p+q)}$ to the above subspace. Consider $U$ as a group action. The orbit types of this action are called the classes of the phenomenon under consideration (e.g., photoelasticity classes, piezoelectricity classes and so on). All symmetry classes of all possible linear constitutive laws were described in \cite{Olive2013,MR3208052}.

For each class, one can consider its fixed point set $\mathsf{V}^{H}\subset V^{\otimes(p+q)}$, a group $K$ with $H\subseteq K\subseteq N(H)$, and the restriction $U$ of the representation $g\mapsto g^{\otimes(p+q)}$ of the group $K$ to $\mathsf{V}^{H}$. Calculating the general form of the one-point and two-point correlation tensors of the corresponding homogeneous and $(K,U)$-isotropic random field and the spectral expansion of the field in terms of stochastic integrals with respect to orthogonal scattered random measures is an interesting research question.

The part of the above question concerning the one-point correlation tensor is almost trivial: it is any tensor lying in the direct sum of all one-dimensional subspaces of $\mathsf{V}^{H}$ where the copies of the trivial representations of $K$ live. To find the general form of the two-point correlation tensor, we need to describe all measurable functions that map $\hat{\mathsf{V}}$ to the set of all Hermitian nonnegative-definite operators on the complexification of the space $\mathsf{V}^{H}$ satisfying the following conditions:
\begin{equation}\label{eq:1}
\begin{aligned}
f(g\mathbf{p})&=(U\otimes U)(g)f(\mathbf{p}),\qquad g\in K,\\
f(g\mathbf{p})&=V(g)f(\mathbf{p}),\qquad g\in Z^c_2.
\end{aligned}
\end{equation}
The first condition easily follows from the very definition of the $(K,U)$-isotropic random field. It is well known that if the random field
under consideration takes values in $\mathsf{V}^H$, then $f(-\mathbf{p})=f(\mathbf{p})^{\top}$. This condition is equivalent to the second condition in \eqref{eq:1}, where $V$ is the direct sum of $\dim\mathsf{S}^2(\mathsf{V}^{H})$ copies of the trivial representation $A_g$ of the group $Z^c_2$ and $\dim\mathsf{\Lambda}^2(\mathsf{V}^{H})$ copies of its non-trivial representation $A_u$.

Consider the three possible cases.

\begin{enumerate}
  \item $K$ is a subgroup of the group $\mathrm{SO}(3)$. In this case \eqref{eq:1} is equivalent to the following condition:
      \[
      f(g\mathbf{p})=[\mathsf{S}^2(U)\otimes A_g\oplus\mathsf{\Lambda}^2(U)\otimes A_u](g)f(\mathbf{p}),\qquad g\in K\times Z^c_2.
      \]
  \item $K$ is a subgroup of $\mathrm{O}(3)$ containing $-I$. As we have seen in proof of Theorem~\ref{th:0}, \eqref{eq:1} is equivalent to
      \[
      f(g\mathbf{p})=\mathsf{S}^2(U)(g)f(\mathbf{p}),\qquad g\in K.
      \]
  \item $K$ is neither a subgroup of the group $\mathrm{SO}(3)$ nor contains $-I$. Both conditions in \eqref{eq:1} must be treated separately.
\end{enumerate}

An example of the second case has been considered here. The remaining cases will be treated elsewhere.

There are two principal uses of the results obtained here. The first one is
to model and simulate any statistically wide-sense homogeneous and
isotropic, linear hyperelastic, random medium. One example is a polycrystal
made of grains belonging to a specific crystal class, while another example
is a mesoscale continuum defined through upscaling of a random material on
scales smaller than the RVE; if the upscaling is conducted on the RVE level,
there is no spatial randomness and the continuum model is deterministic.
Here one would proceed in the following steps:

\begin{itemize}
  \item for a given microstructure, determine the one- and two-point statistics using some experimental and/or image-based computational methods;
  \item calibrate the entire correlation structure of the elasticity TRF;
  \item simulate the realisations of this TRF.
\end{itemize}

The second application of our results is their use as input of a random mesoscale continuum (Fig.~\ref{fig:1}(c)) into stochastic field equations such as SPDEs and SFEs.

\bibliographystyle{plain}

\bibliography{\jobname}

\end{document}